\documentclass[11pt,a4paper]{article}
\usepackage{jheppub}

\usepackage{amsthm,amsbsy,amsfonts,mathrsfs,enumerate,float,wrapfig,amsmath}
\usepackage[utf8]{inputenc}
\usepackage[outline]{contour}
\usepackage{pdflscape}
\usepackage{appendix}
\DeclareUnicodeCharacter{2212}{-}
\usepackage{comment}

\newcommand{\be}{\begin{equation}}
\newcommand{\ee}{\end{equation}}
\newcommand{\bea}{\begin{eqnarray}}
\newcommand{\eea}{\end{eqnarray}}
\newcommand{\ba}{\begin{align}}
\newcommand{\ea}{\end{align}}

\def\K{{\scriptscriptstyle \mathrm{K3}}}
\def\Os{{\scriptscriptstyle \mathrm{O7}}}

\def\Osi{{\scriptscriptstyle \mathrm{O7^i}}}
\def\Dsi{{\scriptscriptstyle \mathrm{D7_i}}}

\def\W{{\scriptscriptstyle \mathrm{W}}}
\def\Ot{{\scriptscriptstyle \mathrm{O3}}}
\def\Ds{{\scriptscriptstyle \mathrm{D7}}}
\def\Dt{{\scriptscriptstyle \mathrm{D3}}}
\def\Et{{\scriptscriptstyle \mathrm{E3}}}
\def\Kk{{\scriptscriptstyle \mathrm{d}}}

\usepackage{xcolor,circuitikz}
\usetikzlibrary{decorations.pathreplacing}

\usepackage{amsmath}
\usepackage{amssymb}
\usepackage{MnSymbol}

\usepackage{array}
\newcolumntype{L}[1]{>{\raggedright\let\newline\\\arraybackslash\hspace{0pt}}m{#1}}
\newcolumntype{C}[1]{>{\centering\let\newline\\\arraybackslash\hspace{0pt}}m{#1}}
\newcolumntype{R}[1]{>{\raggedleft\let\newline\\\arraybackslash\hspace{0pt}}m{#1}}

\usepackage{tikz}
\usetikzlibrary{decorations.pathmorphing}
\usetikzlibrary{decorations.markings}
\usetikzlibrary{decorations.pathreplacing}
\usetikzlibrary{shapes, shapes.geometric, shapes.symbols, shapes.multipart, shapes.callouts, shapes.misc}
\tikzset{snake it/.style={decorate, decoration=snake}}
\tikzset{7brane/.style={circle, draw=black, fill=black,ultra thick,inner sep=1 pt, minimum size=1 pt,}, c/.default={4pt}}
\tikzset{cross/.style={cross out, draw=black,thick, minimum size=2*(#1-\pgflinewidth), inner sep=0pt, outer sep=0pt}, cross/.default={5pt}}
\tikzset{big7brane/.style={circle, draw=black, fill=black,ultra thick,inner sep=2.5 pt, minimum size=1 pt,}, c/.default={4pt}}
\tikzset{u/.style={circle, draw=black, fill=white,inner sep=2 pt, minimum size=2 pt,},f/.style={square, draw=black, fill=white,ultra thick,inner sep=4 pt, minimum size=2 pt,}}
\tikzset{so/.style={circle, draw=black, fill=red, thick,inner sep=2 pt, minimum size=2 pt,},f/.style={square, draw=black, fill=white,ultra thick,inner sep=4 pt, minimum size=2 pt,}}
\tikzset{sp/.style={circle, draw=black, fill=blue,thick,inner sep=2 pt, minimum size=2 pt,},f/.style={square, draw=black, fill=white,ultra thick,inner sep=4 pt, minimum size=2 pt,}}
\tikzset{uf/.style={rectangle, draw=black, fill=white,inner sep=2.75 pt, minimum size=4 pt,}}
\tikzset{spf/.style={rectangle, draw=black, fill=blue, thick,inner sep=2.5 pt, minimum size=4 pt,}}
\tikzset{sof/.style={rectangle, draw=black, fill=red, thick,inner sep=2.5 pt, minimum size=4 pt,}}
\usetikzlibrary{positioning}
\usepackage{mathtools}
\usepackage{multirow}
\tikzset{hasse/.style={circle, fill,inner sep=2pt}}

\usepackage{todonotes}
\usepackage{xstring}
\usepackage{afterpage}
\usepackage{placeins}
\usepackage{adjustbox}
\usepackage{caption}
\usepackage{subcaption}
\title{New Type IIB Poly-instanton Effects and Axion Quintessence}

\author[a,b]{Luca Caraffi\,}
\author[b]{,\,Federico Carta\,}
\author[a,b]{,\,Michele Cicoli\,}

\affiliation[a]{Dipartimento di Fisica e Astronomia, Universit\`a di Bologna, via Irnerio 46, Bologna, 40126, Italy}
\affiliation[b]{
INFN, Sezione di Bologna, viale Berti Pichat 6/2, Bologna, 40127, Italy}

\emailAdd{luca.caraffi2@unibo.it}
\emailAdd{federico.carta@bo.infn.it}
\emailAdd{michele.cicoli@unibo.it}

\abstract{Previous studies of poly-instantons in type IIB Calabi-Yau orientifolds found that these effects arise when Euclidean D3-branes wrap rigid cycles with a Wilson line. In this work, we extend this result through a general analysis of instanton zero modes, revealing the existence of new poly-instanton effects for rigid cycles with a deformation, provided that O3-planes are localised on the worldvolume of the Euclidean D3-brane. We provide concrete Calabi-Yau examples with explicit orientifold involution and brane setup. One of them is particularly well suited for realising axion hilltop quintessence. In particular, the double exponential suppression of poly-instanton effects allows to reproduce the observed vacuum energy for natural values of the underlying parameters.}

\begin{document}

\maketitle

\clearpage

\clearpage

\section{Introduction}

Non-perturbative corrections to the low-energy effective action of string compactifications have received considerable attention over the years (see \cite{Blumenhagen:2009qh} for a comprehensive review). In particular, corrections to the 4D superpotential have important implications for moduli stabilisation and phenomenology due to their ability to generate the hierarchies we observe in Nature via exponentially small numbers.

Following the seminal paper \cite{Witten:1996bn}, most of the work has been focused on contributions to the 4D superpotential $W$ arising from Euclidean brane instantons, depending on their fermionic zero modes. The best case involves Euclidean brane instantons wrapping rigid divisors which can yield standard single instanton corrections to $W$. Divisors with additional zero modes from deformations or Wilson lines can instead give rise to non-perturbative corrections to the action of another instanton, resulting in a doubly exponentially suppressed correction to the superpotential.

The need to include these non-perturbative effects, seen as instanton corrections to the gauge kinetic function, has been pointed out for the first time in \cite{Akerblom:2007uc}. Subsequently, ref. \cite{Blumenhagen:2008ji} performed the first computation of these effects in type I toroidal orbifolds with E1-instantons, and named them \emph{poly-instantons}. Poly-instantons have also been shown to manifest as E(-1) corrections to gauge couplings in 8D type IIB orientifolds in \cite{Petersson:2010qu}. Ref. \cite{Blumenhagen:2012kz} analysed then the fermionic zero modes needed to generate poly-instantons in 4D type IIB compactifications on Calabi-Yau (CY) orientifolds involving O7-planes but no O3-planes on the E3 worldvolume. The required configuration involves E3-branes wrapping divisors with a Wilson line but no deformations. Explicit examples of poly-instanton corrections for Wilson divisors in type IIB CY orientifolds with O3-planes on the E3 worldvolume have been then provided in \cite{Lust:2013kt}.

These constructions are particularly promising for phenomenology since type IIB flux compactification is the framework where moduli stabilisation is best understood. Indeed, poly-instanton effects have been widely exploited in phenomenological and cosmological applications of type IIB string compactifications:
\begin{itemize}
\item \textbf{Moduli stabilisation for GUT scenarios:} Ref. \cite{Blumenhagen:2008kq} focused on cases where the flux superpotential vanishes, and used a racetrack with poly-instanton corrections to generate an exponentially small effective $W$ suitable for stabilising the moduli in GUT scenarios with low-energy supersymmetry (a similar moduli stabilisation mechanism based on poly-instantons has been used in \cite{Higaki:2012ba}). 

\item \textbf{Anisotropic compactifications with large extra dimensions:} As shown in \cite{Cicoli:2011yy}, the huge suppression of poly-instanton corrections allows to obtain anisotropic CY compactifications with 2 large extra dimensions and a string scale around the TeV range. Soft supersymmetry breaking for these models has been studied in \cite{Angus:2012dd}. 

\item \textbf{Inflationary models:} The double exponential suppression of poly-instanton corrections is ideal to generate an inflationary plateau for fibred CY models \cite{Cicoli:2011ct}. The first explicit examples of K3-fibred CYs with del Pezzo divisors have been built in \cite{Cicoli:2011it}. Moduli stabilisation and inflation from poly-instantons on Wilson divisors has been instead analysed in \cite{Blumenhagen:2012ue} and then in \cite{Gao:2013rra} for the general case of fluxed instantons with odd moduli, while ref. \cite{Kobayashi:2017jeb} studied poly-instanton axion inflation.

\item \textbf{Modulated reheating and non-gaussianities:} As shown in \cite{Cicoli:2012cy}, poly-instantons can yield non-standard mechanisms for the generation of the primordial density fluctuations with associated large non-gaussianities. In fact, the inflaton coupling to visible sector degrees of freedom, and hence reheating, can be modulated by the oscillations of a field which behaves as a spectator during inflation due to its lightness caused by the fact that its potential is generated by tiny poly-instantons. Non-gaussianities in two-field poly-instanton inflation have also been discussed in \cite{Gao:2013hn}.

\item \textbf{Dark matter and astrophysical lines:} Poly-instantons can explain the small mass ($m\simeq 7$ keV) of an axion dark matter particle decaying into a pair of massless axions which then convert into photons in the magnetic field of galaxy clusters, giving a $3.5$ KeV line \cite{Cicoli:2017zbx}. Moreover, poly-instantons might help to realise ultra-light axions which can play the role of fuzzy dark matter \cite{Cicoli:2021gss}.
    
\item \textbf{Quintessence models:} As stressed in \cite{Cicoli:2021skd}, axions are the best candidates to drive quintessence since they do not mediate any fifth force and their potential is radiatively stable since they enjoy a perturbatively exact shift symmetry. Moreover, due to their double exponential suppression, poly-instanton effects are particularly well suited for yielding axionic potentials with low energy scale. This leads to a natural embedding of axion hilltop quintessence. Indeed, fibred CY models can provide a consistent history of the Universe from inflation to quintessence, where inflation is driven by a K\"ahler modulus as in \cite{Cicoli:2008gp}, while dynamical dark energy is caused by an axion sitting close to the maximum of its potential generated by poly-instanton effects \cite{Cicoli:2024yqh}. Note that poly-instantons have been exploited also to realise saxion quintessence, still for the case of K3-fibred CY compactifications with stabilised moduli \cite{Cicoli:2012tz}. 
\end{itemize}

As can be seen from this very brief review, poly-instanton effects have been argued to lead to promising phenomenological applications, not just when they arise from E3-branes wrapping Wilson divisors, but also when they involve K3 divisors. A practical realisation of this case is provided in \cite{Lust:2013kt} where the E3-brane wraps a Wilson divisor whose volume, however, coincides with the one of a K3 surface when a diagonal del Pezzo divisor is collapsed to zero size.

In this paper, we shall instead take a more general approach, and extend the analysis of \cite{Blumenhagen:2012kz} by focusing on type IIB CY orientifolds with both O7- and O3-planes on the E3 worldvolume, and see which zero mode configuration can lead to non-zero poly-instanton contributions to the superpotential. By means of a generalised version of the Lefschetz fixed-point theorem, we shall show indeed that poly-instantons can arise from E3-branes wrapping divisors which can admit either a Wilson line or a deformation. In the case of E3-branes wrapping a deformation divisor, we shall however make use of a theorem by Nikulin \cite{Nikulin1980,Nikulin1983} to prove that this divisor cannot be a K3 surface. Ultimately, we find that the emergence of poly-instantons in the superpotential depends on two quantities: ($i$) the number of O3-planes lying on the divisor $D_\Et$ wrapped by the E3-brane; ($ii$) the intersection number $k_{\Et\Et\Os}$ which characterises the intersection between $D_\Et$ and the O7-plane.

In light of these general results, when an E3-brane wraps a Wilson divisor, it is not guaranteed that non-zero poly-instantons are generated. On the other hand, poly-instanton corrections to $W$ could arise from deformations divisors which are blow-ups of K3 surfaces, or share the same Hodge numbers as a K3 divisor but feature a non-zero first Chern class.

We illustrate our findings in four explicit examples from the Kreuzer-Skarke (KS) database \cite{Kreuzer:2000xy} where CY threefolds are built as hypersurfaces in toric ambient spaces. On top of a concrete CY threefold, each example features an explicit orientifold involution, brane setup and gauge fluxes which satisfy global consistency conditions like D7-tadpole cancellation and the absence of Freed-Witten (FW) anomalies \cite{Freed:1999vc}. Moreover, we analyse the generation of both single and poly-instanton corrections to the superpotential by checking in detail all fermionic zero modes and their possible lifting by gauge fluxes. Two examples show that the presence of O3-planes forbids the emergence of non-zero poly-instanton from E3-branes wrapped around Wilson and K3-like divisors. On the other hand, two examples allow for non-zero poly-instanton effects for E3-branes wrapped around K3-like and blown-up K3 divisors. 

Among the two explicit examples exhibiting new poly-instanton effects, we analyse the application of one of them to late time cosmology. Assuming, as typical in the type IIB framework, that the axio-dilaton and the complex structure moduli are fixed by $3$-form fluxes, we study in detail the stabilisation of all K\"ahler moduli. We find an LVS-like minimum \cite{Balasubramanian:2005zx} inside the K\"ahler cone and in a region where the effective field theory is under computational control. Interestingly, the double exponential suppression of poly-instanton effects allows the potential of one of the $C_4$-axions to reproduce the observed value of the cosmological constant for natural values of all microscopic parameters. Moreover, the decay constant of the canonically normalised axion is around the GUT scale. This leads to an explicit realisation of axion hilltop quintessence \cite{Cicoli:2024yqh} where the current accelerated expansion of the Universe is driven by an axion slowly evolving around the maximum of its potential. 

This paper is organised as follows. Sec. \ref{sec:sec2} is a concise review of E3-brane instanton and poly-instanton effects in type IIB CY orientifold compactifications. Sec. \ref{sec:Hol_Lef_FP_Theor} exploits a generalised version of the Lefschetz fixed point theorem to derive the possible zero mode configurations which lead to non-zero poly-instantons in the case when the E3 worldvolume contains both O7- and O3-planes. In Sec. \ref{Sec3} we discuss four explicit CY constructions of poly-instanton effects, and Sec. \ref{SecQuint} shows that one of them is well suited for realising axion hilltop quintessence. We present our conclusions in Sec. \ref{Concl}, while App. \ref{sec:Zero_Modes} provides a review of instanton fermionic zero modes and App. \ref{app:A_Kreuzer-Skarke} gives a short introduction to the tools to build CY threefolds via toric geometry.

\section{Review of E3-brane non-perturbative effects}
\label{sec:sec2}

\subsection{Type IIB Calabi-Yau orientifolds with O3/O7-planes}
\label{sec:IIB_Ori}

In this section we provide a brief review of type IIB CY orientifold compactifications with O3/O7-planes, setting the notation used throughout this paper. We therefore focus on type IIB string theory compactified on $\mathbb{R}^{1,3}\times X$, where $X$ is a compact CY threefold, namely a compact K\"ahler algebraic variety of complex dimension $3$ with vanishing canonical bundle $K_X=0$. We shall denote the CY volume as $\mathcal{V}$. The compactification ansatz leads to an effective 4D $\mathcal{N}=2$ theory at energies below the Kaluza-Klein scale $M_{\rm KK}\simeq M_p/\mathcal{V}^{2/3}$. More realistic 4D $\mathcal{N}=1$ models are obtained after orientifolding $X$ via a combination of worldsheet parity, fermionic parity and an isometric holomorphic $\mathbb{Z}_2$ involution $\sigma: X \to X$. There are two classes of orientifolds. In the following, we restrict to the case where the pullback of $\sigma$ on the K\"ahler form $J$ and the holomorphic $(3,0)$-form $\Omega$ is:
\begin{equation}
\sigma^*J=J\qquad\text{and}\qquad \sigma^*\Omega=-\,\Omega\,.
\end{equation}
The orientifold involution induces a splitting in the CY Dolbeault cohomology:
\begin{equation}
H^{p,q}(X,\mathbb{Z})=H^{p,q}_{+}(X,\mathbb{Z})\oplus H^{p,q}_{-}(X,\mathbb{Z})\,,
\end{equation}
where the basis elements of $H^{p,q}_{\pm}(X,\mathbb{Z})$ are $(p,q)$-forms which are eigenvectors of $\sigma^*$ with eigenvalues respectively $\pm 1$. Furthermore, the involution $\sigma$ induces a similar splitting in the Dolbeault cohomology of each divisor and curve of $X$, both those inherited from the ambient space $V$ and those which are autochthonous to $X$. 

The fixed locus of $\sigma$, which we will denote as $I_{\sigma}$, is a union of complex codimension $1$ loci ($4$-cycles) and codimension $3$ loci (points) in $X$. The former are called O7-planes, while the latter are O3-planes. There are two possible variants of O7-planes, respectively denoted by O7$^+$ or O7$^-$, depending on whether they carry positive or negative RR-charge. Similarly, there are four variants of O3-planes, denoted by O3$^{\pm}$ and $\tilde{O3}^{\pm}$, depending on whether they carry positive or negative RR-charge, and vanishing or non-vanishing discrete torsion with respect to the $B$-field and RR $C_2$-form \cite{Witten:1998xy}. In the following, we shall focus only on O7$^-$-planes and O3$^-$-planes. Therefore we shall simply talk about O7- and O3-planes, dropping the superscripts `minus'.

The O7$^-$-planes carry negative RR charge. In particular, in the convention where the positive RR-charge of any D-brane in the double-cover CY is $+1$, the O7-planes carry RR-charge $-8$, and the O3-planes carry fractional charge $-1/2$.\footnote{For an explanation of why fractional RR-charges of O-planes are compatible with Dirac quantisation, see \cite{Tachikawa:2018njr}.}  The existence of orientifold planes in the compact space $X$ causes a RR $C_8$-form tadpole. Indeed, if an O7-plane is wrapped around a divisor $D_\Os$ of $X$, tadpole cancellation requires the presence of D7-branes with opposite RR charge. The geometrical position and the number of D7-branes needed to cancel the tadpole can be derived from:
\begin{equation}
\sum_a N_a\,\left(\hat{D}_a+\hat{D}_a'\right)=8\, \hat{D}_{\Os}\, .
\label{eq:tadpole}
\end{equation}
In \eqref{eq:tadpole}, $N_a$ is the number D7-branes of the $a$-th stack wrapping the divisor $D_a$, $D_a'$ is its orientifold image, and the hat denotes the Poincaré dual $2$-forms. The simplest way to satisfy (\ref{eq:tadpole}) is to have all D7-branes on top of the O7-plane, so that the D7-tadpole is cancelled locally for $N_a=4$. In all the examples presented in this paper, the D7-tadpole will always be cancelled locally. 

On top of the D7-tadpole, there is also a D3-tadpole: 
\begin{equation}
N_\Dt+N_{\Dt'}+ N_{\rm flux} = \frac12\,N_\Ot +\frac16\,\chi(D_\Os)- \sum_a N_a \left(Q_\Ds^a+Q_{\Ds'}^a\right)\,,
\label{eq:D3_Tadpole}
\end{equation}
where the D3-charge of $3$-form background fluxes is: 
\begin{equation}
N_{\rm flux}=\int_X H_3\wedge F_3\,,
\end{equation}
while the D3-charge of a D7-brane wrapping the divisor $D_a$ is given in terms of the Euler characteristic $\chi_0$ of the smoothened divisor $D_a$ and the gauge invariant worldvolume flux $\mathcal{F}_a= F_a-\iota^*_{D_a}\,B$ as:
\begin{equation}
Q_{\Ds}^a = -\frac{\chi_0(D_a)}{24}-\frac12\int_{D_a} \mathcal{F}_a\wedge \mathcal{F}_a\,.
\end{equation}
Under the assumptions that the divisor is smooth, the D7-tadpole is cancelled locally and all worldvolume fluxes are switched off, \eqref{eq:D3_Tadpole} simplifies to:
\begin{equation}
N_{\Dt}+N_{\Dt'} + N_{\rm flux}=\frac12\,N_{\Ot} +\frac12\,\chi(D_{\Os})\,.
\label{QD3}
\end{equation}
Let us comment a bit more on the gauge flux $F_a$ on the worldvolume of the D7-branes wrapping the divisor $D_a$, which needs to satisfy the following quantisation condition to cancel the FW anomaly \cite{Freed:1999vc}:
\begin{equation}
F_a+\frac12\,c_1(K_{D_a})\,\in\, H^2(D_a,\mathbb{Z})\,,
\label{eq:FW}
\end{equation}
where $K_{D_a}$ is the canonical bundle of the divisor $D_a$, and $c_1(K_{D_a})$ is the first Chern class of the line bundle $K_{D_a}$. In particular, in our case we have $c_1(K_{D_a}) = -\iota^*_{D_a}\hat{D}_a$, implying that $F_a$ has to include a half-integer contribution.

Note that the FW anomaly cancellation condition holds also for the worldvolume theory of E3-brane instantons wrapped around a divisor. In what follows we shall be interested in E3-brane configurations with vanishing worldvolume flux $\mathcal{F}_{\Et}$ which can lead to a non-zero $O(1)$ instanton contribution to the superpotential $W$. If the divisor wrapped by the E3 is non-spin, this can be compatible with FW anomaly cancellation only if the pull-back of the $B$-field cancels the half-integer contribution in $F_\Et$. Note that FW anomaly cancellation can potentially affect also the presence of chiral zero modes at the intersection between a D7-stack and an E3-brane, which can kill the instanton contribution to $W$. This issue has to be checked in detail case by case in explicit examples.

\subsection{E3-brane instantons and poly-instantons}
\label{sec:Inst_Poly_Inst}

Having discussed some of the basic ingredients for the geometrical construction of type IIB CY orientifold compactifications, we now move to another crucial ingredient: Euclidean D3-branes. These objects are D3-branes whose worldvolume is completely space-like: they wrap $4$-cycles of the internal space $X$ and are point-like in the effective 4D spacetime. E3-branes behave in the 4D effective theory as instantons. When the same $4$-cycle wrapped by an E3-brane is also wrapped by a stack of D7-branes, the 4D instanton is simply a Yang-Mills instanton for the non-Abelian gauge theory on the D7 worldvolume. This is a stringy realisation of the ADHM construction \cite{Douglas:1995bn}. On the other hand, the E3-brane may also wrap a $4$-cycle that is not wrapped by any D7-branes. The corresponding 4D instanton lacks then a gauge description, and is therefore called a \emph{stringy instanton} \cite{Ibanez:2007rs, Blumenhagen:2009qh}.

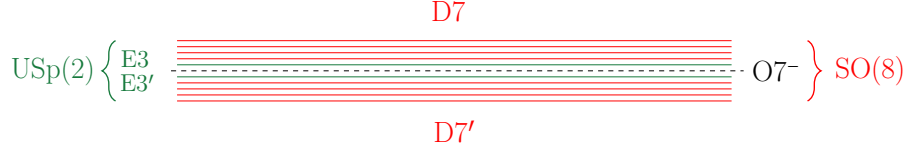
\begin{figure}[!ht]
\centering
\resizebox{0.8\textwidth}{!}{
\begin{circuitikz}
\tikzstyle{every node}=[font=\fontsize{14.8pt}{19.2pt}\selectfont]

\draw [dashed] (21.25,9.875) -- (33.125,9.875);
\node [font=\fontsize{18.2pt}{23.7pt}\selectfont, inner xsep=0.080cm, inner ysep=0.085cm, rounded corners=0.020cm] at (33.8,9.875) {O7$^-$};
\draw [ color={rgb,255:red,23; green,122; blue,61}, draw opacity=1, short] (21.375,10) -- (32.875,10);
\draw [ color={rgb,255:red,255; green,0; blue,0}, draw opacity=1, short] (21.375,10.125) -- (32.875,10.125);
\draw [ color={rgb,255:red,255; green,0; blue,0}, draw opacity=1, short] (21.375,10.25) -- (32.875,10.25);
\draw [ color={rgb,255:red,255; green,0; blue,0}, draw opacity=1, short] (21.375,10.375) -- (32.875,10.375);
\draw [ color={rgb,255:red,23; green,122; blue,61}, draw opacity=1, short] (21.375,9.75) -- (32.875,9.75);
\draw [ color={rgb,255:red,255; green,0; blue,0}, draw opacity=1, short] (21.375,9.625) -- (32.875,9.625);
\draw [ color={rgb,255:red,255; green,0; blue,0}, draw opacity=1, short] (21.375,9.5) -- (32.875,9.5);
\draw [ color={rgb,255:red,255; green,0; blue,0}, draw opacity=1, short] (21.375,9.375) -- (32.875,9.375);
\node [font=\fontsize{18.2pt}{23.7pt}\selectfont, color={rgb,255:red,255; green,0; blue,0}, text opacity=1, fill={rgb,255:red,255; green,255; blue,255}, fill opacity=1, inner xsep=0.080cm, inner ysep=0.085cm, rounded corners=0.020cm] at (27,11.125) {D7};
\node [font=\fontsize{18.2pt}{23.7pt}\selectfont, color={rgb,255:red,255; green,0; blue,0}, text opacity=1, fill={rgb,255:red,255; green,255; blue,255}, fill opacity=1, inner xsep=0.080cm, inner ysep=0.085cm, rounded corners=0.020cm] at (27.125,8.625) {D7$'$};
\node [font=\fontsize{14.8pt}{19.2pt}\selectfont, color={rgb,255:red,23; green,122; blue,61}, text opacity=1, fill={rgb,255:red,255; green,255; blue,255}, fill opacity=1, inner xsep=0.080cm, inner ysep=0.085cm, rounded corners=0.020cm] at (20.5,10.125) {E3};
\node [font=\fontsize{18.2pt}{23.7pt}\selectfont, color={rgb,255:red,23; green,122; blue,61}, text opacity=1, fill={rgb,255:red,255; green,255; blue,255}, fill opacity=1, inner xsep=0.080cm, inner ysep=0.085cm, rounded corners=0.020cm] at (18.8,9.875) {USp(2)};
\node [font=\fontsize{18.2pt}{23.7pt}\selectfont, color={rgb,255:red,255; green,0; blue,0}, text opacity=1, fill={rgb,255:red,255; green,255; blue,255}, fill opacity=1, inner xsep=0.080cm, inner ysep=0.085cm, rounded corners=0.020cm] at (35.75,9.875) {SO(8)};
\draw [ color={rgb,255:red,255; green,0; blue,0}, draw opacity=1, short] (21.375,10.5) -- (32.875,10.5);
\draw [ color={rgb,255:red,255; green,0; blue,0}, draw opacity=1, short] (21.375,9.25) -- (32.875,9.25);
\node [font=\fontsize{14.8pt}{19.2pt}\selectfont, color={rgb,255:red,23; green,122; blue,61}, text opacity=1, fill={rgb,255:red,255; green,255; blue,255}, fill opacity=1, inner xsep=0.080cm, inner ysep=0.085cm, rounded corners=0.020cm] at (20.55,9.625) {E3$'$};

\draw [decorate, decoration={brace, amplitude=8pt}, color={rgb,255:red,23; green,122; blue,61}, thick] (20.1,9.25) -- (20.1,10.5);

\draw [decorate, decoration={brace, amplitude=8pt}, color={rgb,255:red,255; green,0; blue,0}, thick] (34.45,10.5) -- (34.45,9.25);

\end{circuitikz}
}
\caption{$USp(2)$ instanton configuration.}
\label{fig:USp(2)_Setup}
\end{figure}

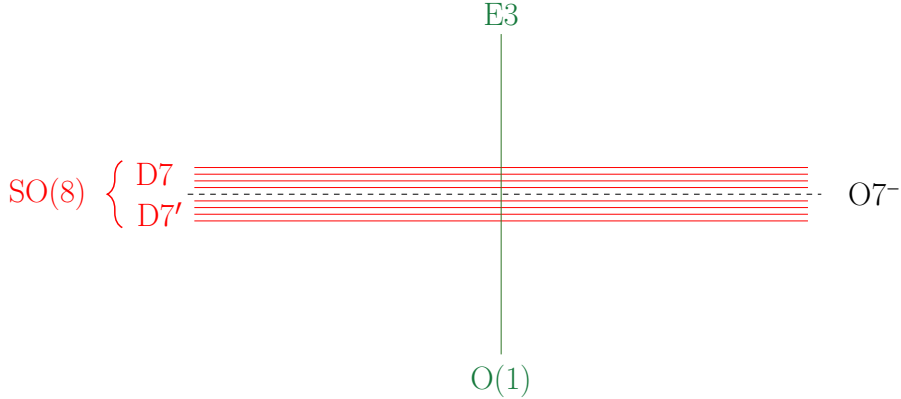
\begin{figure}[!ht]
\centering
\resizebox{0.8\textwidth}{!}{
\begin{circuitikz}
\tikzstyle{every node}=[font=\fontsize{14.8pt}{19.2pt}\selectfont]

\draw [dashed] (3.625,9.875) -- (15.5,9.875);
\node [font=\fontsize{18.2pt}{23.7pt}\selectfont, inner xsep=0.080cm, inner ysep=0.085cm, rounded corners=0.020cm] at (16.5,9.875) {O7$^-$};
\draw [ color={rgb,255:red,255; green,0; blue,0}, draw opacity=1, short] (3.75,10) -- (15.25,10);
\draw [ color={rgb,255:red,255; green,0; blue,0}, draw opacity=1, short] (3.75,10.125) -- (15.25,10.125);
\draw [ color={rgb,255:red,255; green,0; blue,0}, draw opacity=1, short] (3.75,10.25) -- (15.25,10.25);
\draw [ color={rgb,255:red,255; green,0; blue,0}, draw opacity=1, short] (3.75,10.375) -- (15.25,10.375);
\draw [ color={rgb,255:red,255; green,0; blue,0}, draw opacity=1, short] (3.75,9.75) -- (15.25,9.75);
\draw [ color={rgb,255:red,255; green,0; blue,0}, draw opacity=1, short] (3.75,9.625) -- (15.25,9.625);
\draw [ color={rgb,255:red,255; green,0; blue,0}, draw opacity=1, short] (3.75,9.5) -- (15.25,9.5);
\draw [ color={rgb,255:red,255; green,0; blue,0}, draw opacity=1, short] (3.75,9.375) -- (15.25,9.375);
\node [font=\fontsize{18.2pt}{23.7pt}\selectfont, color={rgb,255:red,255; green,0; blue,0}, text opacity=1, fill={rgb,255:red,255; green,255; blue,255}, fill opacity=1, inner xsep=0.080cm, inner ysep=0.085cm, rounded corners=0.020cm] at (3,10.25) {D7};
\node [font=\fontsize{18.2pt}{23.7pt}\selectfont, color={rgb,255:red,255; green,0; blue,0}, text opacity=1, fill={rgb,255:red,255; green,255; blue,255}, fill opacity=1, inner xsep=0.080cm, inner ysep=0.085cm, rounded corners=0.020cm] at (3.10,9.5) {D7$'$};
\draw [ color={rgb,255:red,46; green,117; blue,37}, draw opacity=1, short] (9.5,6.875) -- (9.5,12.875);
\node [font=\fontsize{18.2pt}{23.7pt}\selectfont, color={rgb,255:red,23; green,122; blue,61}, text opacity=1, fill={rgb,255:red,255; green,255; blue,255}, fill opacity=1, inner xsep=0.080cm, inner ysep=0.085cm, rounded corners=0.020cm] at (9.5,13.25) {E3};
\node [font=\fontsize{18.2pt}{23.7pt}\selectfont, color={rgb,255:red,23; green,122; blue,61}, text opacity=1, fill={rgb,255:red,255; green,255; blue,255}, fill opacity=1, inner xsep=0.080cm, inner ysep=0.085cm, rounded corners=0.020cm] at (9.5,6.375) {O(1)};
\node [font=\fontsize{18.2pt}{23.7pt}\selectfont, color={rgb,255:red,255; green,0; blue,0}, text opacity=1, fill={rgb,255:red,255; green,255; blue,255}, fill opacity=1, inner xsep=0.080cm, inner ysep=0.085cm, rounded corners=0.020cm] at (1,9.875) {SO(8)};

\draw [decorate, decoration={brace, amplitude=8pt}, color={rgb,255:red,255; green,0; blue,0}, thick] (2.4,9.25) -- (2.4,10.5);

\end{circuitikz}
}
\caption{$O(1)$ instanton configuration.}
\label{fig:O(1)_Setup}
\end{figure}

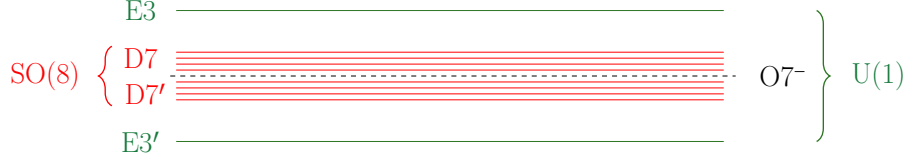
\begin{figure}[!ht]
\centering
\resizebox{0.8\textwidth}{!}{
\begin{circuitikz}
\tikzstyle{every node}=[font=\fontsize{14.8pt}{19.2pt}\selectfont]

\draw [dashed] (3.625,9.875) -- (15.5,9.875);
\node [font=\fontsize{18.2pt}{23.7pt}\selectfont, inner xsep=0.080cm, inner ysep=0.085cm, rounded corners=0.020cm] at (16.5,9.875) {O7$^-$};

\draw [ color={rgb,255:red,255; green,0; blue,0}, draw opacity=1, short] (3.75,10) -- (15.25,10);
\draw [ color={rgb,255:red,255; green,0; blue,0}, draw opacity=1, short] (3.75,10.125) -- (15.25,10.125);
\draw [ color={rgb,255:red,255; green,0; blue,0}, draw opacity=1, short] (3.75,10.25) -- (15.25,10.25);
\draw [ color={rgb,255:red,255; green,0; blue,0}, draw opacity=1, short] (3.75,10.375) -- (15.25,10.375);

\draw [ color={rgb,255:red,255; green,0; blue,0}, draw opacity=1, short] (3.75,9.75) -- (15.25,9.75);
\draw [ color={rgb,255:red,255; green,0; blue,0}, draw opacity=1, short] (3.75,9.625) -- (15.25,9.625);
\draw [ color={rgb,255:red,255; green,0; blue,0}, draw opacity=1, short] (3.75,9.5) -- (15.25,9.5);
\draw [ color={rgb,255:red,255; green,0; blue,0}, draw opacity=1, short] (3.75,9.375) -- (15.25,9.375);

\node [font=\fontsize{18.2pt}{23.7pt}\selectfont, color={rgb,255:red,255; green,0; blue,0}, text opacity=1, fill={rgb,255:red,255; green,255; blue,255}, fill opacity=1, inner xsep=0.080cm, inner ysep=0.085cm, rounded corners=0.020cm] at (3,10.25) {D7};
\node [font=\fontsize{18.2pt}{23.7pt}\selectfont, color={rgb,255:red,255; green,0; blue,0}, text opacity=1, fill={rgb,255:red,255; green,255; blue,255}, fill opacity=1, inner xsep=0.080cm, inner ysep=0.085cm, rounded corners=0.020cm] at (3.10,9.5) {D7$'$};

\draw [ color={rgb,255:red,46; green,117; blue,37}, draw opacity=1, short] (3.75,11.25) -- (15.25,11.25);
\node [font=\fontsize{18.2pt}{23.7pt}\selectfont, color={rgb,255:red,23; green,122; blue,61}, text opacity=1, fill={rgb,255:red,255; green,255; blue,255}, fill opacity=1, inner xsep=0.080cm, inner ysep=0.085cm, rounded corners=0.020cm] at (3.0,11.25) {E3};

\draw [ color={rgb,255:red,46; green,117; blue,37}, draw opacity=1, short] (3.75,8.5) -- (15.25,8.5);
\node [font=\fontsize{18.2pt}{23.7pt}\selectfont, color={rgb,255:red,23; green,122; blue,61}, text opacity=1, fill={rgb,255:red,255; green,255; blue,255}, fill opacity=1, inner xsep=0.080cm, inner ysep=0.085cm, rounded corners=0.020cm] at (3.0,8.5) {E3$'$};

\node [font=\fontsize{18.2pt}{23.7pt}\selectfont, color={rgb,255:red,255; green,0; blue,0}, text opacity=1, fill={rgb,255:red,255; green,255; blue,255}, fill opacity=1, inner xsep=0.080cm, inner ysep=0.085cm, rounded corners=0.020cm] at (1,9.875) {SO(8)};

\draw [decorate, decoration={brace, amplitude=8pt}, color={rgb,255:red,255; green,0; blue,0}, thick] (2.4,9.25) -- (2.4,10.5);

\draw [decorate, decoration={brace, amplitude=8pt, mirror}, color={rgb,255:red,46; green,117; blue,37}, thick] (17.2,8.5) -- (17.2,11.25);

\node [font=\fontsize{18.2pt}{23.7pt}\selectfont, color={rgb,255:red,23; green,122; blue,61}, text opacity=1, fill={rgb,255:red,255; green,255; blue,255}, fill opacity=1, inner xsep=0.080cm, inner ysep=0.085cm, rounded corners=0.020cm] at (18.5,9.875) {U(1)};

\end{circuitikz}
}
\caption{$U(1)$ instanton configuration.}
\label{fig:U(1)_Setup}
\end{figure}

Depending on the behaviour of the cycle $D_\Et$ under the orientifold map, different kinds of instantons can be realised:
\begin{itemize}
\item If $D_\Et$ is pointwise invariant under the orientifold, it supports a $USp(2)$-instanton (see Fig. \ref{fig:USp(2)_Setup}).

\item If $D_\Et$ is transversally, but not pointwise, invariant under the orientifold and it intersects the O7-plane, it supports an $O(1)$ instanton (see Fig. \ref{fig:O(1)_Setup}).

\item If $D_\Et$ is mapped to $D_{\Et'}\neq D_\Et$  under the orientifold, it carries a $U(1)$ instanton (see Fig. \ref{fig:U(1)_Setup}).
\end{itemize}

As in the case of gauge instantons, the presence of such E3-instantons can induce corrections to the 4D effective action. In particular, an E3-brane wrapped around a $4$-cycle can generate exponentially suppressed corrections to the 4D superpotential (as well as to gauge couplings) of the form \cite{Blumenhagen:2008zz}:
\begin{equation}
\label{eq:Sup_Non_Pert}
    W_{\rm np}=\sum_{\Et_i}\,A_i(U)\,e^{-2\pi T_i}+\text{poly-instantons}\,,
\end{equation}
where $U$ are the complex structure moduli. However, E3-instantons do not always contribute to the superpotential in the way outlined in \eqref{eq:Sup_Non_Pert} since an excess of fermionic zero modes for the Dirac operator on the E3 worldvolume can lead to a vanishing contribution to the superpotential \citep{Witten:1996bn}. The counting of fermionic zero modes for E3-instantons is reviewed in App. \ref{sec:Zero_Modes}. Let us now, instead, review the required zero mode configurations to generate non-zero instanton and poly-instanton contributions to the superpotential.

In order to have single E3-instanton corrections as in \eqref{eq:Sup_Non_Pert}, we need exactly two fermionic zero modes which, for an $O(1)$ instanton, are given by the universal zero modes. Hence, the simplest situation where E3-instantons contribute to the superpotential is the one where they wrap a 4-cycle $D_\Et$ which is transversally invariant under the orientifold, intersects the O7 locus and has the additional following properties:
\begin{itemize}
\item $h^{1,0}(D_{\Et})=h^{2,0}(D_{\Et})=0$, i.e. the divisor is rigid\footnote{If the divisor is not rigid, a non-zero worldvolume flux $\mathcal{F}_\Et\in H^{1,1}_-(D_{\Et})$ can be turned on to remove deformation zero modes. In this case, the rigidifying flux needs to be Poincaré dual (with respect to the divisor) to a curve which is rigid in $X$ \cite{Bianchi:2011qh,Bianchi:2012pn,Louis:2012nb}.};

\item $\mathcal{F}_\Et\in H^{1,1}_-(D_{\Et})$, i.e. the orientifold-even part of the flux on the instanton vanishes in order to have an invariant configuration since, for an $O(1)$ instanton, the gauge field on the E3 worldvolume satisfies $\sigma^*\mathcal{F}_\Et = -\mathcal{F}_\Et$;

\item Chiral and vector-like zero modes are absent. This is automatic if $D_{\Et}\cap D_{\Dsi}=\emptyset$, $\forall i$. When this is not the case, chiral zero modes are counted by $\int_X \mathcal{F}_{\Dsi}\wedge \hat{D}_{\Et}\wedge \hat{D}_{\Dsi}$, where $\mathcal{F}_{\Dsi}$ is the gauge flux on the $i$-th stack of D7-branes. Vector-like zero modes are absent if the intersection is a $\mathbb{P}^1$, or generically massive if the intersection is a $T^2$. 
\end{itemize}

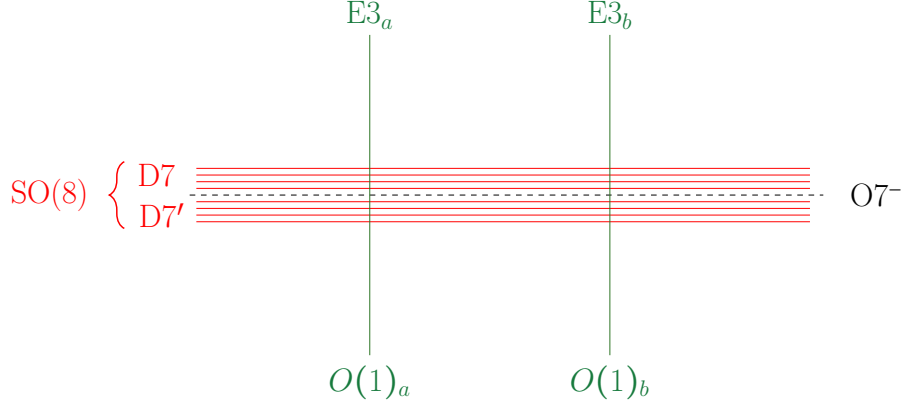
\begin{figure}[!ht]
\centering
\resizebox{0.8\textwidth}{!}{
\begin{circuitikz}
\tikzstyle{every node}=[font=\fontsize{14.8pt}{19.2pt}\selectfont]

\draw [dashed] (3.625,9.875) -- (15.5,9.875);
\node [font=\fontsize{18.2pt}{23.7pt}\selectfont, inner xsep=0.080cm, inner ysep=0.085cm, rounded corners=0.020cm] at (16.5,9.875) {O7$^-$};
\draw [ color={rgb,255:red,255; green,0; blue,0}, draw opacity=1, short] (3.75,10) -- (15.25,10);
\draw [ color={rgb,255:red,255; green,0; blue,0}, draw opacity=1, short] (3.75,10.125) -- (15.25,10.125);
\draw [ color={rgb,255:red,255; green,0; blue,0}, draw opacity=1, short] (3.75,10.25) -- (15.25,10.25);
\draw [ color={rgb,255:red,255; green,0; blue,0}, draw opacity=1, short] (3.75,10.375) -- (15.25,10.375);
\draw [ color={rgb,255:red,255; green,0; blue,0}, draw opacity=1, short] (3.75,9.75) -- (15.25,9.75);
\draw [ color={rgb,255:red,255; green,0; blue,0}, draw opacity=1, short] (3.75,9.625) -- (15.25,9.625);
\draw [ color={rgb,255:red,255; green,0; blue,0}, draw opacity=1, short] (3.75,9.5) -- (15.25,9.5);
\draw [ color={rgb,255:red,255; green,0; blue,0}, draw opacity=1, short] (3.75,9.375) -- (15.25,9.375);
\node [font=\fontsize{18.2pt}{23.7pt}\selectfont, color={rgb,255:red,255; green,0; blue,0}, text opacity=1, fill={rgb,255:red,255; green,255; blue,255}, fill opacity=1, inner xsep=0.080cm, inner ysep=0.085cm, rounded corners=0.020cm] at (3,10.25) {D7};

\node [font=\fontsize{18.2pt}{23.7pt}\selectfont, color={rgb,255:red,255; green,0; blue,0}, text opacity=1, fill={rgb,255:red,255; green,255; blue,255}, fill opacity=1, inner xsep=0.080cm, inner ysep=0.085cm, rounded corners=0.020cm] at (3.10,9.5) {D7$'$};
\draw [ color={rgb,255:red,46; green,117; blue,37}, draw opacity=1, short] (7,6.875) -- (7,12.875);

\node [font=\fontsize{18.2pt}{23.7pt}\selectfont, color={rgb,255:red,23; green,122; blue,61}, text opacity=1, fill={rgb,255:red,255; green,255; blue,255}, fill opacity=1, inner xsep=0.080cm, inner ysep=0.085cm, rounded corners=0.020cm] at (7,13.25) {E3$_a$};
\node [font=\fontsize{18.2pt}{23.7pt}\selectfont, color={rgb,255:red,23; green,122; blue,61}, text opacity=1, fill={rgb,255:red,255; green,255; blue,255}, fill opacity=1, inner xsep=0.080cm, inner ysep=0.085cm, rounded corners=0.020cm] at (7,6.375) {$O(1)_a$};

\draw [ color={rgb,255:red,46; green,117; blue,37}, draw opacity=1, short] (11.5,6.875) -- (11.5,12.875);
\node [font=\fontsize{18.2pt}{23.7pt}\selectfont, color={rgb,255:red,23; green,122; blue,61}, text opacity=1, fill={rgb,255:red,255; green,255; blue,255}, fill opacity=1, inner xsep=0.080cm, inner ysep=0.085cm, rounded corners=0.020cm] at (11.5,13.25) {E3$_b$};
\node [font=\fontsize{18.2pt}{23.7pt}\selectfont, color={rgb,255:red,23; green,122; blue,61}, text opacity=1, fill={rgb,255:red,255; green,255; blue,255}, fill opacity=1, inner xsep=0.080cm, inner ysep=0.085cm, rounded corners=0.020cm] at (11.5,6.375) {$O(1)_b$};

\node [font=\fontsize{18.2pt}{23.7pt}\selectfont, color={rgb,255:red,255; green,0; blue,0}, text opacity=1, fill={rgb,255:red,255; green,255; blue,255}, fill opacity=1, inner xsep=0.080cm, inner ysep=0.085cm, rounded corners=0.020cm] at (1,9.875) {SO(8)};

\draw [decorate, decoration={brace, amplitude=8pt}, color={rgb,255:red,255; green,0; blue,0}, thick] (2.4,9.25) -- (2.4,10.5);

\end{circuitikz}
}
\caption{Poly-instanton configuration.}
\label{fig:polyinstanton}
\end{figure}

Poly-instanton corrections instead arise from a more involved setup involving at least two E3-instantons, denoted by E3$_a$ and E3$_b$. We focus on the simplest case where both E3$_a$ and E3$_b$ are $O(1)$ instantons, as depicted in Fig. \ref{fig:polyinstanton}. The instanton E3$_a$ contributes to the superpotential as the first term in \eqref{eq:Sup_Non_Pert}. On the other hand, the instanton E3$_b$ has to feature, on top of the two universal zero modes $\theta^\alpha$, an additional Wilson line $\gamma^\alpha$ or deformation Goldstino $\chi^\alpha$, requiring either $h_+^{1,0}=1$ or $h_+^{2,0}=1$.

When the properties listed above are satisfied, the non-zero instanton and poly-instanton corrections to the superpotential depend on the two instanton actions $S_a$ and $S_b$ and look like:
\begin{equation}
W_{\rm np}=W_{\rm inst}+W_{\rm poly}=A_a\,e^{-S_a+A_b\, e^{-S_b}}=A_a\,e^{-S_a}+A_a\,A_b\,e^{-S_a -S_b}+\dots\,.
\label{Wnpaction}
\end{equation}
If each E3$_i$-instanton is wrapped around a divisor whose volume is controlled by the real part of the K\"ahler modulus $T_i$, with $i=a,b$, (\ref{Wnpaction}) can also be rewritten as:
\begin{equation}
W_{\rm np}\simeq A_a\,e^{-2\pi T_a}+A_a\,A_b\,e^{-2\pi (T_a+T_b)}\,.
\end{equation}
According to the previous discussion of zero modes, there are two possible topologies that can give rise to poly-instanton corrections the superpotential: Wilson divisors with $h^{1,0}=1$ and $h^{2,0}=0$, and deformation divisors with $h^{1,0}=0$ and $h^{2,0}=1$. A primary example of a deformation divisor is a K3 surface $D_\K$. However, the authors of \cite{Blumenhagen:2012kz} showed that, if the K3 intersects and the orientifold planes intersect on a single isolated curve, then $h^{2,0}(D_{\K})=h_-^{2,0}(D_{\K})=1$, implying that no K3 poly-instantons are allowed. The logic supporting this statement is explained in detail in Sec. \ref{sec:K3_naive_exc}. We will see that this negative result holds also in more general setups with O3-planes on the E3 worldvolume. However, when the E3 worldvolume contains O3-planes, new poly-instanton effects can arise for both K3-like (i.e. divisors with the same Hodge diamond as a K3 surface but with non-zero first Chern class) and blown-up K3 divisors (i.e. divisors with $h^{1,0}=0$ and $h^{2,0}=1$ but $h^{1,1}>20$). For Wilson divisors, ref. \cite{Blumenhagen:2012kz} showed that they can generate a poly-instanton superpotential. Nevertheless, as we shall see in detail in Sec. \ref{sec:Wilson_naive_acc}, in the presence of O3-planes on the E3-instanton, Wilson divisors do not automatically lead to a non-zero poly-instanton contribution to $W$.

\section{New poly-instanton effects}
\label{sec:Hol_Lef_FP_Theor}

\subsection{Generalised Lefschetz fixed-point theorem}

Let $M$ be a complex Riemannian manifold, and $V$ a holomorphic vector bundle over $M$. An \emph{analytic index} is defined as the alternating sum of the dimensions of the cohomology groups associated with a given operator. A standard example is the index of the Dolbeault-Dirac operator $
D=\bar{\partial}+\bar{\partial}^{\dagger}$, twisted by $V$.
The cohomology of this operator is given by the Dolbeault groups $H^{i}(M,V)$. Elements of $H^{i}(M,V)$ can be thought as $V$-valued $(i,0)$-forms. The corresponding analytic index is:
\begin{equation}
\operatorname{ind}(D,V)\equiv \chi(M,V)
\equiv \sum_{i=0}^{n}(-1)^i \dim H^{i}(M,V)\,.
\label{eq:index}
\end{equation}
Index theorems are powerful tools that allow to link analytic indices to topological invariants of the space itself. For the index \eqref{eq:index} the corresponding index theorem is the Riemann-Roch theorem, which allows to compute $\chi(M,V)$ as:
\begin{equation}
\chi(M,V)=\int_M\operatorname{ch}(V)\,\operatorname{Td}(TM)\,,
\label{eq:indexth}
\end{equation}
where $\operatorname{ch}(V)$ is the Chern character of $V$ and $\operatorname{Td}(TM)$ the Todd class.

In type IIB string compactifications, manifolds are usually equipped with an orientifold. The orientifold is a combination of left-moving fermion number, worldsheet parity and a geometric involution $\sigma$ acting on spacetime. After taking a quotient by the orientifold action, the cohomology group $H^i(M,V)$ splits in:
\begin{equation}
    H^{i}(M,V)=H_+^i(M,V)\oplus H_-^i(M,V)\, .
\end{equation}
The index \eqref{eq:indexth} must then be generalised to its $\mathbb{Z}_2$-equivariant version as:
\begin{equation}
\chi^\sigma(M,V)=\sum_{i=1}^n (-1)^i\left(\operatorname{dim}H_+^i(M,V)-\operatorname{dim}H_-^i(M,V)\right)\ .
\end{equation}
The quantity $\chi^\sigma(M,V)$ is called the \emph{holomorphic Lefschetz number}. The equivariant version of the Riemann-Roch theorem allows to link this quantity to the topological data of the manifold $M$ and the set $I^\sigma$ of fixed points of the involution $\sigma$ restricted to $M$ (which we shall take as the transversally invariant divisor wrapped by the E3-instanton). $I^\sigma$ is a disjoint union of curves $M_i^\sigma$ and points O3$^j$: namely, $I^\sigma=\sum_i\, M_i^\sigma\cup\sum_j {\rm O3}^j$. The holomorphic Lefschetz theorem states:
\begin{flalign}
\chi^\sigma(M,V)&=\int_{I^\sigma}\frac{\operatorname{ch}_\sigma(V)\operatorname{Td}(T\,I^{\sigma})}{\operatorname{ch}_\sigma(\Lambda_{-1}\overline{N}_{I^\sigma})}\equiv\sum_i\chi_1^i(M,V)+\sum_j\chi_2^j(M,V) \nonumber \\
&=\sum_i\int_{M_i^\sigma}\frac{\operatorname{ch}_\sigma(V)\operatorname{Td}(T M_i^{\sigma})}{\operatorname{ch}_\sigma(\Lambda_{-1}\overline{N}_{M_i^\sigma})}+\sum_j\int_{{\rm O3}^j}\frac{\operatorname{ch}_\sigma(V)\operatorname{Td}(T\,{\rm O3}^j)}{\operatorname{ch}_\sigma(\Lambda_{-1}\overline{N}_{\Ot^j})}\,,
\label{eq:gen_lef}
\end{flalign}
where now $\Lambda_{-1}\overline{N}_A\equiv \sum_k\Lambda^K\overline{N}_A$ is a formal alternating sum of exterior powers of the complex conjugate of the normal bundle to a generic submanifold $A$, namely $\overline{N}_A$. 

The main task of this section is to explain how to simplify \eqref{eq:gen_lef} and reduce the computation of the holomorphic Lefschetz number $\chi^\sigma(M,V)$ to intersection data, as well as fixed points counting. For the following discussion, let $M=D_\Et$ be a divisor of the CY threefold $X$ wrapped by an E3-instanton. Let us further assume that $D_\Et$ intersects transversally at least an O7-plane at some curves $M_i^\sigma$. In addition, $D_\Et$ might contain ${\rm O3}^j$ points where O3-planes are localised. The general case involving both O7 and O3-planes has already been considered in \cite{Cvetic:2010ky}.

The two contributions $\chi_1^i$, and $\chi_2^i$ are now discussed in turns. For what regards $\chi_1^i$, its denominator is given by:
\begin{flalign}
\operatorname{ch}_\sigma(\Lambda_{-1}\overline{N}_{M_i^\sigma})&=\operatorname{ch}_\sigma(\mathcal{O}-\overline{N}_{M_i^\sigma})=1-\operatorname{ch}_\sigma(\overline{N}_{M_i^\sigma})\\
    &\overset{\sigma^*(N_{M_i^\sigma})=-N_{M_i^\sigma}}{=}1+\operatorname{ch}(\overline{N}_{M_i^\sigma})=1+(1-c_1
(N_{M_i^\sigma})+\dots)=2-c_1(N_{M_i^\sigma})+\dots\,
\nonumber
\end{flalign}
The numerator of $\chi_1^i$ is composed instead by the Todd class:
\begin{equation}
    \operatorname{Td}(TM_i^\sigma)=1+\frac{1}{2}c_1(M_i^\sigma)+\frac{1}{12}(c_1^2(M_i^\sigma)+c_2(M_i^\sigma))+\frac{1}{24}c_1(M_i^\sigma)c_3(M_i^\sigma)+\dots
\end{equation}
and by the holomorphic Chern character of $V$. Therefore we can expand the denominator and rewrite $\chi_1^i$ as:
\begin{flalign}
\chi_1^i(D_\Et,V)&=\int_{M_i^\sigma}\frac{\operatorname{ch}_\sigma(V)(1+\frac{1}{2}c_1(M_i^\sigma)+\dots)}{2-c_1(N_{M_i^\sigma})+\dots} \nonumber \\
&\simeq\int_{M_i^\sigma}\operatorname{ch}_\sigma(V)\left(1+\frac{1}{2}c_1(M_i^\sigma)+\dots\right)\left(\frac{1}{2}\left(1+\frac{1}{2}c_1(N_{M_i^\sigma})+\dots\right)\right) \nonumber \\
&=\int_{M_i^\sigma}\operatorname{ch}_\sigma(V)\frac{1}{2}\left(1+\frac{1}{2}c_1(M_i^\sigma)+\frac{1}{2}c_1(N_{M_i^\sigma})+\dots\right).
\end{flalign}
Restricting to the case $V=\mathcal{O}_\Et$, we get $\operatorname{ch}_\sigma(V)=\operatorname{ch}_\sigma(\mathcal{O}_\Et)=1$, obtaining:
\begin{flalign}
\chi_1^i(D_\Et,\mathcal{O}_\Et)&=\int_{M_i^\sigma}\left(\frac{1}{4}c_1(M_i^\sigma)+\frac{1}{4}c_1(N_{M_i^\sigma})\right)=\frac{1}{4}\chi(M_i^\sigma)-\frac{1}{4}M_i^\sigma\cdot M_i^\sigma\,,
\end{flalign}
where $M_i^\sigma\cdot M_i^\sigma$ is the self-intersection number in the divisor $D_\Et$, i.e. $M_i^\sigma\cdot M_i^\sigma=\int_{D_\Et} \check{M}_i^\sigma\wedge \check{M}_i^\sigma$, with $\check{M}_i^\sigma$ Poincaré dual of the curve $M_i^\sigma$ in the divisor $D_\Et$. At this stage, we use the adjunction formula twice:
\begin{flalign}
c_1(M_i^\sigma)&=c_1(D_\Et){\huge|}_{M_i^\sigma}-c_1(N_{M_i^\sigma}){\huge|}_{M_i^\sigma}=(c_1(X){\huge|}_{D_\Et}-c_1(N_{D_\Et}){\huge|}_{D_\Et}){\huge|}_{M_i^\sigma}-c_1(N_{M_i^\sigma}){\huge|}_{M_i^\sigma} \nonumber \\
&\overset{c_1(N_{D_\Et})=-\hat{D}_{\Et}}{=}-\hat{D}_{\Et}{\huge|}_{M_i^\sigma}-c_1(N_{M_i^\sigma}){\huge|}_{M_i^\sigma}\,.
\end{flalign}
With this last step we can finally rewrite the holomorphic Lefschetz number as:
\begin{equation}
\chi_1^i(D_\Et,\mathcal{O}_\Et)=-\frac{1}{4}\int_{M_i^\sigma}\hat{D}_{\Et}+\frac{1}{4}\int_{M_i^\sigma}(c_1(N_{M_i^\sigma})-c_1(N_{M_i^\sigma}))=-\frac{1}{4}\int_{M_i^\sigma}\hat{D}_{\Et}\,,
\end{equation}
reproducing the result of \cite{Blumenhagen:2012kz}. This expression can be further clarified by rewriting it as:
\begin{equation}
\chi_1^i(D_\Et,\mathcal{O}_\Et)=-\frac{1}{4}\int_{D_\Et\cap D_{{\rm O7}^i}}\hat{D}_{\Et}=-\frac{1}{4}\int_X \hat{D}_{\Et}\wedge\hat{D}_{\Et}\wedge\hat{D}_\Osi=-\frac{1}{4}\,k_{\Et\Et\Osi}\,.
\end{equation}
The treatment of $\chi_2^j(D_\Et,\mathcal{O}_\Et)$ requires addressing some mathematical technicalities. Indeed, even if the numerator is trivial since $\operatorname{ch}_\sigma(V)=\operatorname{ch}_\sigma(\mathcal{O}_\Et)=1$ and $\operatorname{Td}(T\,{\rm O3}^j)=1$, the denominator is more complicated. In fact, for a single point ${\rm O3}^j$, it turns out that $\overline{N}_{{\rm O3}^j}=\overline{T_{{\rm O3}^j}D_\Et}$.\footnote{This comes directly from the definition of the normal bundle via sequences. For a submanifold $Y\subset D_\Et$ we have the following short exact sequence:
\begin{equation}
0\rightarrow TY\rightarrow TD_\Et|_{Y}\rightarrow N_Y\rightarrow0\ .
\end{equation}
For $Y={\rm O3}^j$, this is a point in the CY. Therefore $TY=0$ and the sequence becomes:
\begin{equation}
    0\rightarrow0\rightarrow T_{{\rm O3}^j}D_\Et\rightarrow N_{{\rm O3}^j}\rightarrow0\,,
\end{equation}
and therefore, by exactness of the sequence, we have $N_{{\rm O3}^j}=T_{{\rm O3}^j}$.}

For a general 2D complex manifold $E$, the formal alternated sum of exterior products can be decomposed as:
\begin{equation}
\Lambda_{-1}E\overset{\operatorname{dim}_{\mathbb{C}}E=2}{=}\Lambda^0E-\Lambda^1E+\Lambda^2 E\,.
\end{equation}
Decomposing $E$ into $\sigma^*$-line-eigenbundles under $\sigma^*$ as $E=L_1\oplus L_2$, it follows that:
\begin{equation}
\Lambda^0(L_1\oplus L_2)=\mathbb{C}\,,\qquad
\Lambda^1(L_1\oplus L_2)=L_1\oplus L_2\,,\qquad
\Lambda^2(L_1\oplus L_2)=L_1\otimes L_2\,.
\end{equation}
Since the Chern character is additive with respect to the direct sum and multiplicative with respect to the tensor product, we have:
\begin{equation}
\operatorname{ch}_\sigma(\Lambda_{-1}E)=1-(\operatorname{ch}_\sigma(L_1)+\operatorname{ch}_\sigma(L_2))+(\operatorname{ch}_\sigma(L_1)\times\operatorname{ch}_\sigma(L_2))\,.
\label{ch}
\end{equation}
Given that the Chern character of an eigenbundle is its corresponding eigenvalue, i.e $\operatorname{ch}_\sigma(L_i)=\lambda_i$, $i=1,2$, (\ref{ch}) simplifies to:
\begin{equation}
    \operatorname{ch}_\sigma(\Lambda_{-1}E)=1 -(\lambda_1+\lambda_2)+(\lambda_1\times\lambda_2)=(1-\lambda_1)(1-\lambda_2)\,.
\end{equation}
The task reduces now to finding the eigenvalues $\lambda_1$ and $\lambda_2$ of the eigenbundles $L_1$ and $L_2$. 

Coming back to the case under consideration, we split $\overline{T_{\Ot^j}D_\Et}\cong\mathbb{C}^2= L_1\oplus L_2=\mathbb{C}\oplus\mathbb{C}$. As stated before, the map $\sigma:X\rightarrow X$ is an involution acting on the whole CY and, in particular, can be restricted to the divisor by keeping the involution property $\sigma{\huge|}_{D_{\Et}}:\,D_\Et\rightarrow D_\Et$. Define $\sigma{\huge|}_{D_{\Et}}\equiv \sigma_\Et$. This restricted involution $\sigma_\Et$ can have the ${\rm O3}^j$ fixed points localised on $D_\Et$, namely $\sigma_\Et({\rm O3}^j)={\rm O3}^j$. We can now push-forward this map to the tangent space $T_{{\Ot}^j}D_\Et$ and get $*\sigma_\Et:\,T_{{\Ot}^j}D_{\Et}\rightarrow T_{{\Ot}^j}D_{\Et}$ which corresponds to the differential of $\sigma_\Et$, and so $*\sigma_\Et=d\sigma_\Et$. Due to the chain rule, also this map is an involution, implying $d\sigma_\Et^2=\mathbb{I}$. Thus its possible eigenvalues are:
\begin{equation}
\lambda_{d\sigma_\Et}=\pm1\,.
\end{equation}
Locally around a fixed point, due to the local linearisation theorem\footnote{Local linearisation (or Bochner linearisation) theorem states that for a smooth manifold $M$ and a compact group acting on $M$, which in our case is $G=\{\mathbb{I},\sigma_\Et\}$, with a fixed point under $\sigma_\Et$, $\sigma_\Et(P)=P$, given: 
\begin{itemize}
\item An open neighbourhood $U\subset M$ of $p$ and a neighbourhood $Z$ of the origin of $T_PM$,
\item A diffeomorphism $\phi:\,U\rightarrow Z$, $\forall x\in U$ and $\forall \sigma_\Et \in G$,
\end{itemize}
it follows that $\phi(\sigma_\Et(x))=d\sigma_\Et|_{P}(\phi(x))$. Thus one can choose a coordinate system around the fixed point $P$ such that the non-linear map $\sigma_\Et$ becomes linear.}, the holomorphic group action can be seen as its push-forward. Therefore, if we take $v\in T_{{\Ot}^j}D_\Et$ and if $v$ is an eigenvector with eigenvalue $\lambda_{d\sigma_\Et}=1$, there is a whole invariant 1D locus under the differential of the involution $d\sigma_\Et$, and so an entire invariant 1D curve under the involution $\sigma_\Et$. However, since in this case we have only an invariant point and not a curve, by contradiction, we conclude that the only possible eigenvalue is $\lambda_{d\sigma_\Et}=-1$. Then the action of $\sigma_\Et$ on the line bundles is given just by $\lambda_1=\lambda_2=-1$, which are real, and so not affected by complex conjugation of the tangent space. This leads to:
\begin{equation}
\operatorname{ch}_\sigma(\Lambda_{-1}\overline{T_{{\Ot}^j}D_\Et}) =(1-(-1))(1-(-1))=2\times2=4\,.
\end{equation}
The final expression for $\chi_2^j(D_\Et,\mathcal{O}_\Et)$ is:
\begin{equation}
\chi_2^j(D_\Et,\mathcal{O}_\Et)=\frac{1}{4}\,.
\end{equation}
All this analysis leads to a generalised version of the holomorphic Lefschetz theorem \cite{Cvetic:2010ky,Lust:2013kt}:
\begin{equation}
\boxed{
\chi^\sigma(D_\Et,\mathcal{O}_\Et)=-\frac14\,\sum_i\,k_{\Et\Et\Osi}+\frac14\,N_{\Ot}^{\Et}\,,
}
\end{equation}
where $N_{\Ot}^{\Et}$ is the number of O3-planes on the divisor $D_\Et$.

\subsection{Consequences of the generalised Lefschetz fixed-point theorem}

Our setting includes an E3-instanton wrapping a divisor $D_\Et$ which is transversally invariant under the orientifold involution. Moreover, $D_\Et$ intersects the fixed-point locus of the orientifold at isolated curves $M_i^\sigma$, together with the possibility of additional points where O3-planes are localised. This E3-instanton is an $O(1)$ instanton and by applying the generalised Lefschetz fixed point theorem, we find \cite{Blumenhagen:2012kz}:
\begin{eqnarray}
\chi^\sigma(D_\Et,\mathcal{O}_\Et)&=&-\frac{1}{4}\int_{\sum_i M_i^\sigma\cup\sum_j {\rm O3}^j}\hat{D}_\Et \nonumber \\
&=&-\frac{1}{4}\int_X \hat{D}_\Et \wedge \hat{D}_\Et \wedge \hat{D}_\Os+\frac14\,N_{\Ot}^{\Et}=-\frac{1}{4}\,k_{\Et\Et\Os}+\frac14\,N_{\Ot}^{\Et}\,,
\end{eqnarray}
where $D_\Os$ is the pointwise invariant divisor where the O7-plane is localised, and $N_\Ot^{\Et}$ is the number of O3-planes on the divisor $D_\Et$. Given the Lefschetz fixed point theorem definition:
\begin{equation}
\chi^\sigma(D_\Et,\mathcal{O}_\Et)=\sum_{i=0}^{2}(-1)^i\left(h_+^{i,0}-h_-^{i,0}\right)\,,
\end{equation}
$\chi^\sigma(D_\Et,\mathcal{O}_\Et)$ must be integer, resulting in the following four possible cases:
\begin{itemize}
\item $k_{\Et\Et\Os}=N_{\Ot}^\Et=0$: This leads to a vanishing Lefschetz number $\chi^\sigma(D_\Et,\mathcal{O}_\Et)= 0$;

\item $k_{\Et\Et\Os}=0$ and $=N_{\Ot}^\Et>0$: The number of O3-planes in points $p_i\in D_\Et$, $i=1,\dots,N_{\Ot}^\Et$ should be $N_{\Ot}^\Et=4\,n$, $n\in\mathbb{N}$. This leads to a non-vanishing Lefschetz number $\chi^\sigma(D_\Et,\mathcal{O}_\Et)= n >0$;

\item $k_{\Et\Et\Os}\neq 0$ and $N_{\Ot}^\Et=0$: The intersection number should respect $k_{\Et\Et\Os}=4\alpha$ with $\alpha\in \mathbb{Z}\smallsetminus\{0\}$. This leads to a non-vanishing Lefschetz number $\chi^\sigma(D_\Et,\mathcal{O}_\Et)= -\alpha\neq 0$;

\item $k_{\Et\Et\Os}\neq 0$ and $=N_{\Ot}^\Et>0$: The number of O3-planes in points $p_i\in D_\Et$, $i=1,\dots,N_{\Ot}^\Et$, should respect $N_{\Ot}^\Et-k_{\Et\Et\Os}=4\,\beta$, $\beta\in\mathbb{Z}$, with $\beta$ giving the value of the Lefschetz number, $\chi^\sigma(D_\Et,\mathcal{O}_\Et)=\beta$. Note that $\chi^\sigma(D_\Et,\mathcal{O}_\Et)$ can be either zero or non-zero, depending on the exact values of $N_{\Ot}^\Et$ and $k_{\Et\Et\Os}$.
\end{itemize}
Let us apply the consequences of the generalised Lefschetz fixed-point theorem to the two divisor topologies that are candidates for generating non-zero poly-instanton contributions to the superpotential.

\subsubsection{Poly-instantons on Wilson divisors}
\label{sec:Wilson_naive_acc}

In this section we focus on rigid divisors with a Wilson line and a generic number of independent curve classes, which we denote as $D_\W$ and call \emph{Wilson divisors}. Hence, their Hodge diamond is:
\begin{equation}
    \begin{pmatrix}
        h^{0,0}\\h^{1,0}\\h^{2,0}\\h^{1,1}
    \end{pmatrix}=    \begin{pmatrix}
        1\\1\\0\\h^{1,1}(D_\W)
    \end{pmatrix}\,.
\end{equation}
From this Hodge vector and from the request of a poly-instanton divisor candidate with $h_+^{0,0}=1$ and $h_-^{0,0}=h^{2,0}_\pm=0$, we can derive two conditions on the Hodge numbers:
\begin{flalign}
    &h^{1,0}=h_+^{1,0}+h_-^{1,0}=1\,,\\
    & \chi^\sigma(D_\W,\mathcal{O}_\W)=h_+^{0,0}-h_-^{0,0}-h_+^{1,0}+h_-^{1,0}+h_+^{2,0}-h_-^{2,0}=1-h_+^{1,0}+h_-^{1,0}\,,
\end{flalign}
which give the following conditions on the cohomology classes $H_\pm^{2,0}(D_\W)$:
\begin{equation}
h_+^{1,0}+h_-^{1,0}=1\qquad\text{and}\qquad h_+^{1,0}-h_-^{1,0}=1-\chi^\sigma(D_\W,\mathcal{O}_\W)\,.
\end{equation}
Using the generalised Lefschetz fixed point theorem, we obtain:
\begin{flalign}
&h_+^{1,0}=1-\frac12\,\chi^\sigma(D_\W,\mathcal{O}_\W)=1-\frac18\left(N_\Ot^\W-k_{\W\W\Os}\right)\,, \\
&h_-^{1,0}=\frac12\,\chi^\sigma(D_\W,\mathcal{O}_\W) =\frac18\left(N_\Ot^\W-k_{\W\W\Os} \right)\,,
\end{flalign}
The fact that the two Hodge numbers need to be either $0$ or $1$, leads just to two consistent cases:
\begin{enumerate}
\item Case with poly-instantons:
\begin{equation}
   N_{\Ot}^{\W}=k_{\W\W\Os}\qquad\Rightarrow\qquad \begin{cases}
       h_+^{1,0}=1\,,\\
       h_-^{1,0}=0\,,
   \end{cases}
\end{equation}
which is the suitable topology for generating a non-zero poly-instanton contribution to the superpotential. 

\item Case without poly-instantons:
\begin{equation}
    N_{\Ot}^{\W}=8+k_{\W\W\Os}\qquad\Rightarrow\qquad \begin{cases}
       h_+^{1,0}=0\,,\\
       h_-^{1,0}=1\,,
   \end{cases}
\end{equation}
which does not feature any Wilson poly-instanton contribution to the superpotential.
\end{enumerate}
Let us point out that ref. \cite{Blumenhagen:2012kz} focused on the case with no O3-planes on the E3 worldvolume, $N_{\Ot}^{\W}=0$, which can lead to non-zero Wilson poly-instantons depending on $k_{\W\W\Os}$ (for $k_{\W\W\Os}=0$ the contribution is indeed non-zero, whereas for $k_{\W\W\Os}=-8$ the effect vanishes). If the E3 worldvolume contains O3-planes, $N_{\Ot}^{\W}>0$, the case with $k_{\W\W\Os}=-8$ is inconsistent, while the case with $k_{\W\W\Os}=0$ does not lead anymore to non-zero poly-instantons. On the other hand, new non-zero Wilson poly-instanton configurations could arise for $N_{\Ot}^{\W}=k_{\W\W\Os}>0$. An example with $N_{\Ot}^{\W}=k_{\W\W\Os}=2$ can be found in \cite{Lust:2013kt}.

\subsubsection{Poly-instantons on deformation divisors}
\label{sec:K3_naive_exc}

Let us now discuss the case of divisors with a single deformation, no Wilson lines and an arbitrary number of independent curve classes. We shall denote them as $D_\Kk$ and call them \emph{deformation divisors}. A famous example is given by K3 divisors corresponding to the particular case with $h^{1,1}(D_\Kk)=20$. Their Hodge diamond looks like:
\begin{equation}
    \begin{pmatrix}
        h^{0,0}\\h^{1,0}\\h^{2,0}\\h^{1,1}
    \end{pmatrix}=    \begin{pmatrix}
        1\\0\\1\\h^{1,1}(D_\Kk)
    \end{pmatrix}\,.
\end{equation}
From this Hodge vector and from the request to obtain a poly-instanton superpotential with $h_+^{0,0}=1$ and $h_-^{0,0}=h_\pm^{1,0}=0$, we can derive two conditions on the Hodge numbers:
\begin{flalign}
    &h^{2,0}=h_+^{2,0}+h_-^{2,0}=1\,,\\
    & \chi^\sigma(D_\Kk,\mathcal{O}_{\Kk})=h_+^{0,0}-h_-^{0,0}-h_+^{1,0}+h_-^{1,0}+h_+^{2,0}-h_-^{2,0}=1+h_+^{2,0}-h_-^{2,0}\,,
\end{flalign}
that lead to the following conditions on the cohomology classes $H_\pm^{2,0}(D_{\Kk})$:
\begin{equation}
h_+^{2,0}+h_-^{2,0}=1\qquad\text{and}\qquad
h_+^{2,0}-h_-^{2,0}=\chi^\sigma(D_{\Kk},\mathcal{O}_{\Kk})-1\,.
\end{equation}
Recalling the generalised Lefschetz fixed point theorem, we end up with:
\begin{flalign}
&h_+^{2,0}=\frac12\,\chi^\sigma(D_{\Kk},\mathcal{O}_{\Kk})=\frac18\left(N_{\Ot}^{\Kk}-k_{\Kk\Kk\Os} \right)\,,
\label{h+} \\
&h_-^{2,0}=1-\frac12\,\chi^\sigma(D_{\Kk},\mathcal{O}_{\Kk})= 1-\frac18\left(N_{\Ot}^{\Kk}-k_{\Kk\Kk\Os} \right)\,.
\label{h-}
\end{flalign}
The fact that the two Hodge numbers need to be either $0$ or $1$, leads just to two consistent cases:
\begin{enumerate}
\item Case with poly-instantons:
\begin{equation}
   N_{\Ot}^{\Kk}=8+k_{\Kk\Kk\Os}\qquad\Rightarrow\qquad \begin{cases}
       h_+^{2,0}=1\,,\\
       h_-^{2,0}=0\,,
   \end{cases}
\end{equation}
which is the suitable topology for generating a non-zero poly-instanton contribution to the superpotential. 

\item Case without poly-instantons:
\begin{equation}
    N_{\Ot}^{\Kk}=k_{\Kk\Kk\Os}\qquad\Rightarrow\qquad \begin{cases}
       h_+^{2,0}=0\,,\\
       h_-^{2,0}=1\,,
   \end{cases}
\end{equation}
which does not feature any poly-instanton contribution to the superpotential.
\end{enumerate}
Note that ref. \cite{Blumenhagen:2012kz} focused on the case with no O3-planes on the E3 worldvolume, $N_{\Ot}^{\Kk}=0$, which can lead to non-zero poly-instantons depending on $k_{\Kk\Kk\Os}$ (for $k_{\Kk\Kk\Os}=-8$ the contribution is indeed non-zero, whereas for $k_{\Kk\Kk\Os}=0$ the effect vanishes). In the presence of O3-planes on the E3-brane, $N_{\Ot}^{\Kk}>0$, the case with $k_{\Kk\Kk\Os}=-8$ is inconsistent, while the case with $k_{\Kk\Kk\Os}=0$ could now lead to non-zero poly-instantons. Moreover, new non-zero poly-instanton configurations could arise for $N_{\Ot}^{\Kk}=8+k_{\Kk\Kk\Os}>0$.

The case with $k_{\Kk\Kk\Os}=0$ actually corresponds to the example where the deformation divisor is a K3 surface, $D_\Kk\equiv D_\K$. Indeed, using the adjunction formula, it is straightforward to realise that $k_{\K\K\Os}$ must vanish:
\begin{equation}
k_{\K\K\Os} = \int_X \hat{D}_\K\wedge \hat{D}_\K \wedge \hat{D}_\Os = \int_{D_\K} \hat{D}_\K \wedge \hat{D}_\Os = -\int_{D_\K} c_1(D_\K) \wedge \hat{D}_\Os=0\,,
\end{equation}
since, when restricted on $D_\K$, $c_1(D_\K)=0$ given that the K3 is a CY two-fold. Hence (\ref{h+}) and (\ref{h-}) reduce to:
\begin{equation}
h_+^{2,0}=\frac18\,N_{\Ot}^{\K} \qquad\text{and}\qquad h_-^{2,0}= 1-\frac18\,N_{\Ot}^{\K}\,.
\end{equation}
implying that the emergence of non-zero poly-instantons depends on the presence of O3-planes as follows:
\begin{enumerate}
\item Case without O3-planes:
\begin{equation}
   N_{\Ot}^{\K}=0\qquad\Rightarrow \qquad\begin{cases}
       h_+^{2,0}=0\,,\\
       h_-^{2,0}=1\,,
   \end{cases}
\end{equation}
which is the case considered in \cite{Blumenhagen:2012kz} where K3 poly-instantons do not contribute to the superpotential because of anti-holomorphic fermionic zero modes.

\item Case with $8$ O3-planes:
\begin{equation}
N_{\Ot}^{\K}=8\qquad\Rightarrow \qquad\begin{cases}
       h_+^{2,0}=1\,,\\
       h_-^{2,0}=0\,,
   \end{cases}
\end{equation}
which might now seem to allow K3 as a suitable topology for generating a poly-instanton correction to the superpotential. However, when $N_{\Ot}^{\K}=8$, the K3 does not intersect any O7-plane, resulting in a $USp$-instanton that does not contribute to the superpotential. Indeed, Nikulin studied involutions of K3 surfaces and showed that, in the symplectic case, the fixed locus consists just of $8$ isolated points \cite{Nikulin1980}, while, in the non-symplectic case, the fixed point set is either empty or a disjoint union of non-singular curves \cite{Nikulin1983}.
\end{enumerate}

\section{Explicit Calabi-Yau examples with O3-planes}
\label{Sec3}

In this section we provide four explicit CY examples of poly-instanton constructions. The CY threefolds are taken from the KS database and their properties are analysed via the software \texttt{CYTools} \cite{Demirtas:2022hqf}. More precisely, all CY threefolds are realised as the zero-locus of a single hypersurface equation in a toric weak-Fano ambient space, following the construction explained in App. \ref{app:A_Kreuzer-Skarke}. The first example involves a Wilson line divisor which has the wrong zero modes to generate poly-instanton effects. The last three examples feature instead deformation divisors. Out of those three examples, one has no poly-instantons, one features a poly-instanton contribution on a K3-like divisor (with the same Hodge numbers as a K3 but with a non-zero first Chern class), and the last one exhibits two poly-instanton effects on two blown-up K3 divisors (with $h^{2,0}=1$ but $h^{1,1}>20$).

\subsection{An example with no poly-instantons on a Wilson divisor}

Let us now discuss an explicit example involving a Wilson divisor. This polytope is the number $684$ of the KS database at $h^{1,1}=9$.\footnote{Here we refer to the ordering provided by the \texttt{fetch\_polytopes} function of \texttt{CYTools}.} Its vertices read:
\begin{equation}
\begin{pmatrix}
    v_1\\v_2\\v_3\\v_4\\v_5\\v_6\\v_7
\end{pmatrix}=
\begin{pmatrix}
    \{ 1&  0&  0&  0\}\\
    \{-3&  0& -2& -2\}\\
    \{ 1&  0&  0&  2\}\\
    \{ 1&  0&  2&  0\}\\
    \{ 0&  1&  0&  0\}\\
    \{ 1&  1&  1&  0\}\\
    \{ 0& -1&  1&  0\}
\end{pmatrix}\,.
\end{equation}
This polytope is favourable and each divisor of the ambient space descends with multiplicity one to a divisor of the CY hypersurface $X$. Its GLSM charge matrix is:
\begin{center}
\begin{tabular}{|c|c|c|c|c|c|c|c|c|c|c|c|c|}
\hline
$x_1$ & $x_2$ & $x_3$ & $x_4$ & $x_5$ & $x_6$ & $x_7$ & $x_8$ & $x_9$ & $x_{10}$ & $x_{11}$ & $x_{12}$ & $x_{13}$ \\
\hline
    1 & 0 & 0 & 0 & 1 & 0 & 0 & 0 & 0 & 1 & -1 & 0 & 0 \\
     0 & 1 & 0 & 0 & 1 & 0 & 0 & 0 & -2 & 1 & 1 & 0 & 0 \\
     0 & 0 & 1 & 0 & -1 & 0 & 0 & 0 & 2 & -1 & 1 & 0 & 0 \\
     0 & 0 & 0 & 1 & -1 & 0 & 0 & 0 & 0 & -1 & -1 & 0 & 0 \\
     0 & 0 & 0 & 0 & -1 & 1 & 0 & 0 & 0 & 0 & -1 & 0 & 0 \\
     0 & 0 & 0 & 0 & 1 & 0 & 1 & 0 & -1 & 1 & 0 & 0 & 0 \\
     0 & 0 & 0 & 0 & 0 & 0 & 0 & 1 & 1 & 0 & 0 & 0 & 0 \\
     0 & 0 & 0 & 0 & -1 & 0 & 0 & 0 & 1 & -1 & 0 & 1 & 0 \\
     0 & 0 & 0 & 0 & 0 & 0 & 0 & 0 & 0 & 0 & 1 & 0 & 1 \\
     \hline
NdP$_{14}$ &dP$_{8}$& dP$_{8}$  &  dP$_{5}$ & NdP$_{9}$  & NdP$_{9}$  &   &   & dP$_3$&  K3& dP$_2$  & dP$_2$&W   \\
        \hline
\end{tabular}
\end{center}
where K3 denotes a K3 surface, dP$_n$ are del Pezzo divisors (with $h^{1,1}=1+n$ and $n\leq 8$), NdP$_n$ refers to `non-del Pezzo' divisors which are rigid and have $h^{1,1}=1+n$ but with $n\geq 9$, and W denotes a Wilson divisor.

After constructing the CY manifold as explained in App. \ref{app:A_Kreuzer-Skarke}, the GLSM turns out to be identical to that of the ambient polytope. In this geometry, there are $13$ toric divisors, $9$ of which are linearly independent since $h^{1,1}(X)=9$. One of them, $D_{13}$, defined by $x_{13}=0$, is a Wilson divisor, $D_{13}=D_\W$. The Hodge diamonds for the CY divisors are given in Tab. \ref{tab:Diamond_1}. The Stanley-Reisner (SR) ideal in the Delaunay triangulation is:\footnote{The Delaunay triangulation decomposes polytope faces in triangles whose smallest angle is maximised.}
\begin{align}
{\rm SR}=&\{x_{1} \, x_{2}, x_{1} \, x_{3}, x_{1} \, x_{4}, x_{1} \, x_{12}, x_{2} \, x_{3}, x_{2} \, x_{4}, x_{2} \, x_{8}, x_{2} \, x_{11}, x_{2} \, x_{12}, x_{3} \, x_{4}, x_{3} \, x_{7}, x_{3} \, x_{9}, \nonumber\\
    &x_{3} \, x_{11}, x_{4} \, x_{7}, x_{4} \, x_{8}, x_{4} \, x_{13}, x_{5} \, x_{11}, x_{6} \, x_{10}, x_{6} \, x_{13}, x_{7} \, x_{8}, x_{7} \, x_{11}, x_{7} \, x_{12}, x_{8} \, x_{9}, \nonumber\\
    &x_{9} \, x_{12}, x_{11} \, x_{13}, x_{12} \, x_{13}, x_{1} \, x_{5} \, x_{9}, x_{1} \, x_{5} \, x_{10}, x_{1} \, x_{9} \, x_{13}, x_{2} \, x_{5} \, x_{10}, x_{5} \, x_{7} \, x_{9}, \nonumber\\
    &x_{5} \, x_{7} \, x_{10}, x_{5} \, x_{8} \, x_{10}, x_{5} \, x_{8} \, x_{12}, x_{7} \, x_{9} \, x_{13}\}\,,
\end{align}
while the K\"ahler cone conditions are:
\begin{eqnarray}
&&t_1 - t_8 + t_{12} > 0\,,\qquad t_6 - t_4 + t_{12} > 0\,,\qquad t_7 - t_1 > 0\,,\qquad t_4 + t_8 - t_{12} > 0 \nonumber\\
&&t_7 + t_8  -t_1 - t_{13} > 0\,,\qquad t_3 - t_8 - t_{12} > 0\,, \qquad
t_7-t_2 - t_6 > 0\,,\qquad t_1 + t_2 - 2t_7 > 0\,, \nonumber \\
&&t_6 - t_4 > 0\,, \qquad t_6 + t_{13} > 0\,,\qquad t_8 + t_{12}-t_3-t_6 > 0\,.
\end{eqnarray}

\begin{table}[H]
    \centering
\begin{tabular}{|c|c|c|c|c|c|c|c|c|c|c|c|c|c|c|}
\hline
& $D_1$ & $D_2$ & $D_3$ & $D_4$ & $D_5$ & $D_6$ & $D_7$ & $D_8$ & $D_9$ & $D_{10}$ & $D_{11}$ & $D_{12}$ & $D_{13}$\\
     \hline
    $h^{0,0}$ & 1 & 1 & 1 & 1 & 1 & 1 & 1 & 1 & 1 & 1 & 1 & 1 & 1\\

    $h^{1,0}$ & 0 & 0 & 0 & 0 & 0 & 0 & 2 & 2 & 0 & 0 & 0 & 0 & 1\\

    $h^{2,0}$ & 0 & 0 & 0 & 0 & 0 & 0 & 0 & 0 & 0 & 1 & 0 & 0 & 0\\

    $h^{1,1}$ & 15 & 9 & 9 & 6 & 10 & 10 & 2 & 3 & 4 & 20 & 3 & 3 & 10\\
        \hline
\end{tabular}\,.
\caption{Hodge diamonds for all CY divisors.}
\label{tab:Diamond_1}
\end{table}

The orientifold involution considered in this case is given by:
\begin{equation}
    \sigma:\,x_8\rightarrow -x_8\,,
\end{equation}
which leaves all lattice points of the polytope invariant. The fixed-point set of the toric ambient four-fold intersected with the CY hypersurface is:
\begin{align}
\operatorname{Fix}(\sigma)=&\{D_8,D_9,P_{2,7,13},P_{1,7,13},P_{4,11,12},P_{1,5,6,7},P_{2,5,6,7},P_{3,5,6,12},P_{4,5,6,12}\}\,,\\
\#\operatorname{Fix}(\sigma)=&\{1,1,4,4,1,1,1,1,1\}\,,
\end{align}
where $P_{i,j,k,(s)} \equiv D_i\cap D_j\cap D_k\left(\cap \,D_s\right)$ denotes the intersection of three/four divisors. Note, in particular, that the intersection of four divisors is a point in the ambient space which belongs also to the CY hypersurface. Consequently, there are two O7-planes on the divisors $D_8$ and $D_9$. Furthermore, $8$ O3-planes sit on $D_{13}$, implying $N^\W_\Ot=8$.

There is no intersection between $D_{13}$ and $D_9$ but the Wilson divisor intersects the other O7 on $D_8$ since $k_{1,8,13}\neq 0$. Hence, an E3-instanton on $D_{13}$ would be transversally invariant, resulting in a perfect candidate for an $O(1)$ instanton. However, no poly-instanton contribution to the superpotential is generated. In fact, the intersection numbers between the O7-planes and the Wilson divisor relevant for the zero mode counting are:
\begin{equation}
k_{8,13,13}= k_{9,13,13}=0\,.
\end{equation}
As discussed in Sec. \ref{sec:Wilson_naive_acc}, this implies $\chi(D_{13},\mathcal{O}_{D_{13}})=2$ with $h^{1,0}_+(D_{13})=0$ and $h^{1,0}_-(D_{13})=1$, which is the wrong zero mode configuration to generate poly-instanton effects. Interestingly, if the O3-planes were ignored, this example might seem to give rise to non-zero poly-instantons since the holomorphic Lefschetz number would naively be $\chi(D_{13},\mathcal{O}_{D_{13}})=0$ with $h^{1,0}_+(D_{13})=1$ and $h^{1,0}_-(D_{13})=0$.

\subsection{An example with no poly-instantons on a K3-like divisor}

Let us focus on the polytope number $22$ of the KS database with $h^{1,1}=2$. Its vertices read:
\begin{equation}
\begin{pmatrix}
    v_1\\v_2\\v_3\\v_4
\end{pmatrix}=
\begin{pmatrix}
     \{1,  0,  0,  0\}\\
     \{-3,  1, -2, -1\}\\
       \{ 0,  0,  0,  1\}\\
       \{ 0,  0,  1,  0\}\\
       \{-2, -1,  0,  0\}\\
       \{ 0,  1,  0,  0\}
\end{pmatrix}\,,
\end{equation}
and the GLSM charge matrix looks like:
\begin{equation}
\label{eq:GLSM_1}
\begin{tabular}{|c|c|c|c|c|c|}
\hline
$x_1$ &$x_2$ & $x_3$ & $x_4$ &$x_5$ & $x_6$\\
     \hline
      5&  1&  1&  2&  1&  0\\
      -3& -1& -1& -2&  0&  1\\
      \hline
      & K3-$\ell$ & K3-$\ell$ & & & NdP$_8$\\
     \hline
\end{tabular}\,
\end{equation}
where K3-$\ell$ stays for `K3-like' and indicates a divisor which has the same Hodge diamond as a K3 surface but with a non-zero first Chern class. The Hodge numbers of all CY divisors are listed in Tab. \ref{tab:Diamond_2} and the SR ideal in the Delaunay triangulation is given by:
\begin{equation}
{\rm SR}=\{x_1x_5x_6,x_2x_3x_4\}\,.
\end{equation}

\begin{table}[H]
    \centering
\begin{tabular}{|c|c|c|c|c|c|c|c|}
\hline
     & $D_1$ & $D_2$ & $D_3$ & $D_4$ & $D_5$ & $D_6$\\
     \hline
    $h^{0,0}$ & 1 & 1 & 1 & 1 & 1 & 1\\

    $h^{1,0}$ & 0 & 0 & 0 & 0 & 0 & 0\\

       $h^{2,0}$ & 27 & 1 & 1 & 3 & 2 & 0\\

        $h^{1,1}$ & 182 & 20 & 20 & 40 & 29 & 9\\
        \hline
\end{tabular}\,.
\caption{Hodge diamond for all CY divisors.}
\label{tab:Diamond_2}
\end{table}

There are $6$ toric divisors of which $2$ are linearly independent since $h^{1,1}(X)=2$. Among those $6$ divisors, $2$ are K3-$\ell$, namely $D_2$ and $D_3$, and belong to same homology class. Hence, from now on, we shall focus just on $D_2$. The intersection form in the basis $\{D_2,D_6\}$ is:
\begin{equation}
\label{eq:Int_1}
I=D_2D_2D_6-D_2D_6D_6+D_6D_6D_6\,,
\end{equation}
and the CY volume becomes:
\begin{equation}
    \mathcal{V}=\frac{1}{6}\left(3t_2^2t_6-3t_2t_6^2+t_6^3\right)\,,
\end{equation}
while the K\"ahler cone is defined by $2t_2>5t_6>0$. The intersection form \eqref{eq:Int_1} shows that $D_2$ is not an actual K3 surface since its first Chern class is non-zero: 
\begin{equation}
\int_{D_2\cap D_6}c_1(D_2)=-\int_{X}\hat{D}_2\wedge \hat{D}_2\wedge \hat{D}_6\neq0\qquad\Rightarrow\qquad c_1(D_2)\neq 0\,.
\end{equation}
The K\"ahler cone conditions for such a CY manifold are $t_2>t_6>0$. Let us focus on the orientifold involution: 
\begin{equation}
\sigma:\,x_6\leftrightarrow -x_6\,,
\end{equation}
with fixed points:
\begin{align}
{\rm Fix}(\sigma)=&\{D_5,D_6,D_1\cap D_2\cap D_3\}\,,\\
\#{\rm Fix}(\sigma)=&\{1,1,2\}\,.
\end{align}
Hence there are two O7-planes localised on $D_5$ and $D_6$, while $2$ O3-planes lie at the intersection locus $D_1\cap D_2\cap D_3$. The K3-$\ell$ divisor $D_2$ intersects the O7-planes and it is transversally invariant, and so it is a promising candidate for an $O(1)$ E3-instanton that generates poly-instanton effects. However, its zero mode configuration would be the wrong one. In fact, the relevant intersection numbers for the generalised Lefschetz fixed point theorem turn out to be:
\begin{equation}
k_{225}=k_{226}=1\,.
\end{equation}
Following the general considerations of Sec. \ref{sec:K3_naive_exc}, one would obtain $\chi^\sigma(D_2,\mathcal{O}_{D_2})=0$ with $h^{2,0}_+(D_2)=0$ and $h^{2,0}_-(D_2)=1$, which does not allow for non-zero poly-instanton effects.

For completeness, let us point out that this CY geometry features a rigid divisor, $D_6$, which corresponds to the location of an O7-plane. An E3-brane wrapping this divisor would, therefore, yield a $USp$ instanton which does not contribute to the superpotential. However, if the D7-tadpole is cancelled by having $4$ D7-branes (and $4$ image D7-branes) on $D_6$, a non-zero superpotential would arise from gaugino condensation within the pure $SO(8)$ theory living on $D_6$:\footnote{It is easy to check that an appropriate choice of the $B$-field compatible with FW anomaly cancellation allows for a vanishing gauge flux on $D_6$.}
\begin{equation}
W_{\rm np}=A_6\,e^{-\frac{\pi}{3}\,T_6}\,,
\end{equation}
where $T_6=\tau_6+i\, \theta_6$ is the complexified K\"ahler modulus with:
\begin{equation}
\tau_6\equiv{\rm Vol}(D_6)=\frac12 \int_X J\wedge J\wedge\hat{D}_6 = \frac12\left(t_2-t_6\right)^2\qquad\text{and}\qquad \theta_6=\int_{D_6}C_4\,.
\end{equation}
Note that $D_5$ is instead a deformation divisor, and so matter in the adjoint representation, if not lifted by fluxes, would in general forbid gaugino condensation on a stack of D7-branes wrapped around $D_5$ (as might be required by D7 tadpole cancellation).

One could wonder whether an orientifold involution which flips the sign of a different coordinate could allow for non-zero poly-instantons on the K3-like divisor $D_2$. However, this does not seem to be possible in this CY geometry. For example, one could also consider the orientifold involution defined by the action:
\begin{equation}
    \sigma:\, x_1 \rightarrow -x_1\,,
\end{equation}
which yields $h_+^{1,2}=h_-^{1,1}=0$ and the following fixed-point set:
\begin{align}
    \operatorname{Fix}(\sigma)=&\{D_1, \, D_2\cap D_3\cap D_5, \, D_2\cap D_3\cap D_6\}\,,\\
    \#\operatorname{Fix}(\sigma)=&\{1, \, 1, \, 1\}\,,
\end{align}
The divisor where the O7-plane is localised corresponds to $D_1$ which intersects $D_2$ that is transversally invariant. The relevant intersection number is $k_{122}=2$ which would again lead to a vanishing equivariant holomorphic Euler characteristic, $    \chi^\sigma(D_2,\mathcal{O}_{D_2}) = 0$ with $h^{2,0}_+(D_{2})=0$ and $h^{2,0}_-(D_{2})=1$. As explained above, this zero mode configuration does not generate any poly-instanton correction to the superpotential.

\subsection{An example with poly-instantons on a K3-like divisor}
\label{sec:Flux_Naive_Exc}

Another example is given by the polytope number $161$ of the KS database which is favourable and features $h^{1,1}=3$. Its vertices are given by:
\begin{equation}
\begin{pmatrix}
    v_1\\v_2\\v_3\\v_4\\v_5\\v_6\\v_7
\end{pmatrix}=
\begin{pmatrix}
     \{ 1&  0&  0&  0\}\\
       \{-2& -1&  0&  0\}\\
       \{-2&  1& -1& -1\}\\
       \{-1&  0& -1&  1\}\\
       \{ 0&  0&  0&  1\}\\
       \{ 0&  0&  1&  0\}\\
       \{ 0&  1&  0&  0\}
\end{pmatrix}\,,
\end{equation}
and the GLSM charge matrix is:
\begin{equation}
\label{eq:GLSM_3}
\begin{tabular}{|c|c|c|c|c|c|c|}
\hline
     $x_1$ &$x_2$ & $x_3$ & $x_4$ &$x_5$ & $x_6$ & $x_7$\\
     \hline
       1&  0&  0&  1& -1&  1&  0\\
        0&  1&  1& -4&  5& -3&  0\\
        0&  1&  0& -2&  2& -2&  1\\
        \hline
        & & K3-$\ell$ & NdP$_9$ & NdP$_9$ & & dP$_8$\\
        \hline
\end{tabular}
\end{equation}
while the SR ideal looks like:
\begin{equation}
{\rm SR}=\{x_3x_5,x_1x_2x_7,x_1x_4x_6,x_2x_5x_7,x_3x_4x_6\}\,.
\end{equation}
There are $7$ prime toric divisors, of which only $3$ are linearly independent since $h^{1,1}(X)=3$. The Hodge diamonds for all divisors are listed in Tab. \ref{tab:Diamond_3}.

\begin{table}[H]
    \centering
\begin{tabular}{|c|c|c|c|c|c|c|c|c|}
\hline
     & $D_1$ & $D_2$ & $D_3$ & $D_4$ & $D_5$ & $D_6$ & $D_7$\\
\hline
    $h^{0,0}$ & 1 & 1 & 1 & 1 & 1 & 1 & 1\\
 
    $h^{1,0}$ & 0 & 0 & 0 & 0 & 0 & 0 & 0\\

       $h^{2,0}$ & 24 & 2 & 1 & 0 & 0 & 2 & 0\\

        $h^{1,1}$ & 164 & 29 & 20 & 10 & 10 & 30 & 9\\
        \hline
\end{tabular}\,.
\caption{Hodge diamonds for all CY divisors.}
\label{tab:Diamond_3}
\end{table}

Among those $7$ divisors, one of them is K3-like, namely
$D_3$, obtained by setting $x_3=0$. Indeed, as can be seen from Tab. \ref{tab:Diamond_3}, $D_3$ shares the same Hodge numbers as a K3 surface but its first Chern class is non-zero. This can be inferred from the intersection polynomial which in the basis $\{D_3,D_4,D_7\}$ takes the form:
\begin{equation}
I=D_3D_4D_7+D_3D_3D_7-D_4D_4D_7-D_3D_7D_7-D_4D_7D_7+D_7D_7D_7\,.
\label{eq:Int_2}
\end{equation}
The fact that $k_{337}\neq 0$ then implies that $D_3$ is only K3-like:
\begin{equation}
\int_{D_3\cap D_7}c_1(D_3)=-\int_X\hat{D}_3\wedge \hat{D}_3\wedge \hat{D}_7=-1 \quad\Rightarrow \quad c_1(D_3)\neq 0\,.
\end{equation}
The volume of this K3-like divisor is:
\begin{equation}
\tau_3 = \frac12 \int_X J\wedge J \wedge \hat{D}_3 = \frac{t_7}{2} \left(2 t_3 + 2 t_4 - t_7 \right),
\end{equation}
and the fact that $h^{2,0}(D_3)=1$ makes it a suitable candidate for generating poly-instanton effects. The overall CY volume becomes:
\begin{equation}
\mathcal{V}=\frac16 \left(6 t_3t_4t_7+ 3 t_3^2t_7- 3 t_3t_7^2-3 t_4^2t_7- 3 t_4t_7^2+ t_7^3\right).
\end{equation}
A good approximation of the CY K\"ahler cone is obtained by taking the union of the K\"ahler cones of all ambient spaces which give the same CY hypersurface. Following this procedure for our example, we find:
\begin{equation}
t_3-t_4-t_7>0\,,\qquad t_7>0\,,\qquad t_4>0\,.
\end{equation}
This CY geometry features $3$ rigid divisors, $D_4$, $D_5$ and $D_7$, which can all, in principle, give rise to non-perturbative corrections to $W$. If this is indeed the case, at leading non-perturbative order $W$ would already depend on $3$ linearly independent K\"ahler moduli, rendering any potential poly-instanton effect negligible. In what follows, we shall therefore choose an orientifold involution and a gauge flux configuration which will allow just for $T_5$-dependent instanton contributions to $W$. The volumes of the divisors $D_5$ and $D_7$ are:
\begin{equation}
\tau_5 = \frac12 \int_X J\wedge J \wedge \left(\hat{D}_3-\hat{D}_4\right) = 2\,t_4\, t_7\,,
\end{equation}
and:
\begin{equation}
\tau_7 = \frac12 \int_X J\wedge J \wedge \hat{D}_7 = \frac12\left(t_3^2 + t_7^2 - t_4^2 + 2 t_3 t_4 -2t_3 t_7 - 2 t_4 t_7\right).
\end{equation}
Note that only $D_5$ is shrinkable (for $t_4\to 0$) without forcing the overall volume also to collapse to zero. Moreover, the intersection curves between pairs of divisors relevant for our discussion can be computed as described in App. \ref{app:A_Kreuzer-Skarke} and are:
\begin{eqnarray}
D_3\cap D_4 &=& T^2\,,\qquad D_3\cap D_5 = \emptyset\,,\qquad D_3\cap D_7 = T^2   \nonumber \\  
D_4\cap D_5 &=& T^2\,,\qquad D_4\cap D_7 = \mathbb{P}^1\,,\qquad D_5\cap D_7 = \mathbb{P}^1\,.
\label{eq:Inter_Curv_1}
\end{eqnarray}
We shall consider the orientifold involution:
\begin{equation}
\sigma:\, x_4 \rightarrow -x_4\,,
\end{equation}
whose fixed-point set is given by:
\begin{align}
\operatorname{Fix}(\sigma)=&\{D_4, \, D_1\cap D_3 \cap D_7, \, D_1 \cap D_2 \cap D_3, \, D_1\cap D_5\cap D_6\}\,, \nonumber \\
\#\operatorname{Fix}(\sigma)=&\{1, \, 3, \, 5, \, 4\}\,.
\label{eq:Fix_1}
\end{align}
There is an O7-plane localised on $D_4$ together with several O3-planes. We shall cancel the D7-tadpole by placing $4$ D7-branes (and $4$ image D7-branes) on top of the O7-plane in $D_4$. Let us analyse all potential non-perturbative effects in details:
\begin{itemize}
\item \textbf{dP$_8$ divisor $D_7$:} The rigid divisor $D_7$ is transversally invariant and intersects $D_4$ in a $\mathbb{P}^1$, as can be seen from (\ref{eq:Inter_Curv_1}). An E3-brane wrapping $D_7$ would therefore lead to an $O(1)$ instanton that contributes to $W$ if the total gauge flux on $D_7$ is vanishing, with the exception of potentially non-zero contributions on curves which are trivial in the CY. The general structure of the flux on the E3-brane wrapping $D_7$ imposed by FW anomaly cancellation is (ignoring fluxes on trivial curves):
\begin{equation}
\mathcal{F}_7 = \iota^*_{D_7}\left(\frac12 \hat{D}_7 + f_{73} \hat{D}_3+f_{74}\hat{D}_4+f_{77}\hat{D}_7\right)-\iota^*_{D_7}\,B\,,\quad f_{7i} \in \mathbb{Z}\,\, \forall\,i=3,4,7\,.
\end{equation}
Choosing $B=\frac12\hat{D}_3$, it is easy to check that the reduction of $\frac12\left(\hat{D}_7-\hat{D}_3\right)$ on $D_7$ is an integral form which can be set to zero by an appropriate choice of the integer coefficients $f_{73}$, $f_{74}$ and $f_{77}$, resulting in $\mathcal{F}_7=0$. Finally, we have also to ensure the absence of any chiral and vector-like zero mode at the intersection between the E3-brane wrapping $D_7$ and the D7-stack on $D_4$. Vector-like zero modes are absent since $D_4\cap D_7=\mathbb{P}^1$ \cite{Blumenhagen:2012kz}, while the number of chiral intersections is counted by:
\begin{equation}
I_{47} = \int_X \hat{D}_4\wedge \hat{D}_7\wedge \left(\mathcal{F}_4 - \mathcal{F}_7\right) =  \int_X \hat{D}_4\wedge \hat{D}_7\wedge \mathcal{F}_4\,,
\label{I47}
\end{equation}
where the flux $\mathcal{F}_4$ on the D7-branes wrapping $D_4$ can be expanded as:
\begin{equation}
\mathcal{F}_4 = \iota^*_{D_4}\left(f_{43} \hat{D}_3+f_{44}\hat{D}_4+f_{47}\hat{D}_7\right),\quad f_{4i} \in \mathbb{Z}\,\, \forall\,i=3,4,7\,.
\label{F4}
\end{equation}
Note that FW anomaly cancellation requires a half-integer contribution of the form $F_4\supset \iota^*_{D_4}\frac12\hat{D}_4$ but the pullback on $D_4$ of $\frac12\left(\hat{D}_4-\hat{D}_3\right)$ is again an integral form, justifying the expansion of $\mathcal{F}_4$ in (\ref{F4}) just in terms of integral fluxes. Plugging (\ref{F4}) in (\ref{I47}) we end up with:
\begin{equation}
I_{47} = f_{43} - f_{44} - f_{47} = 0\qquad\Leftrightarrow\qquad f_{43} = f_{44} + f_{47}\,,
\label{I47new}
\end{equation}
which is the condition on the gauge fluxes on the D7-branes wrapped on $D_4$ to generate a superpotential of the form:
\begin{equation}
W_{\rm np}\supset A_7\,e^{-2\pi T_7}\,.    
\label{Wnp7}
\end{equation}

\item \textbf{NdP$_9$ divisor $D_5$:} The divisor $D_5$ is another ideal candidate to generate a single $O(1)$ instanton correction to $W$ since it is rigid, transversally invariant and intersects the O7-plane on a $T^2$, as can be seen from (\ref{eq:Inter_Curv_1}). The gauge flux on an E3-brane wrapping $D_5$ can be made purely trivial since the pullback on $D_5$ of $\frac12\left(\hat{D}_5-\hat{D}_3\right)$ is an integral form. The number of chiral intersections between the E3-instanton on $D_5$ and the D7-branes on $D_4$ is:
\begin{equation}
I_{45} = \int_X \hat{D}_4\wedge \hat{D}_5\wedge \left(\mathcal{F}_4 - \mathcal{F}_5\right) = 2\, f_{47}  = 0 \,,
\label{I45}
\end{equation}
which is an additional condition on the fluxes on the D7-stack wrapping $D_4$ to give rise to another instanton term in $W$:
\begin{equation}
W_{\rm np}\supset A_5\,e^{-2\pi T_5} = A_5\,e^{-2\pi\left(T_3-T_4\right)}\,.    
\label{Wnp5}
\end{equation}
Combining (\ref{I45}) with (\ref{I47new}), we finally have:
\begin{equation}
f_{43} = f_{44} \qquad\text{and}\qquad f_{47} = 0 \,,
\label{fluxon4}
\end{equation}
Let also point out that vector-like zero modes at the $D_5$-$D_4$ intersection can be lifted by turning on appropriate fluxes on curves of $D_4$ and $D_5$ which are trivial in the CY.

\item \textbf{NP$_9$ divisor $D_4$:} The third rigid divisor $D_4$ corresponds to the locus of the O7-plane and it is wrapped by a stack of $4$ D7-branes (plus their images). If the gauge flux on this stack of branes vanishes, i.e. the condition (\ref{fluxon4}) is satisfied trivially for $f_{43}=f_{44}=0$, the resulting supersymmetric field theory on the D7-branes wrapping $D_4$ is a pure $SO(8)$ gauge theory that undergoes gaugino condensation. Hence the non-perturbative superpotential would feature a third contribution of the form:
\begin{equation}
W_{\rm np}\supset A_4\,e^{-\frac{\pi}{3} T_4} \,. 
\label{Wnp4}
\end{equation}
The total superpotential, given by the sum of the three contributions (\ref{Wnp7}), (\ref{Wnp5}) and (\ref{Wnp4}), would then depend on all $3$ K\"ahler moduli, $T_3$, $T_4$ and $T_7$, making any potential poly-instanton contribution negligible. To consider a situation where poly-instantons can be relevant, we need therefore to focus on the case with $\mathcal{F}_4\neq 0$ where a non-zero flux breaks $SO(8)\to SU(4)\times U(1)$. If $SU(4)$ is not a pure SYM theory anymore, matter fields in general prevent the formation of a gaugino condensate. 

As we shall see below, non-zero poly-instantons require $f_{47}=0$. Under this condition, the net chirality of the zero mode spectrum in the antisymmetric representation of $U(4)$ vanishes:
\begin{equation}
I_{U(4)}^A = 2\int_X \hat{D}_4\wedge\hat{D}_4\wedge\mathcal{F}_4  = - 2\,f_{47} = 0 \,.
\end{equation}
Moreover, the number of vector-like zero modes turns out to be proportional to the number of chiral intersections with the E3-instanton on $D_7$, $n_{\rm vec}\propto I_{47} = \left(f_{43}-f_{44}\right)$. Hence, $A_4=0$ requires $f_{43}\neq f_{44}$ which would then imply also $A_7= 0$.

Note that in this case the anomalous $U(1)$ would generate a D-term potential with an associated moduli-dependent FI-term of the form \cite{Haack:2006cy,Cicoli:2011yh}:
\begin{equation}
\xi_4 \simeq \frac{1}{4\pi\mathcal{V}}\int_X J    \wedge \hat{D}_4 \wedge \mathcal{F}_4  = \left(f_{43}- f_{44}\right)\frac{t_7}{4\pi\mathcal{V}}\,.
\label{FIterm}
\end{equation}

\item \textbf{K3-like divisor $D_3$:} The non-rigid divisor $D_3$ is transversally invariant and intersects the O7-plane in a $T^2$, as can be seen from \eqref{eq:Inter_Curv_1}, and as indicated by the non-vanishing intersection number $k_{347}\neq 0$. Moreover, \eqref{eq:Fix_1} shows that there are $8$ O3-planes on $D_3$ which, together with the fact that $k_{334}=0$, implies $h_+^{2,0}(D_3)=1$ and $h_-^{2,0}(D_3)=0$ for $\chi^\sigma(D_3,\mathcal{O}_{D_3})=2$. This is indeed the right zero more configuration to generate a $T_3$-dependent poly-instanton correction to the superpotential due to an E3-brane wrapping $D_3$. The choice of $B$-field $B=\frac12\hat{D}_3$ ensures that the flux $\mathcal{F}_3$ on this E3-instanton can be made purely trivial. Moreover, $f_{47}=0$ leads to no chiral intersections between the E3-instanton on $D_3$ and the D7-branes on $D_4$:
\begin{equation}
I_{34} = \int_X \hat{D}_3\wedge \hat{D}_4\wedge \left(\mathcal{F}_4 - \mathcal{F}_3\right) = f_{47}  = 0\,.
\label{I34}
\end{equation}
Given that $\mathcal{F}_7=0$ and $\mathcal{F}_5$ is non-zero only on curves which are trivial in the CY, there are also no chiral intersections with the E3-branes on $D_7$ and $D_5$. In principle, there could however still be vector-like zero modes at the intersection between $D_3$ and $D_4$, and $D_3$ and $D_7$ since, as can be seen from (\ref{eq:Inter_Curv_1}), the corresponding intersection locus is in both cases a $T^2$. For a $2$-torus, $h^0(T^2,K_{T^2}^{1/2})=h^1(T^2,K_{T^2}^{1/2})=1$, indicating the presence of two extra vector-like zero modes, which would cause the poly-instanton contribution to vanish. However these modes can be lifted by turning on fluxes on curves which are trivial in the CY but restrict to a non-trivial Wilson line along the $T^2$ \cite{Blumenhagen:2007bn,Blumenhagen:2008zz}. Since $h^{1,1}(D_3)=20$, $h^{1,1}(D_4)=11$ and $h^{1,1}(D_7)=9$, are all greater than $h^{1,1}(X)=3$, there is enough topological freedom to satisfy this condition. 
\end{itemize}
This detailed analysis of all potential non-perturbative contributions to the superpotential leads to two qualitatively different situations, depending on the gauge flux on the D7-branes wrapping $D_4$:
\begin{enumerate}
\item For $f_{43}=f_{44}$ and $f_{47}=0$, the leading non-perturbative superpotential does not involve any poly-instanton effect and looks like:
\begin{equation}
W_{\rm np}\simeq A_4\,e^{-\frac{\pi}{3} T_4} + A_5\,e^{-2\pi T_5} + A_7\,e^{-2 \pi T_7} \,. 
\label{Wtot1}
\end{equation}
\item  For $f_{43}\neq f_{44}$ and $f_{47}=0$, poly-instanton corrections to the superpotential become relevant since $W$ takes the form:\footnote{Subtleties can arise from singular points lying on the K3-like divisor but we checked that in our case this divisor is smooth. Generically, these singularities, if present, are of the star-crossing type, and the fermionic zero modes are counted by replacing the divisor with its normalisation \cite{Gendler:2022qof}.} 
\begin{equation}
W_{\rm np}= A_5\, e^{-2\pi T_5 + A_3\,e^{-2\pi T_3}}
\simeq A_5\, e^{-2\pi T_5}+ A_5\,A_3\, e^{-2\pi T_5}\,e^{-2\pi T_3}\,.
\label{Wnp}
\end{equation}
\end{enumerate}

\subsection{An example with poly-instantons on two blown-up K3 divisors}
\label{sec:Ex_Blow-Up}

We now study the polytope number $95$ of the KS database at $h^{1,1}=5$. Its vertices are:
\begin{equation}
\begin{pmatrix}
    v_1\\v_2\\v_3\\v_4\\v_5\\v_6\\v_7
\end{pmatrix}=
\begin{pmatrix}
     \{ 0&  1&  0&  0\}\\
       \{-2& 0&  0&  -1\}\\
       \{1&  0& 0& 0\}\\
       \{0&  0& 0&  1\}\\
       \{ 0&  0&  1&  0\}\\
       \{ 1&  -1&  -1&  1\}\\
       \{ 2&  -2&  -1&  0\}
\end{pmatrix}\,.
\end{equation}
This polytope is favourable and the GLSM charge matrix is:

\begin{equation}
\begin{tabular}{|c|c|c|c|c|c|c|c|c|}
\hline
$x_1$ &$x_2$ & $x_3$ & $x_4$ &$x_5$ & $x_6$ & $x_7$ & $x_8$ & $x_9$\\
\hline
1 &  0&  -1&  0&  0&  0&  0&  0 & 1\\
0 &  1&  2&  1&  0&  0&  0& 0 & 0\\
0 &  0&  0&  -1&  1&  1& 0& 0 & -1\\
0 &  0&  0&  0&  1&  0&  1& 0 & -2\\
0 &  0&  1&  0& 0&  0&  0& 1 & 0\\
\hline
& b-K3 & NdP$_{10}$ & dP$_7$ & b-K3 & NdP$_{12}$ & dP$_8$ & W & W \\
\hline
\end{tabular}
\end{equation}
This CY geometry contains $9$ prime toric divisors, $5$ of which are linearly independent since $h^{1,1}(X)=5$. Tab. \ref{tab:Diamond_4} lists the Hodge numbers for all divisors, showing that $D_3$, $D_4$, $D_6$ and $D_7$ are all rigid divisors which could yield standard single instanton corrections to $W$. Moreover, there are four candidate divisors which could generate poly-instanton effects: the two Wilson divisors $D_8$ and $D_9$, and two deformation divisors, $D_2$ and $D_5$, denoted as b-K3, which stays for `blown-up K3', since they share the same Hodge diamond as a K3 surface but with larger $h^{1,1}$.

\begin{table}[H]
    \centering
\begin{tabular}{|c|c|c|c|c|c|c|c|c|c|c|}
\hline
     & $D_1$ & $D_2$ & $D_3$ & $D_4$ & $D_5$ & $D_6$ & $D_7$ & $D_8$ & $D_9$\\
     \hline
    $h^{0,0}$ & 1 & 1 & 1 & 1 & 1 & 1 & 1 & 1 & 1\\

    $h^{1,0}$ & 0 & 0 & 0 & 0 & 0 & 0 & 0 & 1 & 1\\

       $h^{2,0}$ & 2 & 1 & 0 & 0 & 1 & 0 & 0 & 0 & 0\\

        $h^{1,1}$ & 31 & 21 & 11 & 8 & 22 & 13 & 9 & 2 & 2\\
        \hline
\end{tabular}\,.
\caption{Hodge diamonds for all Calabi-Yau divisors.}
\label{tab:Diamond_4}
\end{table}

The intersection curves between pairs of divisors relevant for our discussion can be computed by point counting techniques, as described in App. \ref{app:A_Kreuzer-Skarke}, and are:
\begin{eqnarray}
\label{eq:Int_Curv_1}
D_1\cap D_2 &=& \mathcal{C}_3\,,\quad D_1\cap D_3 = \mathcal{C}_3\,,\quad D_1\cap D_4 = T^2\,,\quad   D_1\cap D_5 = T^2\,,\quad D_1\cap D_6 = T^2\,, \nonumber \\  
D_1\cap D_7 &=& \mathbb{P}^1\,,\quad D_1\cap D_8 = \mathbb{P}^1\cup \mathbb{P}^1\,,\quad   D_1\cap D_9 = \emptyset\,, \quad D_2\cap D_3 = \mathbb{P}^1\,,\quad D_2\cap D_4 = \emptyset\,,\nonumber \\
D_2\cap D_5 &=& T^2\,,\quad D_2\cap D_6 = \mathbb{P}^1\,,\quad D_2\cap D_7 = T^2\,,\quad   D_2\cap D_8 = T^2\,,\quad D_2\cap D_9 = \mathbb{P}^1\cup \mathbb{P}^1\,,\nonumber \\
D_3\cap D_4 &=& \mathbb{P}^1\,,\quad D_3\cap D_5 = \mathbb{P}^1\,,\quad D_3\cap D_6 =\mathbb{P}^1\,,\quad   D_3\cap D_7 = \mathbb{P}^1\,,\quad D_3\cap D_8 = \emptyset\,, \nonumber \\ 
D_3\cap D_9 &=& \mathbb{P}^1\cup\mathbb{P}^1\cup\mathbb{P}^1\cup\mathbb{P}^1\,,\quad D_4\cap D_5 = T^2\,,\quad D_4\cap D_6 =T^2\,,\quad D_4\cap D_7 =\emptyset\,, \nonumber \\
D_4\cap D_8 &=& T^2\,,\quad D_4\cap D_9 = \emptyset\,,\quad D_5\cap D_6 = \mathbb{P}^1\cup \mathbb{P}^1\,,\quad D_5\cap D_7 =\emptyset\,,\quad   D_5\cap D_8 = \mathbb{P}^1\cup\mathbb{P}^1\,, \nonumber \\
D_5\cap D_9 &=& T^2\,,\quad D_6\cap D_7 = T^2\,,\quad D_6\cap D_8 = \mathbb{P}^1\cup \mathbb{P}^1\,,\quad D_6\cap D_9 =\mathbb{P}^1\cup \mathbb{P}^1\,,\quad   D_7\cap D_8 = \emptyset\,,\nonumber \\
D_7\cap D_9 &=& T^2\,, \quad D_8\cap D_9 = \emptyset\,,
\label{Intersect}
\end{eqnarray}
where intersections denoted as $\mathcal{C}_n$ are curves of genus $n$. The SR ideal for the first triangulation in \texttt{CYTools} reads:
\begin{equation}
{\rm SR}=\{ x_1 x_9, x_2 x_4, x_3  x_8, x_4  x_7, x_4  x_9,  x_5 x_7,  x_7  x_8, x_8  x_9,  x_1 x_5 x_6,  x_2 x_3 x_6\}\,,
\end{equation}
while the intersection polynomial in the basis $\{D_2,D_3,D_6,D_7,D_8\}$,  is:
\begin{eqnarray}
I&=&D_2D_3D_7+D_2D_6D_7+3D_2D_6D_8+D_3D_6D_7-D_2D_2D_2-D_2D_2D_6 \nonumber \\
&+& D_2D_2D_7+3D_2D_2D_8-2D_2D_3D_3-D_2D_6D_6-D_2D_7D_7-3D_2D_8D_8 \nonumber \\
&-& D_3D_3D_3-D_3D_3D_6-D_3D_3D_7-D_3D_6D_6-D_3D_7D_7 \nonumber\\
&-&3D_6D_6D_6+D_6D_6D_7-D_6D_7D_7-6D_6D_8D_8+D_7D_7D_7\,,
\label{IntPolyn}
\end{eqnarray}
and the K\"ahler cone conditions are:
\begin{eqnarray}
&& t_2 - 2t_8 > 0\,,\qquad t_2 + 3t_3 + t_6 - t_7 > 0\,,\qquad t_6 -t_3> 0\,, \nonumber \\
&& t_7 > 0\,,\qquad t_7 + 3t_8 -t_2 - t_6 > 0\,,\qquad t_6 < 0\,.
\end{eqnarray}
We focus on the orientifold involution:
\begin{equation}
\sigma:\,x_6\rightarrow-x_6\,.
\end{equation}
whose fixed-point set is given by:
\begin{align}
    \operatorname{Fix}(\sigma)=&\{D_6\,, \, P_{259}\,, \, P_{279}\,, \, P_{1237}, \, P_{1235}\,, \,P_{1258}\,,\, P_{1458}, \, P_{1345}\,\}\,,\\
    \#\operatorname{Fix}(\sigma)=&\{1, \, 2, \, 2, \, 1, \, 1, \, 1, \,1, \,1\}\,,
\end{align}
where $P_{i\dots}$ denotes the intersection of the respective divisors. Consequently, an O7-plane lies on $D_6$, and the number of O3-planes on the Wilson and blown-up K3 divisors is:
\begin{equation}
N_\Ot^{D_2}=7\,,\qquad N_\Ot^{D_5}=6\,,\qquad N_\Ot^{D_8}=2\,, \qquad N_\Ot^{D_9}=4\,.
\end{equation}
As in the previous examples, the D7-tadpole is cancelled by placing $4$ D7-branes (and their $4$ orientifold image D7-branes) on top of the O7-plane on $D_6$. Choosing the $B$-field as $B = \frac{1}{2}\hat{D}_1$, the gauge fluxes in the worldvolume theories of all wrapped D7- and E3-branes can be set to zero in a way compatible with FW anomaly cancellation, i.e.  $\mathcal{F}_i=0$ $\forall i=2,...,9$. Let us analyse the possibility to generate different non-perturbative contributions to the superpotential:
\begin{itemize}
\item \textbf{Rigid divisors $D_3$, $D_4$, $D_6$ and $D_7$:} The D7-branes on $D_6$ support a pure $SO(8)$ theory that undergoes gaugino condensation. All the other rigid divisors are transversally invariant under the orientifold involution, and they all intersect the O7-plane on $D_6$, as can be seen from (\ref{Intersect}). Thus, an E3-brane wrapped on each of these $4$-cycles yields an $O(1)$ instanton that contributes to $W$. Indeed, the fact that all gauge fluxes are zero guarantees the absence of chiral intersections with the D7-branes on $D_6$. Moreover, the intersection of $D_3$ with $D_6$ is a $\mathbb{P}^1$, signalling also the absence of vector-like zero modes. On the other hand, both $D_4$ and $D_7$ intersect $D_6$ on a $T^2$, indicating the need to turn on trivial fluxes to lift vector-like zero modes.

\item \textbf{Blown-up K3 divisors $D_2$ and $D_5$:} Both of them are transversally invariant and intersect the O7-plane on $D_6$, as can be seen from (\ref{Intersect}). Hence an E3-brane wrapped on them can potentially yield poly-instanton effects. Following the general analysis performed in Sec. \ref{sec:K3_naive_exc}, we find that both blown-up K3 divisors have indeed the right zero modes to support poly-instanton effects since $\chi(D_2,\mathcal{O}_{D_2})=   \chi(D_5,\mathcal{O}_{D_5})=2$ with:
\begin{equation}
N_{\Ot}^{D_2} = 7=8+k_{226} = 8-1\qquad\Rightarrow\qquad \begin{cases}
       h_+^{2,0}(D_2)=1\,,\\
       h_-^{2,0}(D_2)=0\,,
   \end{cases}
\end{equation}
and:
\begin{equation}
N_{\Ot}^{D_5} = 6=8+k_{556}=8-2\qquad\Rightarrow\qquad \begin{cases}
       h_+^{2,0}(D_5)=1\,,\\
       h_-^{2,0}(D_5)=0\,,
   \end{cases}
\end{equation}
where the intersection number $k_{556}$ follows from:
\begin{equation}
D_5 = D_6+D_7\qquad\Rightarrow\qquad  k_{556} = k_{666}+2 k_{667} + k_{677}  = -3 + 2 -1 = -2\,.
\end{equation}
Given that all $2$-form fluxes are zero, no chiral zero modes are present. The only obstacle against the generation of poly-instanton effects could therefore come from vector-like zero modes. However, as can be seen from \eqref{eq:Int_Curv_1}, the intersections of the two blown-up K3 divisors, $D_2$ and $D_5$, with the $4$ rigid divisors, $D_3$, $D_4$, $D_6$ and $D_7$, are either an empty set, a $\mathbb{P}^1$, a $T^2$ or a disjoint unions of two $\mathbb{P}^1$ curves. In the $T^2$ case, vector-like zero modes need to be lifted by turning on appropriate trivial gauge fluxes, while in all other cases these fermionic zero modes are simply absent. 

\item \textbf{Wilson divisors $D_8$ and $D_9$:} Both Wilson divisors intersect the O7-plane and are transversally invariant under the orientifold involution but have the wrong zero mode structure to generate poly-instanton effects. In fact, following the general analysis of Sec. \ref{sec:Wilson_naive_acc}, we have $\chi(D_8,\mathcal{O}_{D_8})= \chi(D_9,\mathcal{O}_{D_9})=2$ with:
\begin{equation}
N_{\Ot}^{D_8}=2=8+k_{688}=8-6\qquad\Rightarrow\qquad \begin{cases}
       h_+^{1,0}(D_8)=0\,,\\
       h_-^{1,0}(D_8)=1\,,
   \end{cases}
\end{equation}
and:
\begin{equation}
N_{\Ot}^{D_9}=4= 8+k_{699}=8-4\qquad\Rightarrow\qquad \begin{cases}
h_+^{1,0}(D_9)=0\,,\\
h_-^{1,0}(D_9)=1\,,
\end{cases}
\end{equation}
where $k_{699}=-4$ follows from $D_9 = 2D_2+D_8 -D_3-D_6-2 D_7$ and the intersection polynomial (\ref{IntPolyn}). 
\end{itemize}
Putting all these results together, the non-perturbative superpotential looks like:
\begin{eqnarray}
W_{\rm np} &=&  A_3\,e^{-2\pi T_3+ A_2\,e^{-2\pi T_2} + A_5\,\,e^{-2\pi T_5}} + A_4\,e^{-2\pi T_4+ B_2\,e^{-2\pi T_2} + B_5\,\,e^{-2\pi T_5}} \nonumber \\
&+& A_7\,e^{-2\pi T_7+ C_2\,e^{-2\pi T_2} + C_5\,\,e^{-2\pi T_5}}+A_6\,e^{-\frac{\pi}{3} T_6}\,. 
\end{eqnarray}

\section{Axion quintessence from new poly-instanton effects}
\label{SecQuint}

In this section we discuss a phenomenological application of the explicit CY example of Sec. \ref{sec:Flux_Naive_Exc}. We shall realise an axionic quintessence model for the case when poly-instanton effects are relevant.

A crucial prerequisite for any application to phenomenology is moduli stabilisation. We shall assume, as usual in type IIB, that $3$-form fluxes fix the complex structure structure moduli and the axio-dilaton at $g_s\ll 1$, and show that the system admits an LVS-like minimum for the K\"ahler moduli where the volume is exponentially large in terms of the small modulus $\tau_5$. A large CY volume allows to organise the scalar potential $V$ as an expansion in inverse powers of $\mathcal{V}$ as follows:
\begin{itemize}
\item \textbf{Leading order}: The leading order contribution to $V$ is a D-term scalar potential. For definiteness we shall focus on the case with $f_{43}= f >0$ and $f_{44}=0$. This flux on the D7-branes wrapped on $D_4$ breaks $SO(8)$ to $U(4)$. Accordingly, the adjoint representation of $SO(8)$ decomposes into representations of $SU(4)$ $\textbf{R}_q$ with charge $q$ under the diagonal $U(1)$ as $\textbf{28}\to \textbf{16}_0 \oplus \textbf{6}_{+2} \oplus \textbf{6}_{-2}$. Recalling the expression of the FI-term (\ref{FIterm}), the associated D-term potential looks like:
\begin{equation}
V_D =\frac{g^2}{2}\left(\sum_i q_i |\varphi_i|^2-\xi_4\right)^2\simeq \frac{1}{2\tau_4}\left(2\sum_j |\varphi_{+2,j}|^2-2\sum_k|\varphi_{-2,k}|^2- \frac{f}{4\pi}\frac{t_7}{\mathcal{V}}\right)^2\,,
\end{equation}
where $\varphi_i$ are canonically normalised charged matter fields. A leading order supersymmetric stabilisation requires a vanishing D-term potential which is obtained at:
\begin{equation}
\varphi_{-2,k}=0\,\,\forall k\qquad\text{and}\qquad \sum_j |\varphi_{+2,j}|^2  = \frac{f}{8\pi}\,\frac{t_7}{\mathcal{V}}\,.
\label{Dvev}
\end{equation}

\item \textbf{Subleading order}: Different contributions to $V$ arise at subleading order. These are given by: ($i$) the standard LVS potential which includes the $\mathcal{O}(\alpha'^3)$ term proportional to the CY Euler number $\zeta$ and two non-perturbative effects generated by the single instanton in (\ref{Wnp}) with prefactor $A_5$; ($ii$) F-term potential for the charged matter fields $\varphi_{+2,j}$ due to soft supersymmetry breaking which gives them a mass of order the gravitino mass $m_{3/2}\simeq \sqrt{g_s}\,|W_0|/\mathcal{V}$; ($iii$) $t_7$-dependent string loop corrections.

The general expression of the LVS potential is:
\begin{equation}
V_{\rm LVS} \simeq 4\pi^2 g_s A_5^2 K^{55} \frac{e^{-4\pi\tau_5}}{\mathcal{V}^2}- 8\pi g_s |W_0| A_5 \cos\left(2\pi\theta_5\right)\tau_5\frac{e^{-2\pi\tau_5}}{\mathcal{V}^2} + \frac{3\,\zeta\,W_0^2}{4 \sqrt{g_s}\mathcal{V}^3}\,,
\label{VLVS}
\end{equation}
where, without loss of generality, we take $W_0=-|W_0|$ and $A_5$ real and positive. The $55$ element of the inverse K\"ahler metric reads:
\begin{equation}
K^{55} = 4 \tau_5^2 - 4\mathcal{V} \sum_{i=1}^{h^{1,1}} k_{55i}\,r_i\,,
\end{equation}
where $r_i$ are the K\"ahler coordinates in a divisor basis which includes $D_5$. Choosing $\{D_3,D_5,D_7\}$ as the new basis, it is straightforward to realise that: 
\begin{equation}
t_3 = r_3+r_5\,,\qquad t_4=-r_5\,,\qquad t_7=r_7\,.
\end{equation}
Moreover, from the intersection polynomial (\ref{eq:Int_2}) we find:
\begin{equation}
k_{553} = 0\,,\qquad k_{555}=0\,,\qquad k_{557} = - 2\,, 
\end{equation}
which imply:
\begin{equation}
K^{55} =  4 \tau_5^2 + 8\,\mathcal{V} \,t_7\simeq 8\,\mathcal{V} \,t_7\,.
\end{equation}
On the other hand, the F-term potential for the charged matter fields at the VEV (\ref{Dvev}) becomes (for $M_{\rm soft}^2= \lambda\, m_{3/2}^2)$: 
\begin{equation}
V_{\rm soft} \simeq m_{3/2}^2 \sum_j |\varphi_{+2,j}|^2 \simeq g_s\,\tilde{f}\,\frac{|W_0|^2\,t_7}{\mathcal{V}^3}\qquad\text{with}\qquad \tilde{f}\equiv \frac{\lambda\,f}{8\pi}\,.
\label{Vsoft}
\end{equation}
Additionally, string loop corrections to the K\"ahler potential are expected to generate another contribution to the scalar potential of $t_7$ of the form \cite{vonGersdorff:2005bf,Berg:2005ja,Berg:2007wt,Cicoli:2007xp}:
\begin{equation}
V_{\rm loop} \simeq g_s\,c_{\rm loop} \,\frac{|W_0|^2}{\mathcal{V}^3\,t_7}\,,
\label{Vloop}
\end{equation}
where we included a generic coefficient $c_{\rm loop}$. 

After minimising the axion $\theta_5$ at $\langle\theta_5\rangle = k_5\in\mathbb{Z}$, the total potential $V=V_{\rm LVS} + V_{\rm soft} + V_{\rm loop}$ becomes:
\begin{equation}
V\simeq  \frac{g_s\,|W_0|^2}{ \mathcal{V}^3}\left[\frac{3\,\zeta}{4 g_s^{3/2}} - X \,\tau_5 +\frac12\left(X^2+ \tilde{f}\right) t_7 + \frac{c_{\rm loop}}{t_7}\right],
\end{equation}
where:
\begin{equation}
X\equiv 8\pi\,A_5\,e^{-2\pi\tau_5}\frac{\mathcal{V}}{|W_0|}\,.    
\end{equation}
Let us first stabilise $t_7$ at:
\begin{equation}
t_7 =\frac{\sqrt{2\,c_{\rm loop}}}{X}\left(1+\frac{\tilde{f}}{X^2}\right)^{-1/2}\,.
\end{equation}
Integrating out $t_7$, we remain with:
\begin{equation}
V\simeq  \frac{g_s\,|W_0|^2}{ \mathcal{V}^3}\left[\frac{3\,\zeta}{4 g_s^{3/2}} - X \left(\tau_5 - \sqrt{2 c_{\rm loop} }\left(1+\frac{\tilde{f}}{X^2}\right)^{1/2}\right)  \right].
\end{equation}
Solving now $\partial V/\partial \tau_5=0$, we obtain:
\begin{equation}
\tau_5 - \sqrt{2 c_{\rm loop} }\left(1+\frac{\tilde{f}}{X^2}\right)^{1/2} = \frac{1}{2\pi}- \sqrt{2 c_{\rm loop} }\left(1+\frac{\tilde{f}}{X^2}\right)^{-1/2}\frac{\tilde{f}}{X^2}\,.
\end{equation}
In the limit:
\begin{equation}
\frac{\tilde{f}}{X^2}\ll\frac{1}{2\pi \sqrt{2 c_{\rm loop} }}\ll 1\,,
\label{limit}
\end{equation}
the extremum lies approximately at:
\begin{equation}
\langle t_7\rangle\simeq \frac{\sqrt{2\,c_{\rm loop}}}{X}\qquad\text{and}\qquad\langle\tau_5\rangle\simeq \sqrt{2\,c_{\rm loop}}\qquad\Rightarrow\qquad \langle\tau_5\rangle \simeq X\,\langle t_7\rangle\,.    
\label{extr}
\end{equation}
After integrating out $\tau_5$, the scalar potential becomes:
\begin{equation}
V\simeq  V_0\left( 
\frac{a}{X^3} -\frac{1}{X^2} \right)\qquad V_0\equiv  \frac{g_s}{2\pi |W_0|} \left(8\pi\,A_5\right)^3 e^{-6\pi\tau_5}\qquad a \equiv \frac{3\pi\zeta}{2 g_s^{3/2}}\,.
\end{equation}
This potential has an AdS minimum at:
\begin{equation}
\langle X \rangle \simeq \frac{3a}{2} = \frac{9\pi\zeta}{4 g_s^{3/2}}\,,
\end{equation}
which can be uplifted to a Minkowski vacuum by the addition of an appropriate hidden sector. Given that our CY threefold $X$ features $h^{1,2}(X)=103$ and $h^{1,1}(X)=3$, the corresponding Euler number is $\chi(X) = 2\left(h^{1,1}(X)-h^{1,2}(X)\right)=-200$. Hence the coefficient $\zeta$ becomes \cite{Becker:2002nn}:
\begin{equation}
\zeta = -\frac{\zeta(3)\chi(X)}{2(2\pi)^3} \simeq 0.48\,,   
\end{equation}
which gives:
\begin{equation}
\langle X \rangle \simeq \frac{3.42}{g_s^{3/2}}\simeq 16.5\qquad\text{for}\qquad g_s\simeq 0.35\,,
\end{equation}
where we have chosen a small value of $g_s$ to guarantee that perturbation theory is under control. Let us also choose a natural value for the coefficient of the loop corrections, $c_{\rm loop}=1$, which yields:
\begin{equation}
\langle \tau_5\rangle \simeq \sqrt{2 c_{\rm loop}} \simeq 1.4\,.
\end{equation}
Note that the assumed regime (\ref{limit}) is easily verified for $f=1$:
\begin{equation}
\frac{\tilde{f}}{X^2} \simeq 10^{-4}\,\lambda \ll \frac{1}{2\pi\sqrt{2 c_{\rm loop} }}\simeq 0.1\ll 1\,,
\end{equation}
since any supersymmetry breaking mechanism yields $\lambda \lesssim 1$. Moreover, the relations (\ref{extr}) give:
\begin{equation}
\langle t_7\rangle \simeq 0.085\qquad\text{and}\qquad \langle t_4\rangle\simeq 8.3\,.
\end{equation}
One might wonder whether stringy corrections could be important since $t_7$ is fixed close to the wall of the K\"ahler cone at $t_7>0$. However, as shown in \cite{AbdusSalam:2020ywo}, stringy corrections are negligible as long as the volume $t$ of any curve respects the bound:
\begin{equation}
t\gg \frac{1}{\sqrt{g_s}\,(2\pi)^2} \simeq 0.04\,.
\label{StringyCorr}
\end{equation}
It also worth pointing out that the expansion parameter of string loop corrections to the K\"ahler potential is small enough for large CY volume since it scales as:
\begin{equation}
\frac{g_s\,c_{\rm loop}}{\mathcal{V}\,\langle t_7\rangle}\simeq \frac{4}{\mathcal{V}} \ll 1\qquad\text{for}\qquad \mathcal{V}\gg 4\,. 
\end{equation}
Let us now estimate the value of the CY volume at the minimum:
\begin{equation}
\mathcal{V}\simeq \frac{X\,|W_0|}{8\pi\,A_5}\,e^{2\pi\tau_5}\simeq \left(\frac{|W_0|}{A_5}\right) 4766\,,
\label{CYmin}
\end{equation}
which depends on the ratio $(|W_0|/A_5)$ that is expected to take a natural $\mathcal{O}(1)$ value. 

\item \textbf{Sub-subleading order}: The potential studied above features a minimum for all moduli except for the 2 axions $\theta_3$ and $\theta_7$ which are expected to be lifted at sub-subleading order. In what follows, we focus on $\theta_3$ whose potential is generated by poly-instanton effects. The shift symmetry of the other axion $\theta_7$ is expected to be broken by even more subdominant non-perturbative effects, and so we shall consider $\theta_7$ as a spectator field which is effectively massless. $T_3$-dependent poly-instantons give rise to a potential for $\theta_3$ of the form (after fixing $\langle\theta_5\rangle=0$):
\begin{equation}
V_{\rm ax} = \Lambda\left[1- \cos(2\pi\theta_3)\right]\quad\text{with}\quad \Lambda\equiv \frac{8\pi |W_0| A_5 A_3}{\mathcal{V}^2}\left(\tau_5+\tau_3\right)\,e^{-2\pi(\tau_5+\tau_3)}\,.
\end{equation}
This potential could be suitable to realise an axion hilltop quintessence model if $\Lambda\simeq 10^{-122}$ in Planck units. Substituting $\tau_5\simeq \sqrt{2}$ and (\ref{CYmin}), we find:
\begin{equation}
\Lambda\simeq \frac{16.5\,|W_0|^2 A_3}{\mathcal{V}^3} \left(\langle\tau_5\rangle+\tau_3\right)\,e^{-2\pi\tau_3}\,,
\end{equation}
where, at fixed $t_4\simeq 8.3$ and $t_7\simeq 0.085$, both $\tau_3$ and $\mathcal{V}$ depend just on $t_3$. Larger values of $t_3$ imply larger values of $\mathcal{V}$ and $\tau_3$ and, in turn, smaller values of the vacuum energy. The best case scenario to reproduce the observed value of $\Lambda$ corresponds to:
\begin{equation}
\langle t_3\rangle\simeq 480\quad \Rightarrow\quad \langle \mathcal{V}\rangle\simeq 10170\,,\quad \langle \tau_3\rangle\simeq 41.7\,,\quad \Lambda\simeq 7.5\time 10^{-124} \,|W_0|^2 A_3\,. 
\end{equation}
Matching $\Lambda\simeq 10^{-122}$ and $\mathcal{V}\simeq 10170$ requires:
\begin{equation}
A_5\simeq 0.47\,|W_0| \qquad\text{and}\qquad A_3 \simeq 13.3\,|W_0|^{-2}\,,
\end{equation}
showing that the observed value of the vacuum energy can be nicely reproduced for natural values of the underlying parameters, since $W_0\simeq 5$ would give $A_5\simeq 2.3$ and $A_3\simeq 0.5$. Let us finally comment on the decay constant of the canonically normalised axion $\phi_3$. The CY volume can be approximated as:
\begin{equation}
\mathcal{V}\simeq \frac12\,t_3^2 t_7\simeq \frac{1}{\sqrt{2}}\sqrt{\tau_7}\tau_3\,,
\end{equation}
and so the kinetic term for $\theta_3$ scales as:
\begin{equation}
\mathcal{L}_{\rm kin}\supset K_{33}\, \partial_\mu\theta_3\partial^\mu\theta_3 =  \frac{1}{2\langle\tau_3\rangle^2}\,\partial_\mu\theta_3\partial^\mu\theta_3\,.   
\end{equation}
The kinetic terms become canonical for $\theta_3 = \langle\tau_3\rangle\,\phi_3$, giving a quintessence potential of the form (restoring units of $M_p)$:
\begin{equation}
V_{\rm ax} = \Lambda\,M_p^4\,\left[1- \cos\left(\frac{\phi_3}{\mathfrak{f}_3}\right)\right]\qquad\text{with}\qquad    \mathfrak{f}_3 = \frac{M_p}{2\pi \langle\tau_3\rangle}\simeq 10^{16}\,\text{GeV}\,,
\end{equation}
which is large enough to ensure a region close to the maximum where the potential leads to an accelerated expansion \cite{Cicoli:2024yqh,Cicoli:2021skd}.
\end{itemize}

\section{Conclusions}
\label{Concl}

In this work we have investigated new realisations of poly-instanton effects in type IIB CY orientifold compactifications and explored their cosmological implications. Poly-instanton terms in the superpotential of the low-energy 4D effective theory correspond to instanton corrections to the action of a standard E3-instanton. Hence, they feature a double exponential suppression which is very useful to generate some large hierarchies we see in Nature. A promising example is the case of perturbatively flat axions whose potential arises only at poly-instanton order, resulting in a highly suppressed scale of order the current value of the dark energy density. 

Given the potential relevance of poly-instanton effects for moduli stabilisation and axion physics, we performed a general analysis of the instanton zero modes which lead to non-zero poly-instanton corrections to the superpotential. Using a generalised version of the Lefschetz fixed-point theorem, we showed that poly-instantons can arise when an E3-brane wraps a rigid divisor with an additional zero mode which can be either a Wilson line or a deformation. The crucial quantities that determine whether poly-instantons are indeed non-vanishing, are the number of O3-planes localised on the worldvolume of the E3-brane and the topological number $k_{\Et\Et\Os}$ characterising the intersection between the O7-plane and the divisor wrapped by the E3-instanton. Contrary to previous studies, we found that non-zero poly-instantons can arise also for deformation divisors, even if the case of pure K3 surfaces is not allowed. On the other hand, poly-instantons can contribute to the superpotential when the E3-instanton wraps K3-like or blown-up K3 divisors.

We strengthen our general results via four examples where the CY is built as a hypersurface in a toric variety. Each example features a concrete orientifold involution, brane setup and choice of gauge fluxes which are globally consistent. The first two examples illustrate how an explicit setup with a given orientifold involution and intersection numbers can forbid the emergence of poly-instanton corrections when the E3-brane wraps a Wilson or a K3-like divisor. The last two example exhibit instead non-zero poly-instantons for the new cases of E3-instantons wrapping K3-like or blown-up K3 divisors.

For the case of poly-instantons on a K3-like divisor, we analysed in detail K\"ahler moduli stabilisation, including both D- and F-term contributions to the scalar potential. We found an LVS-like minimum inside the K\"ahler cone where the CY volume is large enough, $\mathcal{V}\simeq 10^4$, to trust the $\alpha'$ expansion, and the string coupling is small enough, $g_s\simeq 0.35$, to control loop corrections. The leading contribution to the potential of one of the axions arises from poly-instanton corrections to the superpotential. Interestingly, for natural values of the flux superpotential and the prefactors of non-perturbative effects, the scale of this axion potential turns out to be of order the cosmological constant, $\Lambda\simeq 10^{-122}\,M_p^4$, while its decay constant is around the GUT scale. This setup provides, therefore, an explicit CY embedding of axion hilltop quintessence. 

Our analysis highlights the importance of considering the divisor topology, as well as the full orientifold and intersection data, when searching for poly-instanton effects. Moreover, particular attention has to be paid on the lifting of potential additional chiral or vector-like zero modes by fluxes. The explicit CY constructions presented here provide concrete examples where poly-instanton effects survive all consistency conditions and can have observable cosmological consequences.

There are several directions for future work. A systematic scan of the complete KS database and more general CY constructions, could reveal further classes of geometries supporting poly-instantons. Moreover, given that the path integral involves a sum over all gauge flux configurations compatible with D3-tadpole and FW anomaly cancellation, it is important to check that no choice of gauge fluxes can lift fermionic zero modes \cite{Bianchi:2011qh,Bianchi:2012pn,Cicoli:2012vw,Louis:2012nb}, yielding a fluxed instanton that dominates over unfluxed poly-instantons. Let us also stress that our analysis focused on the case where the E3-instanton intersects the O7-plane on an isolated curve, i.e. a rigid connected curve (which might also consist of multiple rigid disconnected pieces). If instead the intersection is a continuous family of nearby curves, the analysis of instanton zero modes becomes more involved. The expectation is that extra fermionic zero modes, associated to deformations of the curve, should typically prevent a superpotential contribution unless they are lifted. Let us finally mention that it would be interesting to realise  more explicit CY constructions of poly-instanton effects which are relevant, not just for dark energy, but also for applications to inflationary cosmology, reheating and dark matter.

\section*{Acknowledgements}

We would like to thank Mudit Jain and Ling Lin for useful discussions. F.C.\ is supported by the Italian Ministry of Universities and Research (MUR) through the grant “StringGeom” (grant CUP code no.~I33c25000420006).
This article is based upon work from COST Action COSMIC WISPers CA21106, supported by COST (European Cooperation in Science and Technology).

\appendix

\section{E3-instanton zero modes}
\label{sec:Zero_Modes}

Depending on the particular $4$-cycle wrapped by an E3-instanton, a non-perturbative contribution to the F-term superpotential of the 4D EFT could be generated. In this section, the discussion will focus precisely on what are the necessary and sufficient conditions for these non-perturbative corrections to arise. 

The contribution of stringy E3-instantons to the 4D EFT is, in analogy with the gauge theory instanton case:
\begin{equation}
S_{\rm 4D}=\int d\mathcal{M}\,e^{-S_\Et(\mathcal{M})}\,,
\label{eq:Fermi-path-int}
\end{equation}
where $S_\Et$ is the instanton action and $\mathcal{M}$ the instanton moduli space. In a supersymmetric theory, $\mathcal{M}$ is parametrised both by bosonic and fermionic zero modes. Let us schematically denote fermionic zero modes with $\theta_i$. From \eqref{eq:Fermi-path-int} we immediately see that for the grassmannian part of the integral to vanish, each $\theta_i$ should come down from the exponent only once saturating the possibility of the integral to have additional powers of it without vanishing.
This is why the appearance of a certain number of zero modes is a necessary and sufficient condition to get non-perturbative instanton contributions to the superpotential. 

Fermionic zero modes can be of different kinds. This section inspects all of them in the case of E3-instantons, following \cite{Blumenhagen:2009qh}.

\subsubsection*{Universal zero modes}

The first class is the one composed by universal zero modes. They arise from strings with both end-points attached to the same E3-brane, and they are split into two subclasses:
\begin{itemize}
    \item Bosonic zero modes $x^\mu$;
    \item Fermionic zero modes $\theta^\alpha,\overline{\theta}^{\dot{\alpha}}$, $\tau^\alpha$, $\overline{\tau}^{\dot{\alpha}}$.
\end{itemize}
While the former parametrise the position of the E3-brane and are basically Goldstone bosons of the translational symmetry broken by the presence of the brane, the latter are Goldstinos associated to the broken supersymmetries. Depending on the position of the E3-brane with respect to the orientifold, we get different behaviours and different zero modes. There are mainly two cases:
\begin{itemize}
\item E3-brane wrapped on a cycle which is an invariant locus, i.e. $D_{\Et}=D_{\Et'}$;

\item E3-brane wrapped on a non-invariant cycle, i.e. $D_{\Et}\neq D_{\Et'}$.
\end{itemize}

Let us start the discussion of the universal fermionic zero modes from an $\mathcal{N}=2$ setting, before quotienting $X$ by the orientifold involution $\sigma$. The supersymmetry algebra has two subalgebras which will be denoted as $\mathcal{N}=1$, with generators $Q^\alpha$ and $\overline{Q}^{\dot{\alpha}}$, and as $\mathcal{N}=1'$, with generators $Q'^\alpha$ and $\overline{Q'}^{\dot{\alpha}}$. Introducing the orientifold breaks the original supersymmetry algebra into the $\mathcal{N}=1$ subalgebra with generators $Q^\alpha$ and $\overline{Q}^{\dot{\alpha}}$. On the other hand, an instanton wrapping a $\frac{1}{2}$-BPS cycle which is \emph{not} invariant under the orientifold (and therefore would lead to a $U(1)$-instanton) feels locally $\mathcal{N}=2$ supersymmetry. However, due to its localised nature in the 4 extended dimensions, the instanton breaks 4D translational invariance and therefore also half of the supercharges. The preserved supercharges are, however, not those of the $\mathcal{N}=1$ or $\mathcal{N}=1'$ subalgebras, but rather the combination $Q'^\alpha$ and $\overline{Q}^{\dot{\alpha}}$. Associating the Goldstinos $\theta^\alpha$ and $\overline{\theta}^{\dot{\alpha}}$ to the breaking of $\mathcal{N}=1$, and $\tau^\alpha$ and $\overline{\tau}^{\dot{\alpha}}$ to the breaking of $\mathcal{N}=1'$, the fermionic zero modes that appear in the following case are $\theta^\alpha$ and $\overline{\tau}^{\dot{\alpha}}$. If we manage to soak up the anti-holomorphic zero mode $\overline{\tau}^{\dot{\alpha}}$, the grassmannian integral does not vanish and the instanton contributes to the superpotential.

The simplest way to remove the $\overline{\tau}^{\dot{\alpha}}$ zero mode is to consider an E3-brane which does not feel locally the full $\mathcal{N}=2$ supersymmetry, but just a subalgebra with 4 supercharges. Namely, we consider the $4$-cycle wrapped by the E3-brane to be orientifold invariant, namely $D_{\Et}=D_{\Et'}$. One can immediately see how two subcases can arise \cite{Blumenhagen:2009qh,Blumenhagen:2012kz}:
\begin{itemize}
\item If the cycle is pointwise invariant, we have a $USp(2)$ instanton with: 
\begin{equation}
1 \text{ zero mode } \theta^\alpha\,,\qquad
3 \text{ zero modes } \overline{\tau}^{\dot{\alpha}}\,.
\end{equation}
Unless some new mechanism is discovered to soak all anti-holomorphic zero modes while producing a new holomorphic one, this configuration is not suitable for having instanton corrections to the superpotential.

\item If the cycle is transversally invariant, we have an $O(1)$ instanton with:
\begin{equation}
2 \text{ zero modes } \theta^\alpha\,,\qquad
0 \text{ zero modes } \overline{\tau}^{\dot{\alpha}}\,.
\end{equation}
This configuration is suitable for having instanton corrections to the superpotential.
\end{itemize}
One can also consider the case of a stack of multiple E3-branes. The discussion is not much different, in the sense that the only viable contribution is in the form of $O(1)$ instanton
\footnote{Consider a stack of $N$ E3-branes. If the cycle is pointwise invariant, we have a $USp(2N)$ instanton. In this case, we need an even number $2N$ of E3-branes, and we get
$N(2N-1)$ zero modes $\theta^\alpha$ and $N(N+1)$ zero modes $\overline{\tau}^{\dot{\alpha}}$. If the cycle is instead transversally invariant, we have an $O(N)$ instanton. There are
$N(2N+1)$ zero modes $\theta^\alpha$ and $N(2N-1)$ zero modes $\overline{\tau}^{\dot{\alpha}}$.}.

\subsubsection*{Deformation zero modes}

In addition to the universal zero modes, open strings with both ends on the E3-brane can generate further zero modes which can be classified as follows:
\begin{itemize}
\item If the E3-brane is a $U(1)$ instanton, for each complex deformation we have $2$ complex bosonic zero modes, $1$ chiral and $1$ anti-chiral fermionic zero mode;

\item If the E3-brane is an $O(1)$ instanton, the zero modes are counted by the following cohomology groups:
\begin{eqnarray}
\#\gamma^\alpha=\dim( H_+^{1,0}(D_{\Et}))\,,\\
\#\chi^\alpha=\dim( H_+^{2,0}(D_{\Et}))\,,\\
\#(c,\overline{\chi}^{\dot{\alpha}})=\dim( H_-^{2,0}(D_{\Et}))\,,
\end{eqnarray}
Recall that, for a divisor $D_\Et$ wrapped by the E3-instanton, the dimension of $H^{1,0}(D_{\Et})$ counts Wilson-line moduli, while the dimension of $H^{2,0}(D_{\Et})$ gives the number of deformations of the cycle. A rigid divisor features $H^{1,0}(D_\Et)=H^{2,0}(D_{\Et})=0$.
\end{itemize}

\subsubsection*{Charged and vector-like zero modes}

Another kind of zero modes arises from strings stretched between the E3-instanton wrapping the divisor $D_\Et$ and D7-branes wrapping the divisor $D_\Ds$. These are called \emph{charged zero modes} since they are charged under the gauge symmetry on the D7-brane. Let us denote charged zero modes as $\lambda_{\Et\Ds}$ and $\overline{\lambda}_{\Et\Ds}=\lambda_{\Ds\Et}$. The number of charged zero modes is instead denoted by:
\begin{equation}
I_{\Et\Ds}^+=\#\lambda_{\Et\Ds}\,,\qquad\text{and}\qquad
I_{\Et\Ds}^-=\#\overline{\lambda}_{\Et\Ds}\,.
\end{equation}
In practice, $I_{\Et\Ds}^+$ (respectively $I_{\Et\Ds}^-$) counts chiral fields in the representation $(\overline{N}_{\Et},N_\Ds)$ (respectively $(N_{\Et},\overline{N}_\Ds)$). The purely-chiral zero modes are counted by the excess of one chirality with respect to the other. This is encoded in the \emph{net-chiral index} \cite{Blumenhagen:2012kz,Blumenhagen:2009qh,Cicoli:2021dhg}:
\begin{equation}
I_{\Et\Ds}=I_{\Et\Ds}^+-I_{\Et\Ds}^-=\int_{D_{\Et}\cap D_\Ds}\left(\mathcal{F}_\Et-\mathcal{F}_\Ds\right)\,,
\end{equation}
where $\mathcal{F}_i$, $i\in\{{\rm E3},{\rm D7}\}$ are the curvature $2$-forms of the gauge bundles on the E3-brane and on the D7-brane respectively. If we now switch off gauge fluxes, the chiral index vanishes:
\begin{equation}
I_{\Et\Ds}=0\,,
\end{equation}
and so there are no chiral zero modes. There can however still be non-chiral zero modes since $I_{\Et\Ds}=0$ implies just:
\begin{equation}
I_{\Et\Ds}^-=I_{\Et\Ds}^+\,.
\end{equation}
If the two chiral indices are non-zero, charged zero modes pair up forming \emph{vector-like zero modes} counted by the chiral index. One cannot anymore use index theorems to compute the amount of vector-like zero modes. Instead, defining $C\equiv D_{\Et}\cap D_\Ds$, the chiral index is given by \cite{Blumenhagen:2009qh}:
\begin{equation}
\begin{aligned}
I_{\Et\Ds}^-&=\dim\left(H^{1}(C,L_{\Et}\otimes V_\Ds^\vee\otimes K_C^{1/2})\right)\,,\\
I_{\Et\Ds}^+&=\dim\left(H^{0}(C,L_{\Et}\otimes V_\Ds^\vee\otimes K_C^{1/2})\right)\,,
\label{eq:chiral_index}
\end{aligned}
\end{equation}
where $K_C^{1/2}$ is the spin bundle on the intersection curve (if it exists), $L_{\Et}$ is the gauge bundle on $D_{\Et}$ and $V_\Ds^\vee$ the gauge bundle on $D_\Ds$. We recall that for an $O(1)$ instanton, the gauge bundle $L_\Et$ must be such that its connection $\mathcal{F}_\Et$ is orientifold odd, namely $\mathcal{F}_\Et\in H^{1,1}_{-}(D_\Et)$ \citep{Grimm:2011dj}. By turning off fluxes, \eqref{eq:chiral_index} reduce to:
\begin{equation}
I_{\Et\Ds}^-=\dim\left(H^{1}(C,K_C^{1/2})\right)\qquad\text{and}\qquad
I_{\Et\Ds}^+=\dim\left(H^{0}(C,K_C^{1/2})\right)\,,
\end{equation}
which depend only on the topology of the intersection curve. For example, for $C=\mathbb{P}^1$ we have $K_{\mathbb{P}^1}=\mathcal{O}(-2)$ which implies $K_{\mathbb{P}^1}^{1/2}=\mathcal{O}(-1)$. Since in $H^{0/1}(\mathbb{P}^1,K_{\mathbb{P}^1}^{1/2})$ there exist no sections nor $1$-forms respectively, we have:
\begin{equation}
I_{\Et\Ds}^-=\dim\left(H^{1}(\mathbb{P}^1,K_{\mathbb{P}^1}^{1/2})\right)=
I_{\Et\Ds}^+=\dim\left(H^{0}(\mathbb{P}^1,K_{\mathbb{P}^1}^{1/2})\right)=0\,,
\end{equation}
and so no vector-like zero modes are present. The same behaviour can be seen for an intersection curve given by disjoint $\mathbb{P}^1$s.

\subsubsection*{Multi-instanton zero modes}

Multi-instanton zero modes arise from a configuration featuring an E3-brane and its orientifold image with open strings stretched between them. The number of zero modes is computed starting from the following combinations of chiral indices:
\begin{equation}
\#(m,\overline{\mu}^{\dot{\alpha}})=\frac{1}{2}(I_{\Et\Et'}+pI_{\Os\Et})^+\qquad\text{and}\qquad
\#\mu^\alpha=\frac{1}{2}(I_{\Et\Et'}-pI_{\Os\Et})^+\,.
\end{equation}
In our construction E3-branes are wrapped on transversally invariant cycles, and so such modes are absent. For a more detailed discussion of this case, see \cite{Blumenhagen:2009qh}. 

\subsubsection*{Summary}

The fermionic zero modes of a single $O(1)$ E3-instanton wrapping a divisor $D_\Et$ are summarised in Tab. \ref{Summary}. 

\begin{table}[H]
\centering
\begin{tabular}{|c|c|c|}
\hline
$\text{Fermionic zero mode}$ &$\text{Type}$ & $\text{Counted by}$\\
     \hline
     $\theta^\alpha$&  $\text{Universal}$&  2  \\
     \hline
     $\overline{\tau}^{\dot{\alpha}}$&  $\text{Universal}$&  0  \\
     \hline
     $\gamma^\alpha$&  $\text{Deformation}$&  $\dim( H_+^{1,0}(D_{\Et}))$  \\
\hline
$\chi^\alpha$&  $\text{Deformation}$&  $\dim( H_+^{2,0}(D_{\Et}))$  \\
     \hline
     $\overline{\chi}^{\dot{\alpha}}$&  $\text{Deformation}$&  $\dim( H_-^{2,0}(D_{\Et})))-1$  \\
     \hline
     $\lambda_{\Et\,\Ds}$&  $\text{Charged}$&  $\dim(H^{0}(C,L_{\Et}\otimes V_\Ds^\vee\otimes K_C^{1/2}))$  \\
     \hline
     $\overline{\lambda}_{\Et\,\Ds}$&  $\text{Charged}$&  $\dim(H^{1}(C,L_{\Et}\otimes V_\Ds^\vee\otimes K_C^{1/2}))$  \\
     \hline
     $\mu^\alpha$&  $\text{Multi-instanton}$&  $\frac{1}{2}(I_{\Et\Et'}-pI_{\Os\Et})^+$  \\
     \hline
     $\overline{\mu}^{\dot{\alpha}}$&  $\text{Multi-instanton}$&  $\frac{1}{2}(I_{\Et\Et'}+pI_{\Os\Et})^+-1$  \\
     \hline
\end{tabular}
\caption{Fermionic zero modes of an $O(1)$ E3-instanton wrapping the divisor $D_\Et$.}
\label{Summary}
\end{table}

\section{From reflexive polytopes to Calabi-Yau manifolds}
\label{app:A_Kreuzer-Skarke}

In this appendix we provide a minimal introduction to toric geometry, referring the reader to standard textbooks, such as \cite{cox2011toric}, for a deeper treatment.

\paragraph{Definition (Toric variety)}

A \textit{toric variety} $\mathcal{A}$ of dimension $n$ is defined as an algebraic variety containing an algebraic torus $\mathbb{T}=(\mathbb{C}^*)^n\subset \mathcal{A}$ as a dense open subset, together with a natural action $\mathbb{T}\times \mathcal{A}\rightarrow \mathcal{A}$ which acts on the torus as a simple multiplication. One can write $\mathcal{A}$ as:
\begin{equation}
\mathcal{A}=({\mathbb{C}^{n+m}\backslash Z_{\Delta}})/G\,,
\end{equation}
where  $G\simeq (\mathbb{C}^*)^{m}\times H$ is the algebraic torus, together with possibly a non-trivial finite group $H$.

In order to determine uniquely the toric variety it is necessary to specify how $G$ acts and what is the set of points $Z_{\Delta}$. In order to do so let us introduce some simple definitions.

\paragraph{Definition (Cone)}
Let $N\cong \mathbb{Z}^n$ be a lattice and $N_{\mathbb{R}}=N\otimes \mathbb{R}$; then the \textit{cone} $\sigma\subset N_{\mathbb{R}}$ is defined as the set generated by a finite amount of vectors in the lattice $N$:
\begin{equation}
    \sigma=\{\sum_i a_i v_i\,|\, a_i\geq 0\}\,,
\end{equation}
such that $\sigma\cap-\sigma=\{0\}$. 

\paragraph{Definition (Toric fan)}
Let $\mathcal{A}$ be a normal toric variety; the \textit{fan} of $\mathcal{A}$ is defined as the collection of cones in $N_{\mathbb{R}}$ that satisfy two properties:
\begin{itemize}
\item Each face of a cone is a cone;
\item Intersection of two cones is a face of both cones. 
\end{itemize}

\paragraph{Definition (Homogenous coordinates)}
Let $\mathcal{A}$ be a toric variety and $\sigma_i$ a collection of cones of its fan. Denoting as $\sigma(1)$ the set of 1D cones in $N_{\mathbb{R}}$ and defining the vectors in $N$ as $(v_1,\dots,v_n)$, one can associate to each $v_i$ a \textit{homogeneous coordinate} $x_i$ in the quotient construction of $\mathcal{A}$.

\paragraph{Definition (Divisors)}
Formal integer linear combinations of irreducible subvarieties of codimension-$1$ are called \textit{Weil divisors}.

\paragraph{Definition (Toric divisors)}
\textit{Toric divisors} are defined as the codimension-$1$ subvarieties which are $G$-invariant, and can be obtained by setting one of the homogeneous coordinates to $0$:
\begin{equation}
D_{p}\equiv \{x_p=0\}\subset \mathcal{A}\,.
\end{equation}
Our setting will include only toric divisors, and so the adjective `toric' is understood from now on.

\paragraph{Definition (Lattice polytope)}
Let $\{p_1, \dots, p_k\}$ be a finite collection of distinct points in a lattice $N \cong \mathbb{Z}^n$. A \textit{lattice polytope} is defined as the convex hull of these points within the real vector space $N_{\mathbb{R}} = N \otimes \mathbb{R}$; that is, the minimal convex set containing all $p_i$'s. Given a toric fan, one can naturally associate to it a lattice polytope such that the given fan serves as its normal fan.

\paragraph{Definition (Polar dual polytope)}
Given a lattice polytope $\Delta^\circ \subset N_{\mathbb{R}}$ containing the origin in its strict interior, we define its \textit{polar dual} $\Delta \subset M_{\mathbb{R}}$ as the set of points such that: 
\begin{equation}
\label{eq:Pol_Dual}
\Delta \equiv \{q \in M_{\mathbb{R}} \mid \langle p,q \rangle \geq -1, \forall p \in \Delta^\circ\}\,,
\end{equation}
where $\langle p,q \rangle$ is the duality pairing between the points $p \in N_{\mathbb{R}}$ and $q \in M_{\mathbb{R}}$. Due to convexity, the bi-dual of the polytope is the original polytope itself, meaning $((\Delta^\circ)^*)^* = \Delta^\circ$.

\paragraph{Definition (Reflexive polytope)}
A lattice polytope $\Delta^\circ \subset N_{\mathbb{R}}$ containing the origin in its strict interior is called \textit{reflexive} if its polar dual $\Delta \subset M_{\mathbb{R}}$ is also a lattice polytope (i.e. all vertices of $\Delta$ lie in the dual lattice $M$).

An arbitrary toric variety can be singular. It is possible to understand whether this is the case from the toric fan using the following:
\paragraph{Proposition}
Let $\mathcal{A}$ be a toric variety and $\Sigma$ its fan. Denote a top-dimensional cone of $\Sigma$ by $\sigma=<v_{i_1}, \ldots, v_{i_n}>$. Then the volume of the cone is defined as $\mbox{vol}(\sigma)=|\det(v_{i_1}, \ldots, v_{i_n})|$ and the variety is non-singular if and only if all the top-dimensional cones have unit volume.

It is possible to de-singularise a toric variety adding new 1D cones to its fan. This corresponds to performing blow-ups.  It is also possible to partially de-singularise by \textit{performing a triangulation}. This corresponds to refining the structure of the higher dimensional cones of the fan, or equivalently adding curves and surfaces into the original geometry.

\paragraph{Definition (GLSM)}
Let $\mathcal{A}$ be an affine toric variety with a fan $\Sigma$, and let $\Delta^{\circ}$ be the normal polytope of $\Sigma$. Let $p_i, i=1,..., m+4$, be the lattice points of $\Delta^{\circ}$. The points $p_i$ satisfy a set of $m=h^{1,1}(\mathcal{A})$ linear relations of the form:
\begin{equation}
Q_i^b\, p_i=0\,, \qquad b=1,\ldots m\,.
\end{equation}
The coefficients $Q_i^b$ can be rearranged in a matrix called \textit{GLSM charge matrix} since its entries correspond to charges of a gauged linear sigma model.

Moving forward, the previous definitions and propositions will be applied to the KS database. This database catalogues 4D reflexive polytopes which define the fans of the toric varieties that serve as the ambient spaces for our CY manifolds. In what follows, we will explain how to construct these complex 3D manifolds as hypersurfaces within the 4D toric ambient spaces derived from the KS database.

\paragraph{How to build a CY as a hypersurface of a 4D ambient space}
Let $N\cong \mathbb{Z}^4$ be a 4D lattice, $N_{\mathbb{R}}=N\otimes \mathbb{R}$, $M$ the dual lattice and $M_{\mathbb{R}}=M\otimes \mathbb{R}$. Let $\Delta^\circ\subset N$ be a 4D reflexive polytope in the KS database. We then introduce a fine regular star triangulation (FRST) $\mathbf{T}$ of $\Delta^{\circ}$, obtained by excluding all lattice points lying in the interior of the facets of $\Delta^{\circ}$. 

Let $\mathcal{A}$ be a 4D complex toric variety whose fan $\Sigma_{\mathcal{A}}$ is obtained by taking the cones over the unit-volume simplices of the triangulation $\mathbf{T}$. The lattice points of $\Delta^{\circ}$ are in one-to-one correspondence with the rays of $\Sigma_{\mathcal{A}}$. Defining $\mathcal{F}_0$ to be the set of vertices ($0$-faces) of the triangulation:
\begin{equation}
\forall p\,\in\, \mathcal{F}_0\equiv (\Delta^\circ\cap N)\backslash \{0\}\,,
\end{equation}
each point $p$ is associated with a homogeneous toric coordinate $x_p$, hence with toric divisor $D_p\equiv \{x_p=0\}$. On the toric side instead, we can now parametrise $\mathcal{A}$ in terms of homogenous coordinates which correspond to the toric ones, $x_p,\dots,x_{n+4}\in \mathbb{C}^{m+4}\backslash Z$, where $m\equiv h^{1,1}(\mathcal{A})=|\mathcal{F}_0|-4$ and $Z_{\Delta}$ is given by the union over $\{x_{p_1},x_{p_2},\dots,x_{p_k}=0\}\subset \mathbb{C}^{m+4}$ with $\{p_1,\dots,p_k\}\subset \mathcal{F}_0$ such that the cone given by those points $\notin\Sigma$. Furthermore, we define the group of toric scaling relations by:
\begin{equation}
G\equiv \bigg\{\eta\in(\mathbb{C}/\mathbb{Z})^{m+4}:\sum_{p\in\mathcal{F}_0}\eta^p p\in N\bigg\},
\end{equation}
which acts on the homogeneous coordinates as $x_p\rightarrow e^{2\pi i \eta^p}x_p$. Together with the exclusion set $Z_\Delta$, the group $G$ completely specifies the toric variety:
\begin{equation}
\mathcal{A}=(\mathbb{C}^{m+4}\backslash Z_{\Delta})/G\,.
\end{equation}
Recalling the definition of the polar dual of the polytope $\Delta^\circ$ in \eqref{eq:Pol_Dual}, we can show that a basis for the global sections of the anti-canonical bundle of the toric variety is given by the monomials:
\begin{equation}
    m_q\equiv \prod_{p\in\mathcal{F}_0}x_p^{\langle p,q\rangle+1}\,.
\end{equation}
The CY threefold $X$ is then realised as the vanishing locus of a generic linear combination of these monomials:
\begin{equation}
X\equiv \{y=0\}\,,y\equiv \sum_{q\in\Delta\cap M}u_q m_q\,,
\end{equation}
where the $u_q$ are complex parameters. It can be shown that the prime toric divisors of $X$ are obtained by intersecting the CY hypersurface with the divisors of the ambient space, namely $D_p^{X} \equiv D_p \cap X$.

Following the above construction, one obtains a CY threefold from a toric ambient space associated with a reflexive polytope. A fine regular star triangulation partially resolves the singularities of the ambient toric variety, while choosing a generic anti-canonical hypersurface avoids the remaining singularities. Although the KS database contains nearly half a billion reflexive polytopes, each defining a distinct toric ambient space, the total number of admissible triangulations is finite but presently unknown. Consequently, the total number of CY threefolds arising from this construction is also unknown. Moreover, it remains an open question how many of these manifolds are topologically distinct. Indeed, by Wall's theorem, a CY threefold is determined by the following topological data:

\paragraph{Theorem (Wall) \cite{Wall:1966rcd}}
A Calabi-Yau manifold is uniquely determined, up to diffeomorphism, by its Hodge numbers $h^{1,1}$, $h^{1,2}$, its triple intersection numbers $k_{ijk}=\int_{X}\hat{D}_i\wedge \hat{D}_j\wedge \hat{D}_k$ and its second Chern class $c_2(X)$.

The polytope formalism is useful not only because of the presence of an entire database to study, but also because it allows the computation of topological invariants of manifolds, submanifolds and involutions through simple lattice point counting. Below we list the ones used throughout this paper:

\paragraph{Proposition \cite{Khovanskij:1978ag}}
Consider two divisors corresponding to different lattice points lying on the same $1$-face $\mathcal{F}_1$ of the polytope; then their intersection is a curve of genus:
\begin{equation}
    g=\#\mathcal{F}_1^\vee\,,
\end{equation}
where $\#\mathcal{F}_1^\vee$ denotes the number of lattice points interior to the $2$-face dual to $\mathcal{F}_1$.

\paragraph{Proposition \cite{Braun:2017nhi}}
Let $D_{p_1}$ and $D_{p_2}$ be two prime toric divisors corresponding to lattice points $p_1$ and $p_2$ lying on the same $2$-face $\mathcal{F}_2$ of the polytope $\Delta^\circ$. If $p_1$ and $p_2$ are connected by an edge in the chosen triangulation of $\Delta^\circ$, then the intersection $D_{p_1}\cap D_{p_2}\cap X$ is a disjoint union of $n$ rational curves ($\mathbb{P}^1$'s), where:
\begin{equation}
    n = \#\mathcal{F}_2^\vee + 1\,,
\end{equation}
with $\#\mathcal{F}_2^\vee$ denoting the number of lattice points interior to the $1$-face dual to $\mathcal{F}_2$.

\paragraph{Proposition \cite{Batyrev:1994pg}} 
Let $\mathcal{A}$ be a toric variety and $D\subset \mathcal{A}$ a smooth divisor associated with the polytope point $p$. The Hodge numbers of $D$ can be computed as follows:
\begin{itemize}
\item If $p$ is a vertex, $h^{1,0}(D)=0$ and $h^{2,0}(D)=\#\mathcal{F}_0^\vee$, where $\#\mathcal{F}_0^\vee$ denotes the number of interior lattice points of the $3$-face (the facet) dual to the vertex $p$.

\item If $p$ lies on a $1$-face $\mathcal{F}_1$, $h^{1,0}(D)=\#\mathcal{F}_1^\vee$ and $h^{2,0}(D)=0$, where $\#\mathcal{F}_1^\vee$ denotes the number of interior points of the $2$-face dual to $\mathcal{F}_1$.

\item If $p$ lies on a $2$-face $\mathcal{F}_2$, $h^{1,0}(D)=h^{2,0}(D)=0$, and so $D$ is a rigid rational surface.
\end{itemize}

\bibliographystyle{JHEP}

\bibliography{ref}

@article{Grimm:2011dj,
    author = "Grimm, Thomas W. and Kerstan, Max and Palti, Eran and Weigand, Timo",
    title = "{On Fluxed Instantons and Moduli Stabilisation in IIB Orientifolds and F-theory}",
    eprint = "1105.3193",
    archivePrefix = "arXiv",
    primaryClass = "hep-th",
    doi = "10.1103/PhysRevD.84.066001",
    journal = "Phys. Rev. D",
    volume = "84",
    pages = "066001",
    year = "2011"
}

@article{Kreuzer:2000xy,
    author = "Kreuzer, Maximilian and Skarke, Harald",
    title = "{Complete classification of reflexive polyhedra in four-dimensions}",
    eprint = "hep-th/0002240",
    archivePrefix = "arXiv",
    reportNumber = "HUB-EP-00-13, TUW-00-07",
    doi = "10.4310/ATMP.2000.v4.n6.a2",
    journal = "Adv. Theor. Math. Phys.",
    volume = "4",
    pages = "1209--1230",
    year = "2000"
}

@article{Blumenhagen:2012kz,
    author = "Blumenhagen, Ralph and Gao, Xin and Rahn, Thorsten and Shukla, Pramod",
    title = "{A Note on Poly-Instanton Effects in Type IIB Orientifolds on Calabi-Yau Threefolds}",
    eprint = "1205.2485",
    archivePrefix = "arXiv",
    primaryClass = "hep-th",
    reportNumber = "MPP-2012-87",
    doi = "10.1007/JHEP06(2012)162",
    journal = "JHEP",
    volume = "06",
    pages = "162",
    year = "2012"
}

@article{Cicoli:2012tz,
    author = "Cicoli, Michele and Pedro, Francisco G. and Tasinato, Gianmassimo",
    title = "{Natural Quintessence in String Theory}",
    eprint = "1203.6655",
    archivePrefix = "arXiv",
    primaryClass = "hep-th",
    doi = "10.1088/1475-7516/2012/07/044",
    journal = "JCAP",
    volume = "07",
    pages = "044",
    year = "2012"
}

@article{Cicoli:2012cy,
    author = "Cicoli, M. and Tasinato, G. and Zavala, I. and Burgess, C. P. and Quevedo, F.",
    title = "{Modulated Reheating and Large Non-Gaussianity in String Cosmology}",
    eprint = "1202.4580",
    archivePrefix = "arXiv",
    primaryClass = "hep-th",
    reportNumber = "DAMTP-2012-14",
    doi = "10.1088/1475-7516/2012/05/039",
    journal = "JCAP",
    volume = "05",
    pages = "039",
    year = "2012"
}

@article{Cicoli:2011it,
    author = "Cicoli, Michele and Kreuzer, Maximilian and Mayrhofer, Christoph",
    title = "{Toric K3-Fibred Calabi-Yau Manifolds with del Pezzo Divisors for String Compactifications}",
    eprint = "1107.0383",
    archivePrefix = "arXiv",
    primaryClass = "hep-th",
    reportNumber = "DESY-11-103",
    doi = "10.1007/JHEP02(2012)002",
    journal = "JHEP",
    volume = "02",
    pages = "002",
    year = "2012"
}

@article{Petersson:2010qu,
    author = "Petersson, Christoffer and Soler, Pablo and Uranga, Angel M.",
    title = "{D-instanton and polyinstanton effects from type I' D0-brane loops}",
    eprint = "1001.3390",
    archivePrefix = "arXiv",
    primaryClass = "hep-th",
    reportNumber = "IFT-UAM-CSIC-10-02",
    doi = "10.1007/JHEP06(2010)089",
    journal = "JHEP",
    volume = "06",
    pages = "089",
    year = "2010"
}

@article{Akerblom:2007uc,
    author = "Akerblom, Nikolas and Blumenhagen, Ralph and Lust, Dieter and Schmidt-Sommerfeld, Maximilian",
    title = "{Instantons and Holomorphic Couplings in Intersecting D-brane Models}",
    eprint = "0705.2366",
    archivePrefix = "arXiv",
    primaryClass = "hep-th",
    reportNumber = "MPP-2007-57, LMU-ASC-31-07",
    doi = "10.1088/1126-6708/2007/08/044",
    journal = "JHEP",
    volume = "08",
    pages = "044",
    year = "2007"
}

@article{Blumenhagen:2008kq,
    author = "Blumenhagen, Ralph and Moster, Sebastian and Plauschinn, Erik",
    title = "{String GUT Scenarios with Stabilised Moduli}",
    eprint = "0806.2667",
    archivePrefix = "arXiv",
    primaryClass = "hep-th",
    reportNumber = "MPP-2008-60",
    doi = "10.1103/PhysRevD.78.066008",
    journal = "Phys. Rev. D",
    volume = "78",
    pages = "066008",
    year = "2008"
}

@article{Blumenhagen:2008ji,
    author = "Blumenhagen, Ralph and Schmidt-Sommerfeld, Maximilian",
    title = "{Power Towers of String Instantons for N=1 Vacua}",
    eprint = "0803.1562",
    archivePrefix = "arXiv",
    primaryClass = "hep-th",
    reportNumber = "MPP-2008-19",
    doi = "10.1088/1126-6708/2008/07/027",
    journal = "JHEP",
    volume = "07",
    pages = "027",
    year = "2008"
}

@article{Demirtas:2022hqf,
    author = "Demirtas, Mehmet and Rios-Tascon, Andres and McAllister, Liam",
    title = "{CYTools: A Software Package for Analyzing Calabi-Yau Manifolds}",
    eprint = "2211.03823",
    archivePrefix = "arXiv",
    primaryClass = "hep-th",
    month = "11",
    year = "2022"
}

@article{Cicoli:2007xp,
    author = "Cicoli, Michele and Conlon, Joseph P. and Quevedo, Fernando",
    title = "{Systematics of String Loop Corrections in Type IIB Calabi-Yau Flux Compactifications}",
    eprint = "0708.1873",
    archivePrefix = "arXiv",
    primaryClass = "hep-th",
    reportNumber = "DAMTP-2007-75",
    doi = "10.1088/1126-6708/2008/01/052",
    journal = "JHEP",
    volume = "01",
    pages = "052",
    year = "2008"
}

@article{Cicoli:2021skd,
    author = "Cicoli, Michele and Cunillera, Francesc and Padilla, Antonio and Pedro, Francisco G.",
    title = "{Quintessence and the Swampland: The Numerically Controlled Regime of Moduli Space}",
    eprint = "2112.10783",
    archivePrefix = "arXiv",
    primaryClass = "hep-th",
    doi = "10.1002/prop.202200008",
    journal = "Fortsch. Phys.",
    volume = "70",
    number = "4",
    pages = "2200008",
    year = "2022"
}

@article{AbdusSalam:2020ywo,
    author = "AbdusSalam, S. and Abel, S. and Cicoli, M. and Quevedo, F. and Shukla, P.",
    title = {{A systematic approach to K{\"a}hler moduli stabilisation}},
    eprint = "2005.11329",
    archivePrefix = "arXiv",
    primaryClass = "hep-th",
    reportNumber = "IPPP/20/18",
    doi = "10.1007/JHEP08(2020)047",
    journal = "JHEP",
    volume = "08",
    number = "08",
    pages = "047",
    year = "2020"
}

@article{Cicoli:2021gss,
    author = "Cicoli, Michele and Guidetti, Veronica and Righi, Nicole and Westphal, Alexander",
    title = "{Fuzzy Dark Matter candidates from string theory}",
    eprint = "2110.02964",
    archivePrefix = "arXiv",
    primaryClass = "hep-th",
    reportNumber = "DESY-21-153",
    doi = "10.1007/JHEP05(2022)107",
    journal = "JHEP",
    volume = "05",
    pages = "107",
    year = "2022"
}

@article{Cicoli:2017zbx,
    author = "Cicoli, Michele and Diaz, Victor A. and Guidetti, Veronica and Rummel, Markus",
    title = "{The 3.5 keV Line from Stringy Axions}",
    eprint = "1707.02987",
    archivePrefix = "arXiv",
    primaryClass = "hep-th",
    doi = "10.1007/JHEP10(2017)192",
    journal = "JHEP",
    volume = "10",
    pages = "192",
    year = "2017"
}

@article{Kobayashi:2017jeb,
    author = "Kobayashi, Tatsuo and Uemura, Shohei and Yamamoto, Junji",
    title = "{Polyinstanton axion inflation}",
    eprint = "1705.04088",
    archivePrefix = "arXiv",
    primaryClass = "hep-ph",
    reportNumber = "EPHOU-17-008, KUNS-2678, MISC-2017-05",
    doi = "10.1103/PhysRevD.96.026007",
    journal = "Phys. Rev. D",
    volume = "96",
    number = "2",
    pages = "026007",
    year = "2017"
}

@article{Gao:2013rra,
    author = "Gao, Xin and Shukla, Pramod",
    title = "{F-term Stabilization of Odd Axions in LARGE Volume Scenario}",
    eprint = "1307.1141",
    archivePrefix = "arXiv",
    primaryClass = "hep-th",
    reportNumber = "MPP-2013-184",
    doi = "10.1016/j.nuclphysb.2013.11.015",
    journal = "Nucl. Phys. B",
    volume = "878",
    pages = "269--294",
    year = "2014"
}

@article{Gao:2013hn,
    author = "Gao, Xin and Shukla, Pramod",
    title = "{On Non-Gaussianities in Two-Field Poly-Instanton Inflation}",
    eprint = "1301.6076",
    archivePrefix = "arXiv",
    primaryClass = "hep-th",
    reportNumber = "MPP-2013-15",
    doi = "10.1007/JHEP03(2013)061",
    journal = "JHEP",
    volume = "03",
    pages = "061",
    year = "2013"
}

@article{Angus:2012dd,
    author = "Angus, Stephen and Conlon, Joseph P.",
    title = "{Soft Supersymmetry Breaking in Anisotropic LARGE Volume Compactifications}",
    eprint = "1211.6927",
    archivePrefix = "arXiv",
    primaryClass = "hep-th",
    reportNumber = "OUTP-13-01P",
    doi = "10.1007/JHEP03(2013)071",
    journal = "JHEP",
    volume = "03",
    pages = "071",
    year = "2013"
}

@article{Blumenhagen:2012ue,
    author = "Blumenhagen, Ralph and Gao, Xin and Rahn, Thorsten and Shukla, Pramod",
    title = "{Moduli Stabilization and Inflationary Cosmology with Poly-Instantons in Type IIB Orientifolds}",
    eprint = "1208.1160",
    archivePrefix = "arXiv",
    primaryClass = "hep-th",
    reportNumber = "MPP-2012-123",
    doi = "10.1007/JHEP11(2012)101",
    journal = "JHEP",
    volume = "11",
    pages = "101",
    year = "2012"
}

@article{Higaki:2012ba,
    author = "Higaki, Tetsutaro and Kamada, Kohei and Takahashi, Fuminobu",
    title = "{Higgs, Moduli Problem, Baryogenesis and Large Volume Compactifications}",
    eprint = "1207.2771",
    archivePrefix = "arXiv",
    primaryClass = "hep-ph",
    reportNumber = "RIKEN-MP-50, DESY-12-124, TU-915",
    doi = "10.1007/JHEP09(2012)043",
    journal = "JHEP",
    volume = "09",
    pages = "043",
    year = "2012"
}

@article{Gendler:2022qof,
    author = "Gendler, Naomi and Kim, Manki and McAllister, Liam and Moritz, Jakob and Stillman, Mike",
    title = "{Superpotentials from singular divisors}",
    eprint = "2204.06566",
    archivePrefix = "arXiv",
    primaryClass = "hep-th",
    reportNumber = "MIT-CTP/5388",
    doi = "10.1007/JHEP11(2022)142",
    journal = "JHEP",
    volume = "11",
    pages = "142",
    year = "2022"
}

@article{Blumenhagen:2008zz,
    author = "Blumenhagen, Ralph and Braun, Volker and Grimm, Thomas W. and Weigand, Timo",
    title = "{GUTs in Type IIB Orientifold Compactifications}",
    eprint = "0811.2936",
    archivePrefix = "arXiv",
    primaryClass = "hep-th",
    reportNumber = "MPP-2008-144, DIAS-STP-08-15, SLAC-PUB-13466",
    doi = "10.1016/j.nuclphysb.2009.02.011",
    journal = "Nucl. Phys. B",
    volume = "815",
    pages = "1--94",
    year = "2009"
}

@article{Blumenhagen:2007bn,
    author = "Blumenhagen, Ralph and Cvetic, Mirjam and Richter, Robert and Weigand, Timo",
    title = "{Lifting D-Instanton Zero Modes by Recombination and Background Fluxes}",
    eprint = "0708.0403",
    archivePrefix = "arXiv",
    primaryClass = "hep-th",
    doi = "10.1088/1126-6708/2007/10/098",
    journal = "JHEP",
    volume = "10",
    pages = "098",
    year = "2007"
}

@article{Cicoli:2021dhg,
    author = "Cicoli, Michele and Etxebarria, I{\~n}aki Garc{\'\i}a and Quevedo, Fernando and Schachner, Andreas and Shukla, Pramod and Valandro, Roberto",
    title = "{The Standard Model quiver in de Sitter string compactifications}",
    eprint = "2106.11964",
    archivePrefix = "arXiv",
    primaryClass = "hep-th",
    doi = "10.1007/JHEP08(2021)109",
    journal = "JHEP",
    volume = "08",
    pages = "109",
    year = "2021"
}

@article{Witten:1996bn,
    author = "Witten, Edward",
    title = "{Nonperturbative superpotentials in string theory}",
    eprint = "hep-th/9604030",
    archivePrefix = "arXiv",
    reportNumber = "IASSNS-HEP-96-29",
    doi = "10.1016/0550-3213(96)00283-0",
    journal = "Nucl. Phys. B",
    volume = "474",
    pages = "343--360",
    year = "1996"
}

@article{Blumenhagen:2009qh,
    author = "Blumenhagen, Ralph and Cvetic, Mirjam and Kachru, Shamit and Weigand, Timo",
    title = "{D-Brane Instantons in Type II Orientifolds}",
    eprint = "0902.3251",
    archivePrefix = "arXiv",
    primaryClass = "hep-th",
    reportNumber = "MPP-2009-15, UPR-1205-T, SLAC-PUB-13531",
    doi = "10.1146/annurev.nucl.010909.083113",
    journal = "Ann. Rev. Nucl. Part. Sci.",
    volume = "59",
    pages = "269--296",
    year = "2009"
}

@article{Cicoli:2011ct,
    author = "Cicoli, Michele and Pedro, Francisco G. and Tasinato, Gianmassimo",
    title = "{Poly-instanton Inflation}",
    eprint = "1110.6182",
    archivePrefix = "arXiv",
    primaryClass = "hep-th",
    doi = "10.1088/1475-7516/2011/12/022",
    journal = "JCAP",
    volume = "12",
    pages = "022",
    year = "2011"
}

@article{Cicoli:2024yqh,
    author = "Cicoli, Michele and Cunillera, Francesc and Padilla, Antonio and Pedro, Francisco G.",
    title = "{From inflation to quintessence: a history of the universe in string theory}",
    eprint = "2407.03405",
    archivePrefix = "arXiv",
    primaryClass = "hep-th",
    doi = "10.1007/JHEP10(2024)141",
    journal = "JHEP",
    volume = "10",
    pages = "141",
    year = "2024"
}

@article{Cicoli:2011yy,
    author = "Cicoli, M. and Burgess, C. P. and Quevedo, F.",
    title = "{Anisotropic Modulus Stabilisation: Strings at LHC Scales with Micron-sized Extra Dimensions}",
    eprint = "1105.2107",
    archivePrefix = "arXiv",
    primaryClass = "hep-th",
    reportNumber = "DESY-11-073",
    doi = "10.1007/JHEP10(2011)119",
    journal = "JHEP",
    volume = "10",
    pages = "119",
    year = "2011"
}

@article{Balasubramanian:2005zx,
    author = "Balasubramanian, Vijay and Berglund, Per and Conlon, Joseph P. and Quevedo, Fernando",
    title = "{Systematics of moduli stabilisation in Calabi-Yau flux compactifications}",
    eprint = "hep-th/0502058",
    archivePrefix = "arXiv",
    reportNumber = "DAMTP-2005-10, UNH-05-01, UPR-1109-T",
    doi = "10.1088/1126-6708/2005/03/007",
    journal = "JHEP",
    volume = "03",
    pages = "007",
    year = "2005"
}

@article{Freed:1999vc,
    author = "Freed, Daniel S. and Witten, Edward",
    title = "{Anomalies in string theory with D-branes}",
    eprint = "hep-th/9907189",
    archivePrefix = "arXiv",
    journal = "Asian J. Math.",
    volume = "3",
    pages = "819",
    year = "1999"
}

@article{Tachikawa:2018njr,
    author = "Tachikawa, Yuji and Yonekura, Kazuya",
    title = "{Why are fractional charges of orientifolds compatible with Dirac quantization?}",
    eprint = "1805.02772",
    archivePrefix = "arXiv",
    primaryClass = "hep-th",
    reportNumber = "IPMU-18-0067",
    doi = "10.21468/SciPostPhys.7.5.058",
    journal = "SciPost Phys.",
    volume = "7",
    number = "5",
    pages = "058",
    year = "2019"
}

@article{Witten:1998xy,
    author = "Witten, Edward",
    title = "{Baryons and branes in anti-de Sitter space}",
    eprint = "hep-th/9805112",
    archivePrefix = "arXiv",
    reportNumber = "IASSNS-HEP-98-42",
    doi = "10.1088/1126-6708/1998/07/006",
    journal = "JHEP",
    volume = "07",
    pages = "006",
    year = "1998"
}

@article{Douglas:1995bn,
    author = "Douglas, Michael R.",
    editor = "Baulieu, L. and Kazakov, V. and Picco, M. and Windey, Paul and Di Francesco, P. and Douglas, Michael R.",
    title = "{Branes within branes}",
    eprint = "hep-th/9512077",
    archivePrefix = "arXiv",
    reportNumber = "RU-95-92",
    journal = "NATO Sci. Ser. C",
    volume = "520",
    pages = "267--275",
    year = "1999"
}

@article{Ibanez:2007rs,
    author = "Ibanez, L. E. and Schellekens, A. N. and Uranga, A. M.",
    title = "{Instanton Induced Neutrino Majorana Masses in CFT Orientifolds with MSSM-like spectra}",
    eprint = "0704.1079",
    archivePrefix = "arXiv",
    primaryClass = "hep-th",
    reportNumber = "IFT-UAM-CSIC-07-12, CERN-PH-TH-2007-061",
    doi = "10.1088/1126-6708/2007/06/011",
    journal = "JHEP",
    volume = "06",
    pages = "011",
    year = "2007"
}

@article{Batyrev:1994pg,
    author = "Batyrev, Victor V. and Borisov, Lev A.",
    title = "{On Calabi-Yau complete intersections in toric varieties}",
    eprint = "alg-geom/9412017",
    archivePrefix = "arXiv",
    month = "12",
    year = "1994"
}

@article{Khovanskij:1978ag,
author = {Khovanskij, A.G.},
year = {1978},
month = {01},
pages = {},
title = {Newton polyhedra and the genus of complete intersections},
volume = {12},
journal = {Functional Analysis and its Applications}
}

@article{Becker:2002nn,
    author = "Becker, Katrin and Becker, Melanie and Haack, Michael and Louis, Jan",
    title = "{Supersymmetry breaking and alpha-prime corrections to flux induced potentials}",
    eprint = "hep-th/0204254",
    archivePrefix = "arXiv",
    reportNumber = "CALT-68-2379, UMD-PP-02-042, ROM2F-2002-10",
    doi = "10.1088/1126-6708/2002/06/060",
    journal = "JHEP",
    volume = "06",
    pages = "060",
    year = "2002"
}

@article{Berg:2007wt,
    author = "Berg, Marcus and Haack, Michael and Pajer, Enrico",
    title = "{Jumping Through Loops: On Soft Terms from Large Volume Compactifications}",
    eprint = "0704.0737",
    archivePrefix = "arXiv",
    primaryClass = "hep-th",
    doi = "10.1088/1126-6708/2007/09/031",
    journal = "JHEP",
    volume = "09",
    pages = "031",
    year = "2007"
}

@article{vonGersdorff:2005bf,
    author = "von Gersdorff, Gero and Hebecker, Arthur",
    title = "{Kahler corrections for the volume modulus of flux compactifications}",
    eprint = "hep-th/0507131",
    archivePrefix = "arXiv",
    doi = "10.1016/j.physletb.2005.08.024",
    journal = "Phys. Lett. B",
    volume = "624",
    pages = "270--274",
    year = "2005"
}

@article{Cicoli:2011yh,
    author = "Cicoli, Michele and Goodsell, Mark and Jaeckel, Joerg and Ringwald, Andreas",
    title = "{Testing String Vacua in the Lab: From a Hidden CMB to Dark Forces in Flux Compactifications}",
    eprint = "1103.3705",
    archivePrefix = "arXiv",
    primaryClass = "hep-th",
    reportNumber = "DESY-11-042, IPPP-11-13, DCTP-11-26",
    doi = "10.1007/JHEP07(2011)114",
    journal = "JHEP",
    volume = "07",
    pages = "114",
    year = "2011"
}

@article{Haack:2006cy,
    author = "Haack, Michael and Krefl, Daniel and Lust, Dieter and Van Proeyen, Antoine and Zagermann, Marco",
    title = "{Gaugino Condensates and D-terms from D7-branes}",
    eprint = "hep-th/0609211",
    archivePrefix = "arXiv",
    reportNumber = "LMU-ASC-64-06, MPP-2006-123, KUL-TF-06-24",
    doi = "10.1088/1126-6708/2007/01/078",
    journal = "JHEP",
    volume = "01",
    pages = "078",
    year = "2007"
}

@article{Cicoli:2008gp,
    author = "Cicoli, M. and Burgess, C. P. and Quevedo, F.",
    title = "{Fibre Inflation: Observable Gravity Waves from IIB String Compactifications}",
    eprint = "0808.0691",
    archivePrefix = "arXiv",
    primaryClass = "hep-th",
    reportNumber = "DAMTP-2008-59",
    doi = "10.1088/1475-7516/2009/03/013",
    journal = "JCAP",
    volume = "03",
    pages = "013",
    year = "2009"
}

@article{Wall:1966rcd,
    author = "Wall, C. T. C.",
    title = "{Classification problems in differential topology. V}",
    doi = "10.1007/BF01389738",
    journal = "Invent. Math.",
    volume = "1",
    number = "4",
    pages = "355--374",
    year = "1966"
}

@article{Nikulin1980,
  author  = {Nikulin, V. V.},
  title   = {Finite automorphism groups of K{\"a}hler {K3} surfaces},
  journal = {Transactions of the Moscow Mathematical Society},
  volume  = {38},
  pages   = {71--135},
  year    = {1980}
}

@article{Nikulin1983,
  author  = {Nikulin, V. V.},
  title   = {Factor groups of groups of automorphisms of hyperbolic forms with respect to subgroups generated by 2-reflections. Algebro-geometric applications},
  journal = {Journal of Soviet Mathematics},
  volume  = {22},
  number  = {4},
  pages   = {1401--1475},
  year    = {1983},
  doi     = {10.1007/BF01094757}
}

@article{Cvetic:2010ky,
    author = "Cvetic, Mirjam and Garcia-Etxebarria, Inaki and Halverson, James",
    title = "{On the computation of non-perturbative effective potentials in the string theory landscape: IIB/F-theory perspective}",
    eprint = "1009.5386",
    archivePrefix = "arXiv",
    primaryClass = "hep-th",
    reportNumber = "UPR-1219-T, NSF-KITP-10-124",
    doi = "10.1002/prop.201000093",
    journal = "Fortsch. Phys.",
    volume = "59",
    pages = "243--283",
    year = "2011"
}

@article{Lust:2013kt,
    author = {L{\"u}st, Dieter and Zhang, Xu},
    title = "{Four Kahler Moduli Stabilisation in type IIB Orientifolds with K3-fibred Calabi-Yau threefold compactification}",
    eprint = "1301.7280",
    archivePrefix = "arXiv",
    primaryClass = "hep-th",
    reportNumber = "LMU-ASC-06-13, MPP-2013-22",
    doi = "10.1007/JHEP05(2013)051",
    journal = "JHEP",
    volume = "05",
    pages = "051",
    year = "2013"
}

@article{Cicoli:2012vw,
    author = "Cicoli, Michele and Krippendorf, Sven and Mayrhofer, Christoph and Quevedo, Fernando and Valandro, Roberto",
    title = "{D-Branes at del Pezzo Singularities: Global Embedding and Moduli Stabilisation}",
    eprint = "1206.5237",
    archivePrefix = "arXiv",
    primaryClass = "hep-th",
    reportNumber = "DAMTP-2012-47, ZMP-HH-12-10",
    doi = "10.1007/JHEP09(2012)019",
    journal = "JHEP",
    volume = "09",
    pages = "019",
    year = "2012"
}

@article{Bianchi:2011qh,
    author = "Bianchi, Massimo and Collinucci, Andres and Martucci, Luca",
    title = "{Magnetized E3-brane instantons in F-theory}",
    eprint = "1107.3732",
    archivePrefix = "arXiv",
    primaryClass = "hep-th",
    doi = "10.1007/JHEP12(2011)045",
    journal = "JHEP",
    volume = "12",
    pages = "045",
    year = "2011"
}

@article{Bianchi:2012pn,
    author = "Bianchi, Massimo and Collinucci, Andres and Martucci, Luca",
    editor = "Sorokin, Dimitri and Luest, Dieter",
    title = "{Freezing E3-brane instantons with fluxes}",
    eprint = "1202.5045",
    archivePrefix = "arXiv",
    primaryClass = "hep-th",
    reportNumber = "ROM2F-2012-02, CERN-PH-TH-2012-050",
    doi = "10.1002/prop.201200030",
    journal = "Fortsch. Phys.",
    volume = "60",
    pages = "914--920",
    year = "2012"
}

@article{Louis:2012nb,
    author = "Louis, Jan and Rummel, Markus and Valandro, Roberto and Westphal, Alexander",
    title = "{Building an explicit de Sitter}",
    eprint = "1208.3208",
    archivePrefix = "arXiv",
    primaryClass = "hep-th",
    reportNumber = "DESY-12-146, ZMP-HH-12-16",
    doi = "10.1007/JHEP10(2012)163",
    journal = "JHEP",
    volume = "10",
    pages = "163",
    year = "2012"
}

@article{Berg:2005ja,
    author = "Berg, Marcus and Haack, Michael and Kors, Boris",
    title = "{String loop corrections to Kahler potentials in orientifolds}",
    eprint = "hep-th/0508043",
    archivePrefix = "arXiv",
    reportNumber = "MIT-CTP-3671, NSF-KITP-2005-55",
    doi = "10.1088/1126-6708/2005/11/030",
    journal = "JHEP",
    volume = "11",
    pages = "030",
    year = "2005"
}

@article{cox2011toric,
  title={Toric Varieties},
  author={Cox, David and Little, John and Schenck, Henry},
  journal={Graduate Studies in Mathematics},
  volume={124},
  year={2011},
  publisher={American Mathematical Society}
}

@article{Braun:2017nhi,
    author = "Braun, Andreas P. and Long, Cody and McAllister, Liam and Stillman, Michael and Sung, Benjamin",
    title = "{The Hodge Numbers of Divisors of Calabi-Yau Threefold Hypersurfaces}",
    eprint = "1712.04946",
    archivePrefix = "arXiv",
    primaryClass = "hep-th",
    month = "12",
    year = "2017"
}
\end{document}